\documentclass[11pt]{article}

\usepackage{amsmath,amsfonts,amssymb,bbm,epsfig}
\usepackage[textwidth=14.7cm,top=30mm,bottom=35mm]{geometry}
 \usepackage{tabularx}
\usepackage{url}

\numberwithin{equation}{section} 

\renewcommand\slash[1]{\not \! #1}

\begin{document}

 \newcolumntype{L}[1]{>{\raggedright\arraybackslash}p{#1}}
 \newcolumntype{C}[1]{>{\centering\arraybackslash}p{#1}}
 \newcolumntype{R}[1]{>{\raggedleft\arraybackslash}p{#1}}

%\renewcommand{\theequation}{\thesection.\arabic{equation}}
%
%%%%%%%%%%%%%%%%%%%%%%%%%%%%%%%%%%%%%%%%%%
\def\be{\begin{equation}}
\def\ee{\end{equation}}
\def\bea{\begin{eqnarray}}
\def\eea{\end{eqnarray}}
 \newcommand{\ba}{\begin{eqnarray}}
 \newcommand{\ea}{\end{eqnarray}}
\def\rarr{\rightarrow}
\def\nn{\nonumber}
\def\fr{\frac}
\renewcommand\slash[1]{\not \! #1}
\newcommand\qs{\!\not \! q}
\def\del{\partial}
\def\gam{\gamma}
\newcommand\vphi{\varphi}
\def\tr{\mbox{tr}\,}
\newcommand\nin{\noindent}

\def\del{\partial}
\def\pbar{\bar{p}}
%%%%%%%%%%%%%%%%%%%%%%%%%%%%%%%%%%%%%%%%
\begin{titlepage}
\begin{flushright}
MPP-2018-301\\
DESY 19-002
\end{flushright}
%\vfill

\vspace{0.6cm}
\begin{center}
\boldmath
{\LARGE{\bf The Tensor Pomeron and}}\\[.2cm]
{\LARGE{\bf Low-$x$ Deep Inelastic Scattering}}\\
\unboldmath
\end{center}
%\vspace{1.2cm}
\vspace{0.6cm}
\begin{center}
{\bf \Large
Daniel Britzger\,$^{a,1}$, 
Carlo Ewerz\,$^{b,c,d,2}$, Sasha Glazov\,$^{e,3}$,\\
Otto Nachtmann\,$^{b,4}$, Stefan Schmitt\,$^{e,5}$}
\end{center}
\vspace{.2cm}
\begin{center}
$^a$
{\sl
Max-Planck-Institut f\"ur Physik,\\
F\"ohringer Ring 6, D-80805 M\"unchen, Germany}
\\[.5cm]
$^b$
{\sl
Institut f\"ur Theoretische Physik, Universit\"at Heidelberg,\\
Philosophenweg 16, D-69120 Heidelberg, Germany}
\\[.5cm]
$^c$
{\sl
ExtreMe Matter Institute EMMI, GSI Helmholtzzentrum f\"ur Schwerionenforschung,\\
Planckstra{\ss}e 1, D-64291 Darmstadt, Germany}
\\[.5cm]
$^d$
{\sl 
Frankfurt Institute for Advanced Studies,\\
Ruth-Moufang-Stra{\ss}e 1, D-60438 Frankfurt, Germany}
\\[.5cm]
$^e$
{\sl
Deutsches Elektronen-Synchrotron DESY,\\
Notkestra{\ss}e 85, D-22607 Hamburg, Germany}
\end{center}                                                                
\vfill
\vspace{5em}
\hrule width 5.cm
\vspace*{.5em}
{\small \noindent
$^1$ email: britzger@mpp.mpg.de\\
$^2$ email: C.Ewerz@thphys.uni-heidelberg.de\\
$^3$ email: Alexandre.Glazov@desy.de\\
$^4$ email: O.Nachtmann@thphys.uni-heidelberg.de\\
$^5$ email: Stefan.Schmitt@desy.de
}
\newpage
\thispagestyle{empty}
\begin{abstract}
\noindent
The tensor-pomeron model is applied to low-$x$ deep-inelastic 
lepton-nucleon scattering and photoproduction. 
We consider c.\,m.\ energies in the 
range 6 - 318 GeV and $Q^2 < 50 \,\mbox{GeV}^2$. 
In addition to the soft tensor pomeron, 
which has proven quite successful for the description of soft hadronic 
high-energy reactions, we include a hard tensor pomeron. 
We also include $f_2$-reggeon exchange which turns out to be 
particularly relevant for real-photon-proton scattering at c.\,m.\ energies 
in the range up to 30 GeV. 
The combination of these exchanges permits a description 
of the absorption cross sections of real and virtual photons on the proton 
in the same framework. 
In particular, a detailed comparison 
of this two-tensor-pomeron model with the latest HERA data for $x < 0.01$ is made. 
Our model gives a very good description of the transition 
from the small-$Q^2$ regime where the real or virtual photon behaves 
hadron-like to the large-$Q^2$ regime where hard scattering dominates. 
Our fit allows us, for instance, a determination 
of the intercepts of the hard pomeron as $1.3008 \,({}^{+73}_{-84})$, of the 
soft pomeron as $1.0935\, ({}^{+76}_{-64})$, and of the $f_2$ reggeon. 
We find that in photoproduction the hard pomeron does not contribute 
within the errors of the fit. We show that assuming a vector instead of a 
tensor character for the pomeron leads to the conclusion that it must 
decouple in real photoproduction. 
\end{abstract}
\end{titlepage}

%\tableofcontents

\section{Introduction}
\label{Introduction}

In this article we will be concerned with the structure functions 
of deep-inelastic electron- and positron-proton scattering (DIS).
They are given by the absorptive part of the forward virtual Compton 
amplitude, that is, the amplitude for the elastic scattering of a 
virtual photon on a proton. The high-energy, or small 
Bjorken-$x$, behaviour of these structure functions has first been 
observed experimentally in \cite{Adloff:1997mf,Breitweg:1997hz} 
and has since then been subject of extensive experimental and 
theoretical research; see for example \cite{Devenish:2004pb} 
for a review. 

It is not our aim here to address the various theoretical approaches 
to the small-$x$ structure of the proton. We shall concentrate on a 
particular aspect of the approach based on Regge theory. In Regge theory, 
elastic hadron-hadron scattering is dominated, at high energies 
and small angles, by pomeron exchange. The same applies to total 
cross sections which, by the optical theorem, are related to the 
forward scattering amplitudes. For reviews of pomeron physics 
see \cite{Donnachie:2002en,ref7,ref8,Barone:2002cv}. 
In the application of Regge theory the pomeron has often been 
assumed to be describable as a vector exchange. For example, 
the two-pomeron approach to low-$x$ DIS introduced in 
\cite{Donnachie:1998gm,Donnachie:2001he,Donnachie:2004pi} 
makes use of two vector pomerons, a hard one and a soft one. 
However, the assumption of a vector character for the pomeron 
has problems, as we shall also demonstrate again in the present paper. 
In \cite{Ewerz:2013kda} it has been argued that in general 
the pomeron should be a tensor pomeron, that is, an exchange object 
which can be treated effectively as a rank-2 symmetric tensor. 
In the present study we use a two-pomeron model with 
two tensor pomerons, a hard one and a soft one, instead of two 
vector pomerons.\footnote{Obviously, one could add further 
pomeron exchanges with various intercepts, or choose one 
pomeron with a scale-dependent intercept; see for example 
\cite{Dosch:2015jua}. In the present study we will consider 
only the two-pomeron model.} 
With this model we perform a fit to the available data for photoproduction 
in the centre-of-mass energy range $6 < \sqrt{s} < 209 \,\mbox{GeV}$ 
and to the latest HERA data for low-$x$
deep-inelastic lepton-nucleon scattering for centre-of-mass 
energies in the range 225 - 318 GeV and for $x < 0.01$. 
As we will see, the exchange of a tensor pomeron involves 
for the virtual photon $\gamma^*$-pomeron coupling two 
functions which are in essence related to the $\gamma^*$-proton 
cross sections $\sigma_T$ and $\sigma_L$, respectively. 
It is a special aim of our investigations to fit with our model 
simultaneously $\sigma_T$ and $\sigma_L$. Given the 
large kinematic range and the quality of the experimental data 
a successful fit using tensor pomerons will therefore be a nontrivial result. 

In \cite{Ewerz:2013kda} the tensor 
pomeron was introduced for soft reactions and many of its 
properties were derived from comparisons with experiment. 
Further applications of the tensor-pomeron concept were 
given for photoproduction of pion pairs in \cite{Bolz:2014mya} 
and for a number of exclusive central-production reactions in 
\cite{Lebiedowicz:2013ika,Lebiedowicz:2014bea,Lebiedowicz:2016ioh,Lebiedowicz:2016zka,Lebiedowicz:2016ryp,Klusek-Gawenda:2017lgt,Lebiedowicz:2018eui}. 
In \cite{Ewerz:2016onn} the helicity structure of small-$|t|$ proton-proton 
elastic scattering was calculated in three models for the pomeron: 
tensor, vector, and scalar. Comparison with experiment \cite{Adamczyk:2012kn}
left only the tensor pomeron as a viable option. In the present 
paper we go beyond the regime of soft scattering, to DIS. 
In accord with \cite{Donnachie:1998gm} we shall now consider two 
pomerons, but of the tensor type: a soft one, $\mathbbm{P}_1$, 
which is identical to the tensor pomeron of \cite{Ewerz:2013kda}, 
and a hard one, $\mathbbm{P}_0$. From fits to 
the structure functions of DIS, going down in $Q^2$ to 
photoproduction $(Q^2=0)$, we shall be able to extract the 
properties of $\mathbbm{P}_0$ and $\mathbbm{P}_1$ and 
their couplings to virtual photons. Since we shall consider data 
going down in c.\,m.\ energy to around 6 GeV we shall also include 
$f_2$ reggeon ($f_{2R}$) exchange in the theoretical description. 
Following \cite{Ewerz:2013kda}, $f_{2R}$ exchange will also be 
treated as the effective exchange of a symmetric tensor of rank 2. 

A particular aspect relevant to our study concerns real Compton scattering. 
In this regard we discuss further clear evidence against the 
hypothesis that the pomeron has vector character. We show 
that a vector pomeron necessarily decouples in real Compton 
scattering. A tensor pomeron, in contrast, gives a non-vanishing 
contribution and can successfully describe the data. 

Our paper is organised as follows. In section \ref{sec:kin-genrel} 
we review the kinematics of DIS and some general relations for 
the DIS structure functions. In section \ref{sec:structure-in-2Pmodel} 
our ansatz for the exchange of the tensor pomerons and the 
$f_{2R}$ reggeon is introduced. The resulting expressions for 
the real and virtual photon-proton cross sections are derived. 
The vector pomeron and its decoupling in real Compton scattering 
are discussed in section \ref{sec:ComptonAmplitude}. 
Section \ref{sec:CompExp} presents the comparison of our tensor-pomeron 
model with experimental data. We discuss our findings in Section 
\ref{sec:discussion}. 
Section \ref{sec:conclusions} gives our conclusions. Appendix \ref{appA} 
lists the effective propagators and vertices for the two pomerons 
and for the $f_{2R}$ reggeon. In appendix \ref{appVector} we discuss 
the formulae for the case of a vector pomeron. 
In appendix \ref{appParam} we present the parametrisations for the 
coupling functions occurring in our approach. 
In appendices \ref{appfit}, \ref{appfitres}, and \ref{appalt} we give 
details of our fit procedure and of the fit results. 

\section{Kinematics and general relations for structure functions in DIS}
\label{sec:kin-genrel}

\begin{figure}[t]
\begin{center}
\includegraphics[width=0.45\textwidth]{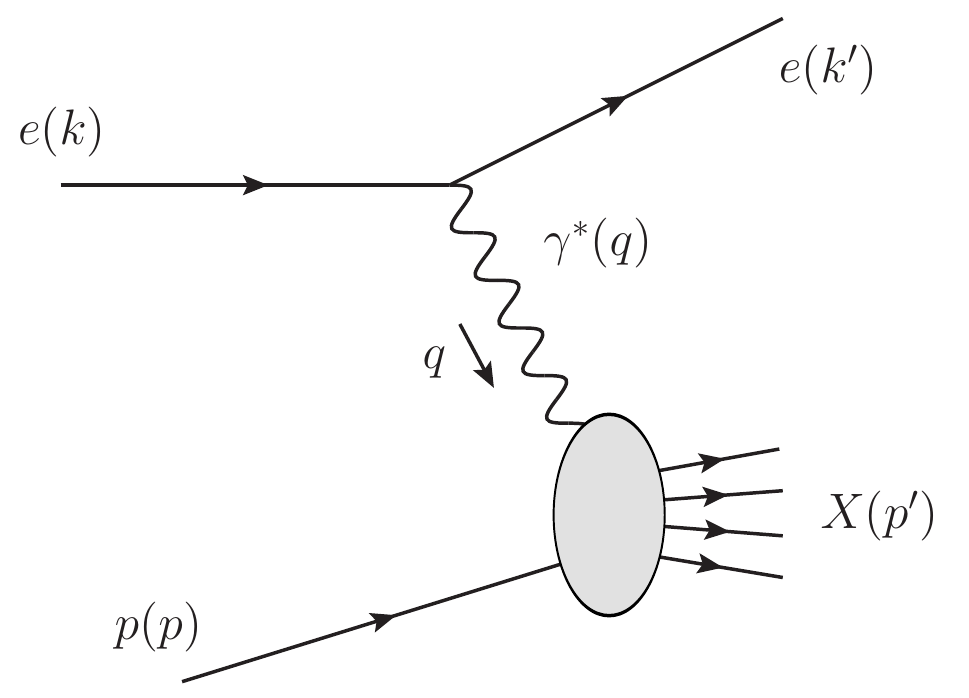}
\caption{Deep inelastic lepton-proton scattering 
\label{figDIS}}
\end{center}
\end{figure}

We want to consider electron- and 
positron-proton inelastic scattering 
(fig.\ \ref{figDIS}) 
\be\label{1.1}
e(k) + p(p) \longrightarrow e(k') + X(p') \,.
\ee
The kinematic variables for the reaction \eqref{1.1} are standard; 
see for instance \cite{Nachtmann:1990ta}: 
\be\label{1.2}
\begin{split}
s &= (p+k)^2 \,,
\\
q &= k-k' \,,
\\
Q^2 &= - q^2 \,,
\\
W^2&=p'^2 = (p+q)^2 \,,
\\
\nu &= \frac{p\cdot q}{m_p} = \frac{W^2 + Q^2 - m_p^2}{2m_p} \,,
\\
x &= \frac{Q^2}{2m_p \nu} = \frac{Q^2}{W^2+Q^2-m_p^2} \,,
\\
y &= \frac{p\cdot q}{p \cdot k} = \frac{W^2 + Q^2 - m_p^2}{s-m_p^2} \,.
\end{split}
\ee
Furthermore, we define the ratio $\epsilon$ of longitudinal and 
transverse polarisation strengths of the virtual photon 
\be\label{1.3}
\epsilon = \frac{2(1-y) - y^2 \delta(W^2,Q^2)}{1+(1-y)^2 + y^2 \delta(W^2,Q^2)} 
\ee
where
\be\label{1.4}
\delta(W^2,Q^2) = \frac{2m_p^2Q^2}{(W^2 + Q^2 -m_p^2)^2} \,.
\ee
For given $W^2>m^2_p$ and $Q^2\geq 0$ the kinematic limits 
for $y$ and $\epsilon$ are 
\be\label{1.5}
0 \le y \le \frac{2}{1+ \sqrt{1+2 \delta(W^2,Q^2)}}
\ee
corresponding to 
\be\label{1.6}
1 \ge \epsilon \ge 0 \,.
\ee
Clearly, for $W^2>m^2_p$ the value $y=0$ ($\epsilon=1$) can only be reached 
for $s\to\infty$; see \eqref{1.2}.

The reaction effectively studied 
in DIS is the absorption of the virtual photon on the proton; 
see fig.\ \ref{figDIS}. 
\begin{figure}[tb]
\begin{center}
\includegraphics[width=0.55\textwidth]{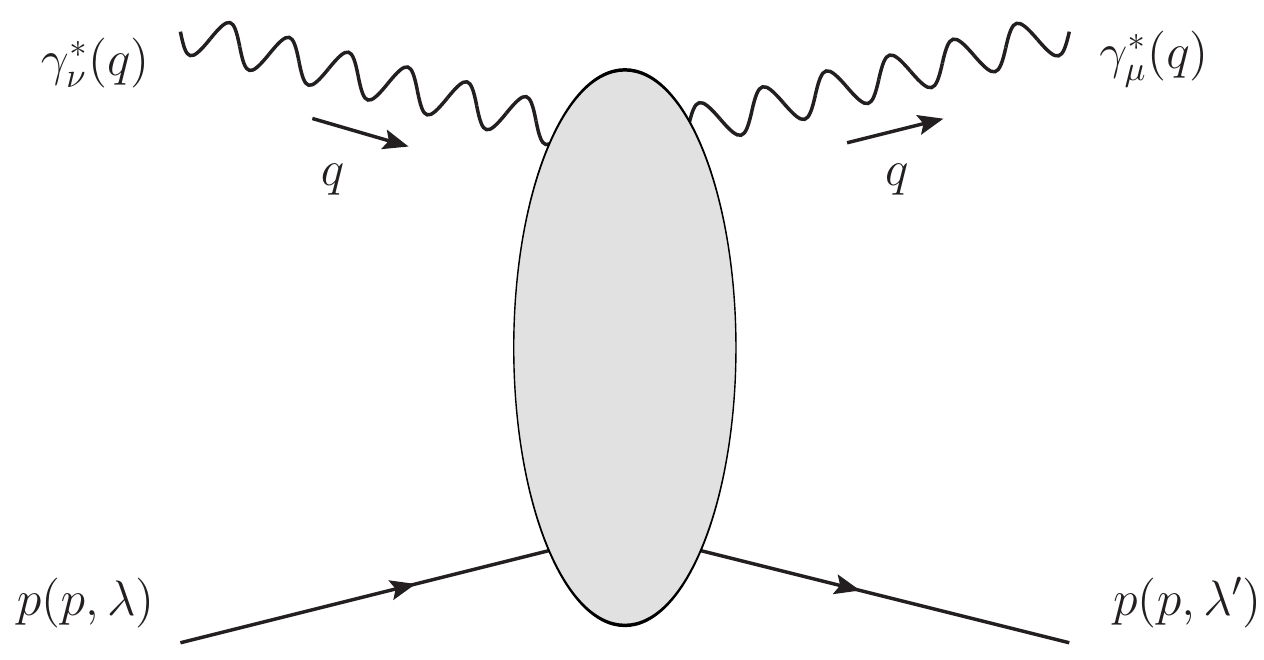}
\caption{Forward virtual Compton scattering on a proton
\label{figvcompton}}
\end{center}
\end{figure}
The total $\gamma^* p$ absorption cross sections 
are related to the absorptive parts of the virtual forward Compton 
scattering amplitude. 
In the following, we shall therefore study the forward virtual 
Compton scattering on a proton, see fig.\ \ref{figvcompton},
\be\label{2.1}
\gamma_\nu^* (q) + p(p,\lambda) \longrightarrow \gamma_\mu^* (q) + p (p,\lambda') \,.
\ee
The momenta are indicated in brackets and $\lambda,\lambda'\in \{1/2,-1/2\}$ 
are the helicity indices of the protons. We define the amplitude for reaction \eqref{2.1} as
\be\label{2.2}
\mathcal{M}_{\lambda' \lambda}^{\mu\nu} (p,q) = \frac{i}{2\pi m_p} \int d^4x \, e^{-iqx} 
\langle p(p,\lambda') | \mathrm{T}^* (J^\mu (0) J^\nu (x)) | p (p,\lambda) \rangle \,.
\ee
Here $m_p$ is the proton mass, $\mathrm{T}^*$ denotes the covariantised 
time-ordered product, and $J^\mu(x)$ is the hadronic part of 
the electromagnetic current. The absorptive part of 
$\mathcal{M}^{\mu\nu}_{\lambda'\lambda}$ \eqref{2.2}, averaged 
over the proton helicities, gives the hadronic tensor and the 
structure functions of DIS,
\be\label{2.3}
\begin{split}
W^{\mu\nu} (p,q) =& 
\sum_{\lambda',\lambda} \frac{1}{2} \,\delta_{\lambda' \lambda} \,
\frac{1}{2i} \left[ \mathcal{M}_{\lambda' \lambda}^{\mu\nu} (p,q) - 
\left(\mathcal{M}_{\lambda \lambda'}^{\nu\mu} (p,q) \right)^* \right] 
\\
=&{}\: W_1(\nu,Q^2) \left( - g^{\mu\nu} + \frac{q^\mu q^\nu}{q^2} \right) 
\\
&+ \frac{1}{m_p^2} \, W_2(\nu,Q^2) \left( p^\mu - \frac{p \cdot q}{q^2} \, q^\mu \right)
\left( p^\nu - \frac{p \cdot q}{q^2} \, q^\nu \right) .
\end{split}
\ee

We shall also use the total $\gamma^*p$ absorption cross sections 
$\sigma_T$ and $\sigma_L$ for transversely and longitudinally 
polarised virtual photons. With $e>0$ the proton charge and 
Hand's convention for the flux factor \cite{Hand:1963bb} these read
\be\label{2.4}
\begin{split}
\sigma_T (W^2,Q^2) &= \frac{2 \pi m_p}{W^2 - m_p^2} \, e^2 W_1(\nu,Q^2) \,,
\\
\sigma_L (W^2,Q^2) &= \frac{2 \pi m_p}{W^2 - m_p^2} \, e^2 
\left[ W_2(\nu,Q^2) \frac{\nu^2 + Q^2}{Q^2} - W_1(\nu,Q^2) \right] \,.
\end{split}
\ee

\section{Structure functions in the tensor-pomeron approach}
\label{sec:structure-in-2Pmodel}

\begin{figure}[ht]
\begin{center}
\includegraphics[width=0.55\textwidth]{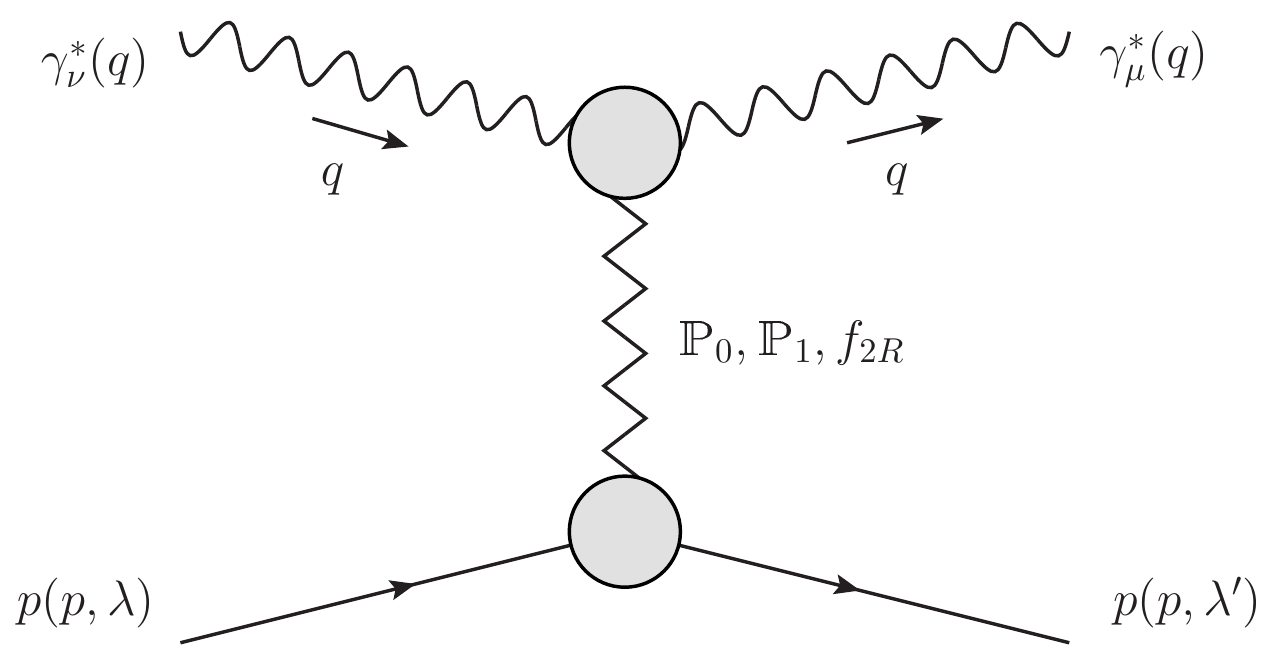}
\caption{Low-$x$ forward virtual Compton scattering with exchange of 
the soft ($\mathbbm{P}_1$) and hard ($\mathbbm{P}_0$) pomeron 
plus the $f_{2R}$ reggeon. 
\label{figcomptonpomex}}
\end{center}
\end{figure}

We shall now assume that for large $W^2$, respectively small $x$, 
the virtual Compton amplitude \eqref{2.2} is dominated by the exchange 
of the two pomerons, $\mathbbm{P}_0$ and $\mathbbm{P}_1$, plus the 
$f_{2R}$ reggeon; see fig.\ \ref{figcomptonpomex}. 
In order to calculate the diagram shown there we need the 
effective propagators for $\mathbbm{P}_0$ and $\mathbbm{P}_1$ 
as well as the vertex functions 
$\mathbbm{P}_j pp$ and ${\mathbbm P}_j\gamma^*\gamma^*$ ($j=0,1$), 
and the analogous quantities for $f_{2R}$. 
Our ans\"atze for these quantities are listed in appendix \ref{appA}. 
It is now straightforward to calculate the analytic expression corresponding 
to the diagram of fig.\ \ref{figcomptonpomex}. Since all three exchanges 
are tensor exchanges, the resulting expressions have a similar structure. 
We find 
\be\label{2.5}
\begin{split}
i \, 2\pi \, m_p e^2 \mathcal{M}_{\lambda' \lambda}^{\mu\nu} (p,q) =& 
\sum_{j=0,1} g^{\mu\mu'} g^{\nu \nu'} 
i \Gamma_{\mu' \nu' \kappa \rho}^{(\mathbbm{P}_j \gamma^*\gamma^*)} (q,q) \,
i \Delta^{(P_j)\,\kappa \rho, \kappa' \rho'} (W^2,0) \, \\
& \qquad
\times \bar{u}(p,\lambda')\, i\Gamma_{\kappa'\rho'}^{(\mathbbm{P}_j pp)} (p,p)\, u(p,\lambda) 
\\
& + g^{\mu\mu'} g^{\nu \nu'}  i \Gamma_{\mu' \nu' \kappa \rho}^{(f_{2R} \gamma^*\gamma^*)} (q,q) \,
i \Delta^{(f_{2R})\,\kappa \rho, \kappa' \rho'} (W^2,0) \, \\
& \qquad
\times \bar{u}(p,\lambda')\, i\Gamma_{\kappa'\rho'}^{(f_{2R} pp)} (p,p) \,u(p,\lambda) \,.
\end{split}
\ee
With the expressions from appendix \ref{appA} we obtain
\be\label{2.6}
\begin{split}
\mathcal{M}_{\lambda' \lambda}^{\mu\nu} (p,q) &= \frac{1}{2\pi m_p} \,
\delta_{\lambda'\lambda} \sum_{j=0,1,2} 
\left[ 2 \hat{a}_j(Q^2) \,\Gamma^{(0) \mu\nu \kappa \rho} (q,-q) 
- \hat{b}_j(Q^2) \,\Gamma^{(2) \mu\nu \kappa \rho} (q,-q) \right]
\\
&\quad \times 
(- i \, 3\beta_{jpp}) (-i W^2 \tilde{\alpha}'_j)^{\epsilon_j} 
\frac{1}{2 W^2} \,(4 p_\kappa p_\rho - g_{\kappa \rho} m_p^2) \,.
\end{split}
\ee
The meaning of the quantities occurring here and in the following 
is summarised in table \ref{tab1}. The detailed behaviour of the 
$\gamma^* \gamma^*$ coupling functions is not predicted by 
the model. They are assumed to be smooth functions of $Q^2$ and 
will be parametrised with the help of spline functions. 
Note that quantities with indices 
$j=0,1,$ and $2$ always refer to the hard pomeron, the soft pomeron, 
and the $f_{2R}$ reggeon, respectively. The tensor functions 
$\Gamma^{(l)\mu\nu\kappa\rho}$ ($l=0,2$) are defined in 
\eqref{A.8a}, \eqref{A.8b}. 
\begin{table}[t]\centering
\renewcommand{\arraystretch}{1.2}
\begin{tabular}{c|c|c|c}
& hard pomeron $\mathbbm{P}_0$ & soft pomeron $\mathbbm{P}_1$ & reggeon $f_{2R}$ \\
\hline
intercept & $\alpha_0(0)=1+\epsilon_0$ & $\alpha_1(0)=1+\epsilon_1$ & $\alpha_2(0)=1+\epsilon_2$ \\
\hline
slope parameter & $\alpha'_0$ & $\alpha'_1$ & $\alpha'_2$ \\
\hline
$W^2$ parameter & $\tilde{\alpha}'_0$& $\tilde{\alpha}'_1$& $\tilde{\alpha}'_2$ \\
\hline
$pp$ coupling parameter & $\beta_{0pp}$ & $\beta_{1pp}$ & $\beta_{2pp}$ \\
\hline
$\gamma^* \gamma^*$ coupling functions & $\hat{a}_0(Q^2)$, $\hat{b}_0(Q^2)$ 
& $\hat{a}_1(Q^2)$, $\hat{b}_1(Q^2)$ & $\hat{a}_2(Q^2)$, $\hat{b}_2(Q^2)$ 
\end{tabular}
\caption{Notation for the parameters of our ansatz with 
hard and soft pomeron and $f_{2R}$ reggeon exchange. 
The propagators and vertices containing these parameters 
are given in detail in appendix \ref{appA}. 
\label{tab1}}
\end{table}
Using \eqref{2.3} we get from \eqref{2.6} 
\be\label{2.7}
\begin{split}
W_{\mu\nu}(p,q) =& \,\frac{1}{2\pi m_p W^2} \sum_{j=0,1,2}  
3 \beta_{jpp} (W^2 \tilde{\alpha}'_j)^{\epsilon_j} \cos\left(\frac{\pi}{2} \, \epsilon_j\right)
\\
&{} \times \left\{ \left(-g_{\mu\nu} + \frac{q_\mu q_\nu}{q^2} \right) 
\Big[ \hat{b}_j(Q^2) \left(4 (p\cdot q)^2 - 2 q^2 m_p^2 \right) \right. 
\\
&  \qquad \qquad \qquad \qquad \quad 
- 2 \hat{a}_j(Q^2) (-q^2) \left( 4 (p\cdot q)^2 - q^2 m_p^2 \right) \Big]
\\
&{} \qquad \left.
+ \left( p^\mu - \frac{p \cdot q}{q^2} \, q^\mu \right)
\left( p^\nu - \frac{p \cdot q}{q^2} \, q^\nu \right) 
(-4q^2) \hat{b}_j(Q^2) 
\right\} ,
\end{split}
\ee
such that 
\be\label{2.8}
\begin{split}
W_1 (\nu,Q^2) =& \,\frac{1}{2\pi m_p W^2} \sum_{j=0,1,2}  
3 \beta_{jpp} (W^2 \tilde{\alpha}'_j)^{\epsilon_j} \cos\left(\frac{\pi}{2} \, \epsilon_j\right)
\\
& \times \left[ \hat{b}_j(Q^2) \left(4 (p\cdot q)^2 + 2 Q^2 m_p^2 \right) 
- 2 Q^2 \hat{a}_j(Q^2) \left(4 (p\cdot q)^2 + Q^2 m_p^2 \right) \right]
\end{split}
\ee
and 
\be\label{2.9}
W_2 (\nu,Q^2) = \,\frac{m_p}{2\pi W^2} \sum_{j=0,1,2} 
3 \beta_{jpp} (W^2 \tilde{\alpha}'_j)^{\epsilon_j} \cos\left(\frac{\pi}{2} \, \epsilon_j\right)
4Q^2 \hat{b}_j(Q^2) \,.
\ee
Writing $W_1$ \eqref{2.8} in terms of the variables $Q^2$ and $W^2$ we get 
\be\label{2.10}
\begin{split}
W_1 (\nu,Q^2) =& \, \frac{(W^2 - m_p^2)^2}{2 \pi m_p W^2} \sum_{j=0,1,2} 
3 \beta_{jpp} (W^2 \tilde{\alpha}'_j)^{\epsilon_j} \cos\left(\frac{\pi}{2} \, \epsilon_j\right)
\\
& \times \left\{ 
\hat{b}_j(Q^2) \left[1+ \frac{2Q^2}{W^2-m_p^2} + \frac{Q^2 (Q^2+2m_p^2)}{(W^2 - m_p^2)^2}\right]
\right.
\\
& \qquad
\left.
- 2 Q^2 \hat{a}_j(Q^2) \left[1+ \frac{2Q^2}{W^2-m_p^2} + \frac{Q^2 (Q^2+m_p^2)}{(W^2 - m_p^2)^2}\right]
\right\} .
\end{split}
\ee
From \eqref{2.9} and \eqref{2.10} we get for $\sigma_T$ and 
$\sigma_L$ \eqref{2.4} with $\alpha_{\rm em}=e^2/(4\pi)$, the fine structure constant, 
\begin{align}
\label{2.11}
\sigma_T(W^2,Q^2)=&\, 4\pi \alpha_{\rm em} \, \frac{W^2-m_p^2}{W^2} \sum_{j=0,1,2} 
3 \beta_{jpp} (W^2 \tilde{\alpha}'_j)^{\epsilon_j} \cos\left(\frac{\pi}{2} \, \epsilon_j\right)
\nn \\
& \times \left\{ 
\hat{b}_j(Q^2) \left[1+ \frac{2Q^2}{W^2-m_p^2} + \frac{Q^2 (Q^2+2m_p^2)}{(W^2 - m_p^2)^2}\right]
\right.
\\
& \qquad
\left.
- 2 Q^2 \hat{a}_j(Q^2) \left[1+ \frac{2Q^2}{W^2-m_p^2} + \frac{Q^2 (Q^2+m_p^2)}{(W^2 - m_p^2)^2}\right]
\right\} ,
\nn\\
\label{2.12}
\sigma_L(W^2,Q^2)=&\, 4\pi \alpha_{\rm em} \, \frac{W^2-m_p^2}{W^2}\,Q^2 \sum_{j=0,1,2} 
3 \beta_{jpp} (W^2 \tilde{\alpha}'_j)^{\epsilon_j} \cos\left(\frac{\pi}{2} \, \epsilon_j\right)
\\
& \times \left\{ 
2 \hat{a}_j(Q^2) \left[1+ \frac{2Q^2}{W^2-m_p^2} + \frac{Q^2 (Q^2+m_p^2)}{(W^2 - m_p^2)^2}\right]
+ \hat{b}_j(Q^2) \frac{2m_p^2}{(W^2 - m_p^2)^2}
\right\} .
\nn
\end{align}
From \eqref{2.11} and \eqref{2.12} we finally get for the structure functions 
$F_2=\nu W_2$ and $F_L$
\begin{align}
\label{2.13}
F_2(W^2,Q^2) 
=& \, \frac{Q^2}{4\pi^2 \alpha_{\rm em}} (1-x) \left[1+2 \delta(W^2,Q^2)\right]^{-1} 
\left[\sigma_T (W^2,Q^2) + \sigma_L(W^2,Q^2) \right]
\nn\\
=& \, \frac{Q^2}{\pi} \, (1-x) \left[1+2 \delta(W^2,Q^2)\right]^{-1} 
\nn\\
&\, \times 
\frac{W^2-m_p^2}{W^2} \sum_{j=0,1,2} 
3 \beta_{jpp} (W^2 \tilde{\alpha}'_j)^{\epsilon_j} \cos\left(\frac{\pi}{2} \, \epsilon_j\right)
\\
&\, \times 
\hat{b}_j(Q^2) \left[1+ \frac{2Q^2}{W^2-m_p^2} + \frac{Q^2 (Q^2+4m_p^2)}{(W^2 - m_p^2)^2}\right] , 
\nn \\
\label{2.14}
F_L(W^2,Q^2) =& \, \frac{Q^2}{4\pi^2 \alpha_{\rm em}} (1-x) \sigma_L(W^2,Q^2) 
\nn\\
=&\, \frac{Q^4}{\pi} \, (1-x) \, \frac{W^2-m_p^2}{W^2} \sum_{j=0,1,2} 
3 \beta_{jpp} (W^2 \tilde{\alpha}'_j)^{\epsilon_j} \cos\left(\frac{\pi}{2} \, \epsilon_j\right)
\\
&\, \times \left\{ 
2 \hat{a}_j(Q^2) \left[1+ \frac{2Q^2}{W^2-m_p^2} + \frac{Q^2 (Q^2+m_p^2)}{(W^2 - m_p^2)^2}\right]
+ \hat{b}_j(Q^2) \frac{2m_p^2}{(W^2 - m_p^2)^2}
\right\} .
\nn
\end{align}

Let us now discuss our results \eqref{2.6}-\eqref{2.14}. 
We first note that with our ansatz for the soft and hard pomeron plus 
$f_{2R}$ reggeon all gauge-invariance relations for the virtual Compton 
amplitude are satisfied. Indeed, we find from \eqref{2.6} and \eqref{A.9b} 
\be\label{2.15}
\begin{split}
q_\mu \mathcal{M}_{\lambda' \lambda}^{\mu\nu} (p,q) &= 0\,,
\\
q_\nu \mathcal{M}_{\lambda' \lambda}^{\mu\nu} (p,q) &= 0\,.
\end{split}
\ee
Also, $\sigma_L(W^2,Q^2)$ vanishes proportional to $Q^2$ for $Q^2\to 0$, 
whereas $\sigma_T(W^2,0)$ gives the pomeron plus $f_{2R}$ reggeon 
part of the total $\gamma p$ cross section for real photons, 
\be\label{2.16}
\begin{split}
\sigma_T(W^2,0) &= \sigma_{\gamma p} (W^2) 
\\
&= 4\pi \alpha_{\rm em} \, \frac{W^2-m_p^2}{W^2} \sum_{j=0,1,2} 
3 \beta_{jpp} (W^2 \tilde{\alpha}'_j)^{\epsilon_j} \cos\left(\frac{\pi}{2} \, \epsilon_j\right) 
\hat{b}_j(0) \,.
\end{split}
\ee
For this soft process the contributions from the soft pomeron $\mathbbm{P}_1$ 
($j=1$) plus $f_{2R}$ reggeon ($j=2$) are expected to dominate. 

For large $Q^2$, on the other hand, we expect the hard pomeron $\mathbbm{P}_0$ 
to give the main contribution to $\sigma_T$ and $\sigma_L$. For 
\be\label{2.17}
W^2 \gg Q^2 \gg m_p^2 
\ee 
we get, therefore, from \eqref{2.11} and \eqref{2.12} the following 
approximate relations: 
\begin{align}
\label{2.18}
\sigma_T(W^2,Q^2) &\cong \, 4\pi \alpha_{\rm em}\,
3 \beta_{0pp} (W^2 \tilde{\alpha}'_0)^{\epsilon_0} \cos\left(\frac{\pi}{2} \, \epsilon_0\right)
\left[ \hat{b}_0(Q^2) - 2 Q^2 \hat{a}_0(Q^2) \right] ,
\\
\label{2.19}
\sigma_L(W^2,Q^2) &\cong \, 4\pi \alpha_{\rm em} \,Q^2 \, 
3 \beta_{0pp} (W^2 \tilde{\alpha}'_0)^{\epsilon_0} \cos\left(\frac{\pi}{2} \, \epsilon_0\right) 
2 \hat{a}_0(Q^2) \,,
\end{align}
and 
\be\label{2.20}
\frac{\sigma_L(W^2,Q^2)}{\sigma_T(W^2,Q^2)} \cong 
\frac{2 Q^2 \hat{a}_0(Q^2)}{\hat{b}_0(Q^2) - 2 Q^2 \hat{a}_0(Q^2)} \,.
\ee
This shows that in the limit \eqref{2.17} $\sigma_L(W^2,Q^2)$ 
determines the function $\hat a_0(Q^2)$ while 
$\sigma_T(W^2,Q^2)+\sigma_L(W^2,Q^2)$ determines 
the function $\hat b_0 (Q^2)$. 

\section{Compton amplitude and vector pomeron}
\label{sec:ComptonAmplitude}

\begin{figure}[ht]
\begin{center}
\includegraphics[width=0.55\textwidth]{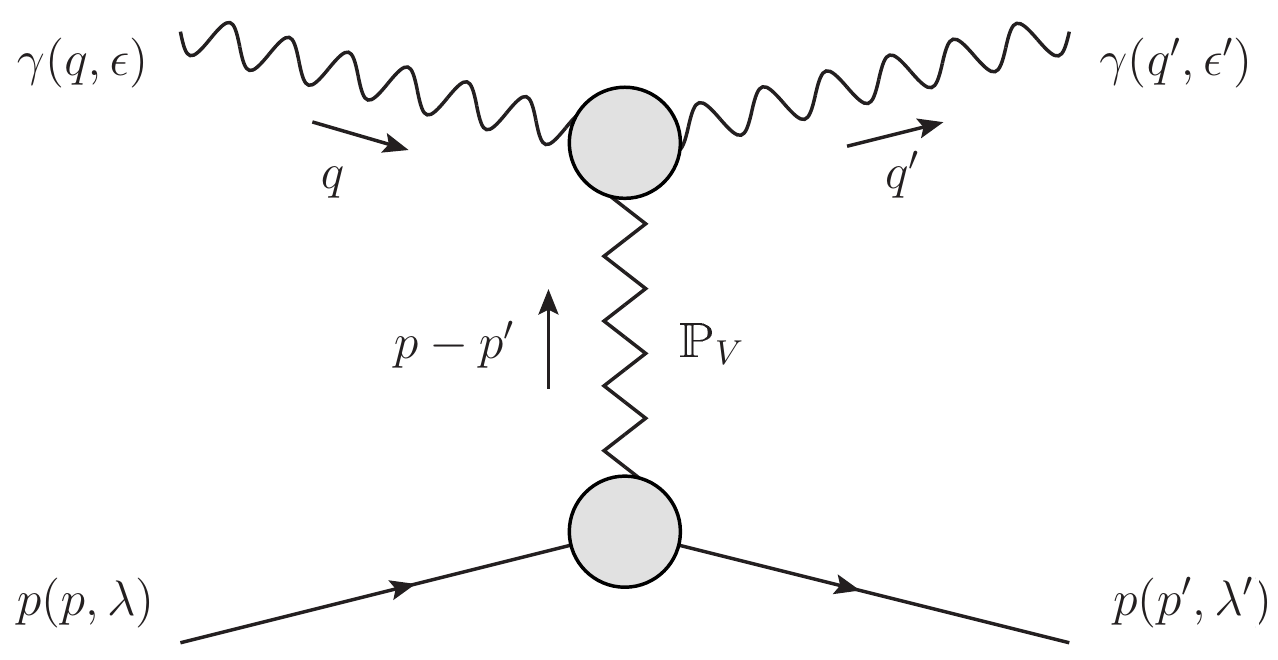}
\caption{Real Compton scattering on a proton with exchange of a vector 
pomeron $\mathbbm{P}_V$. 
\label{fig16}}
\end{center}
\end{figure}

In this section we shall show that for real Compton scattering 
on a proton the exchange of a vector-type pomeron $\mathbbm{P}_V$ gives 
an amplitude that vanishes identically. We investigate the reaction 
\be\label{B.1}
\gamma(q,\varepsilon) + p(p,\lambda) \longrightarrow 
\gamma(q',\varepsilon') + p(p',\lambda') 
\ee
for real photons, $q^2=q'^2=0$, and consider the diagram of fig.\ \ref{fig16} 
with vector pomeron exchange. 
The kinematic variables are the c.\,m.\ energy $W$ and the momentum 
transfer squared, 
\begin{equation}
\begin{split}
\label{W2tdef}
W^2&=(p+q)^2 = (p' + q')^2 \,, \\
t&=(p-p')^2 = (q'-q)^2 \,.
\end{split}
\end{equation}
The $\mathbbm{P}_V pp$ vertex 
and the $\mathbbm{P}_V$ propagator are standard; see e.\,g.\ 
appendix B of \cite{Lebiedowicz:2013ika} and \eqref{B.2}, \eqref{B.3} of the 
present paper. The important task is to find the structure of the 
$\mathbbm{P}_V \gamma\gamma$ vertex. Using the constraints 
of Bose symmetry for the photons, of gauge invariance, and of parity 
conservation in the strong and electromagnetic interactions we derive 
in appendix \ref{appVector} 
for the $\mathbbm{P}_V \gamma\gamma$ vertex function 
the expression  
\be\label{mainB14}
\begin{split}
\Gamma_{\mu\nu\rho}^{(\mathbbm{P}_V \gamma \gamma)} (q',-q) =& {}\,
\hat{A}_2 (t) \left[q'_\mu ( -q'_\nu q_\rho + (q' \cdot q) g_{\nu\rho}) 
- (-q_\mu q'_\rho + (q' \cdot q) g_{\mu\rho}) q_\nu \right]
\\
&- \hat{A}_3 (t) \,q'_\mu q_\nu (q'_\rho - q_\rho) 
\\
&+ \hat{A}_4 (t) \,(- q_\mu q'_\nu + (q' \cdot q) g_{\mu\nu}) (q'_\rho - q_\rho) \,.
\end{split}
\ee
Here $\mu$, $\nu$, and $\rho$ are the Lorentz indices for the outgoing photon, 
the incoming photon, and the vector pomeron $\mathbbm{P}_V$, respectively. 
The $\hat{A}_j(t)$ ($j=2,3,4$) are invariant functions. 

Applying now \eqref{B.2}, \eqref{B.3}, and \eqref{mainB14} to the amplitude for 
reaction \eqref{B.1} we find from the diagram of fig.\ \ref{fig16}  
\be
\label{B.15}
\begin{split}
\langle \gamma(q', \varepsilon'), p(p',\lambda') | \mathcal{T}| \gamma(q,\varepsilon), p(p,\lambda)\rangle^{\mathbbm{P}_V} 
=& - \varepsilon'^{*\mu} \,\Gamma_{\mu\nu\rho}^{(\mathbbm{P}_V \gamma \gamma)} (q', -q) \,
\varepsilon^\nu \Delta^{(\mathbbm{P}_V) \rho \sigma}(W^2,t) 
\\
&\times \bar{u}_{\lambda'} (p') \Gamma_\sigma^{(\mathbbm{P}_V pp)} (p',p) u_\lambda(p)
\\
=&\, 0 \,.
\end{split}
\ee
Here we have used 
\be
\label{B.16a}
q' \cdot \varepsilon' = 0 \,, \qquad q \cdot \varepsilon = 0 
\ee
and 
\be
\label{B.16b}
(q'-q)^\rho \, \bar{u}_{\lambda'}(p') \gamma_\rho u_\lambda (p) = 
(p-p')^\rho \, \bar{u}_{\lambda'}(p') \gamma_\rho u_\lambda (p) = 0 \,.
\ee
The vector pomeron exchange hence gives zero contribution for real 
Compton scattering. 
In particular, this implies that a vector pomeron exchange cannot 
contribute to the total photoabsorption cross section $\sigma_{\gamma p} (W^2)$ 
which is proportional to the absorptive part of the forward Compton 
amplitude. On the other hand, we see from \eqref{2.16} 
that our tensor exchanges give non-zero contributions to $\sigma_{\gamma p}$ 
for $\hat{b}_j(0) \neq 0$. And this will indeed be the case in our fits shown 
in section \ref{sec:CompExp} below. We think that the decoupling of a 
vector pomeron in real Compton scattering is another strong argument 
against treating the pomeron as an effective vector exchange. We note that 
this vector pomeron decoupling is closely related to the famous 
Landau-Yang theorem \cite{Landau:1948kw,Yang:1950rg} which says that a 
massive vector particle cannot decay to two real photons; see appendix \ref{appVector}. 

\section{Comparison with experiment}
\label{sec:CompExp}

In this section we compare our theoretical ansatz for the tensor-pomeron 
and $f_{2R}$-reggeon exchanges, as explained in section 
\ref{sec:structure-in-2Pmodel}, to experiment by making a global fit. 
For this fit we use the HERA inclusive 
DIS data \cite{Abramowicz:2015mha} from four different centre-of-mass 
energies, $\sqrt{s}=225, 251, 300$, and $318\,\mathrm{GeV}$. 
We require 
\be\label{3.1}
Q^2 < 50 \, \mathrm{GeV}^2 \qquad \mathrm{and} \qquad x< 0.01 \,.
\ee
For the photoproduction cross section we use the measurements 
from H1 \cite{Aid:1995bz} at $W=200 \, \mathrm{GeV}$ and ZEUS 
\cite{Chekanov:2001gw} at $W=209 \, \mathrm{GeV}$. In addition, 
we include in the analysis data at intermediate $W$ 
($40\,\mbox{GeV} < W < 150\,\mbox{GeV}$) from 
astroparticle observations \cite{Vereshkov:2003cp} and at low $W$ 
($6\,\mbox{GeV} < W < 19\,\mbox{GeV}$) 
from a tagged-photon experiment at Fermilab \cite{Caldwell:1978yb}. 

The directly measured quantity at HERA is the reduced cross section defined as 
\be\label{3.2}
\sigma_{\rm red} (W^2,Q^2,y) = \frac{Q^4x}{2\pi \alpha_{\rm em}^2 [1 + (1-y)^2]} \, 
\frac{d^2 \sigma}{dx \, dQ^2} (ep \to eX) \,. 
\ee
Expressing this in terms of $\sigma_T$ and $\sigma_L$ \eqref{2.4} we get
\be\label{3.3}
\begin{split}
\sigma_{\rm red} (W^2,Q^2,y) =&{} \, \frac{1+(1-y)^2 + y^2 \delta(W^2,Q^2)}{1+(1-y)^2} 
\, [1+2\delta(W^2,Q^2)]^{-1} \frac{Q^2}{4\pi^2 \alpha_{\rm em}} \, (1-x) 
\\
&
\times 
\left[ \sigma_T(W^2,Q^2) + \sigma_L(W^2,Q^2) - \tilde{f}(W^2,Q^2,y) \,\sigma_L(W^2,Q^2) \right] , 
\end{split}
\ee
where 
\be\label{3.4}
\tilde{f}(W^2,Q^2,y) =1-\epsilon = \frac{y^2[1+2\delta(W^2,Q^2)]}{1+(1-y)^2 + y^2 \delta(W^2,Q^2)} \,.
\ee
Alternatively, we can express $\sigma_{\rm red}$ through the structure functions 
\eqref{2.13}, \eqref{2.14}, 
\be\label{3.5}
\begin{split}
\sigma_{\rm red} (W^2,Q^2,y)=& \,\frac{1+(1-y)^2 + y^2 \delta(W^2,Q^2)}{1+(1-y)^2} 
\\
& \times
\left\{ F_2(W^2,Q^2) - \tilde{f}(W^2,Q^2,y) [1+2\delta(W^2,Q^2)]^{-1} F_L(W^2,Q^2) \right\} . 
\end{split}
\ee

Now we discuss the parameters of our model, cf.\ table \ref{tab1}. 
For the soft pomeron $\mathbbm{P}_1$ 
we take the default values from \eqref{A.2a} for 
\be\label{3.7}
\alpha'_1 = \tilde{\alpha}'_1 = 0.25 \, \mathrm{GeV}^{-2} 
\ee
and leave
\be\label{3.8}
\epsilon_1 = \alpha_1(0) - 1 
\ee
as a fit parameter. The $\mathbbm{P}_1 pp$ coupling parameter $\beta_{1pp}$ is fixed to \eqref{A.7}. 
For our hard pomeron $\mathbbm{P}_0$ we also use, for lack of better 
information, 
\be\label{3.9}
\begin{gathered}
\alpha'_0 = \tilde{\alpha}'_0 = 0.25 \, \mathrm{GeV}^{-2} \,,
\\
\beta_{0pp}= \beta_{1pp} = 1.87 \, \mathrm{GeV}^{-1} 
\end{gathered}
\ee
and leave
\be\label{3.10}
\epsilon_0 = \alpha_0(0) - 1 
\ee
as a fit parameter. The pomeron-$\gamma^* \gamma^*$ coupling functions
\be\label{3.11}
\hat{a}_j(Q^2) \quad \mbox{and} \quad \hat{b}_j(Q^2) \qquad (j=0,1)
\ee
are determined from the fit. These functions are parametrised with 
the help of cubic splines as explained in appendix \ref{appParam}. 
Note that only the products
\be\label{3.12}
\beta_{jpp} \,\hat{a}_j(Q^2) \quad \mbox{and} \quad \beta_{jpp} \,\hat{b}_j(Q^2)
\ee
can be determined from our reaction. 
For $f_{2R}$ exchange we leave $\alpha_2(0) = 1 + \epsilon_2$ as fit 
parameter and use for $\alpha'_2$, $\tilde{\alpha}'_2$, 
$\beta_{2pp}$ the default values from \eqref{A.13a}, \eqref{A.15}, \eqref{A.19a}. 
The function $\hat{b}_2(Q^2)$, 
parametrised according to \eqref{D.2}, is determined from the fit. 
The function $\hat{a}_2(Q^2)$ is set to zero, which is justified in our case 
since for the photoproduction cross section $\hat{a}_2(0)$ does not 
contribute; see \eqref{2.16}. For $Q^2 > 0$, on the other hand, the data 
to which we fit are at sufficiently high $W$ such that the whole contribution 
of the $f_{2R}$ exchange is very small there. With $\hat{a}_2(Q^2) =0$ 
we neglect in essence the possible $f_{2R}$-exchange contribution to $\sigma_L$; 
see \eqref{2.12}. The fit parameters for the pomeron and $f_{2R}$ reggeon properties 
are summarised in table \ref{tab2}. 
\begin{table}[t]\centering
\renewcommand{\arraystretch}{1.2}
\begin{tabular}{c|c|c|c}
& parameter & default value used & fit result\\
\hline
$\mathbbm{P}_0$ & intercept & & $\alpha_0(0)=1+\epsilon_0$\\
& & &$\epsilon_0= 0.3008 \,({}^{+73}_{-84})$\\
& slope parameter & $\alpha'_0 = 0.25 \, \mbox{GeV}^{-2}$ & \\
& $W^2$ parameter & $\tilde{\alpha}'_0 = 0.25 \, \mbox{GeV}^{-2}$ & \\
& $pp$ coupling parameter & $\beta_{0pp} = 1.87 \, \mbox{GeV}^{-1}$ & \\
\hline
$\mathbbm{P}_1$ & intercept & & $\alpha_1(0)=1+\epsilon_1$\\
& & &$\epsilon_1= 0.0935 \,({}^{+76}_{-64})$\\
& slope parameter & $\alpha'_1 = 0.25 \, \mbox{GeV}^{-2}$ & \\
& $W^2$ parameter & $\tilde{\alpha}'_1 = 0.25 \, \mbox{GeV}^{-2}$ & \\
& $pp$ coupling parameter & $\beta_{1pp} = 1.87 \, \mbox{GeV}^{-1}$ & \\
\hline
$f_{2R}$ & intercept & & $\alpha_2(0)=0.485 \,({}^{+88}_{-90})$\\
& slope parameter & $\alpha'_2 = 0.9 \, \mbox{GeV}^{-2}$ & \\
& $W^2$ parameter & $\tilde{\alpha}'_2 = 0.9 \, \mbox{GeV}^{-2}$ & \\
& $pp$ coupling parameter & $\beta_{2pp} = 3.68 \, \mbox{GeV}^{-1}$ & 
\end{tabular}
\caption{Fit values obtained for the pomeron and $f_{2R}$ reggeon intercepts 
and default values used for the other parameters; see appendix \ref{appA}. 
\label{tab2}}
\end{table}
The ans\"atze for the pomeron- and $f_{2R}$ 
reggeon-photon coupling functions are discussed in appendix \ref{appParam}. 
The fit procedure is explained in appendix \ref{appfit} and the fit results 
for the parameters of our model are given in table \ref{tab4} in appendix \ref{appfitres}. 
Further quantities occurring in our formulae are the fine structure constant $\alpha_{\rm em}$, 
the proton mass $m_p$, and $M_0$ used in various places for dimensional reasons. 
We have 
\begin{equation}
\label{4.11a} 
\begin{split}
\alpha_{\rm em} &= 0.0072973525664\,, \\
m_p &=  0.938272 \,\mbox{GeV}\,, \\
M_0 &= 1\,\mbox{GeV}\,. 
\end{split}
\end{equation}

Our global fit has 25 parameters which are, however, not all of the same quality. 
The most important parameters are the three intercepts, $\alpha_0(0) = 1 + \epsilon_0$, 
 $\alpha_1(0) = 1 + \epsilon_1$, and $\alpha_2(0)$; see table \ref{tab2}. Then we have 
the values of the pomeron-$\gamma^* \gamma^*$ and $f_{2R}$-$\gamma^* \gamma^*$ 
coupling functions 
at $Q^2=0$, that is, $\hat{b}_j(0)$ ($j=0,1,2$) and $\hat{a}_j(0)$ ($j=0,1$) 
which give another five parameters. The fall-off of these coupling functions with $Q^2$ 
involves the remaining 17 parameters. Here we have some freedom in choosing e.\,g.\ 
more or fewer spline knots for the functions $\hat{b}_j(Q^2)$ ($j=0,1$). We found it 
convenient to use $N=7$ spline knots; see appendix \ref{appParampom} and table 
\ref{tab4} in appendix \ref{appfitres}. 

\begin{figure}
\begin{center}
  \includegraphics[width=0.6\textwidth]{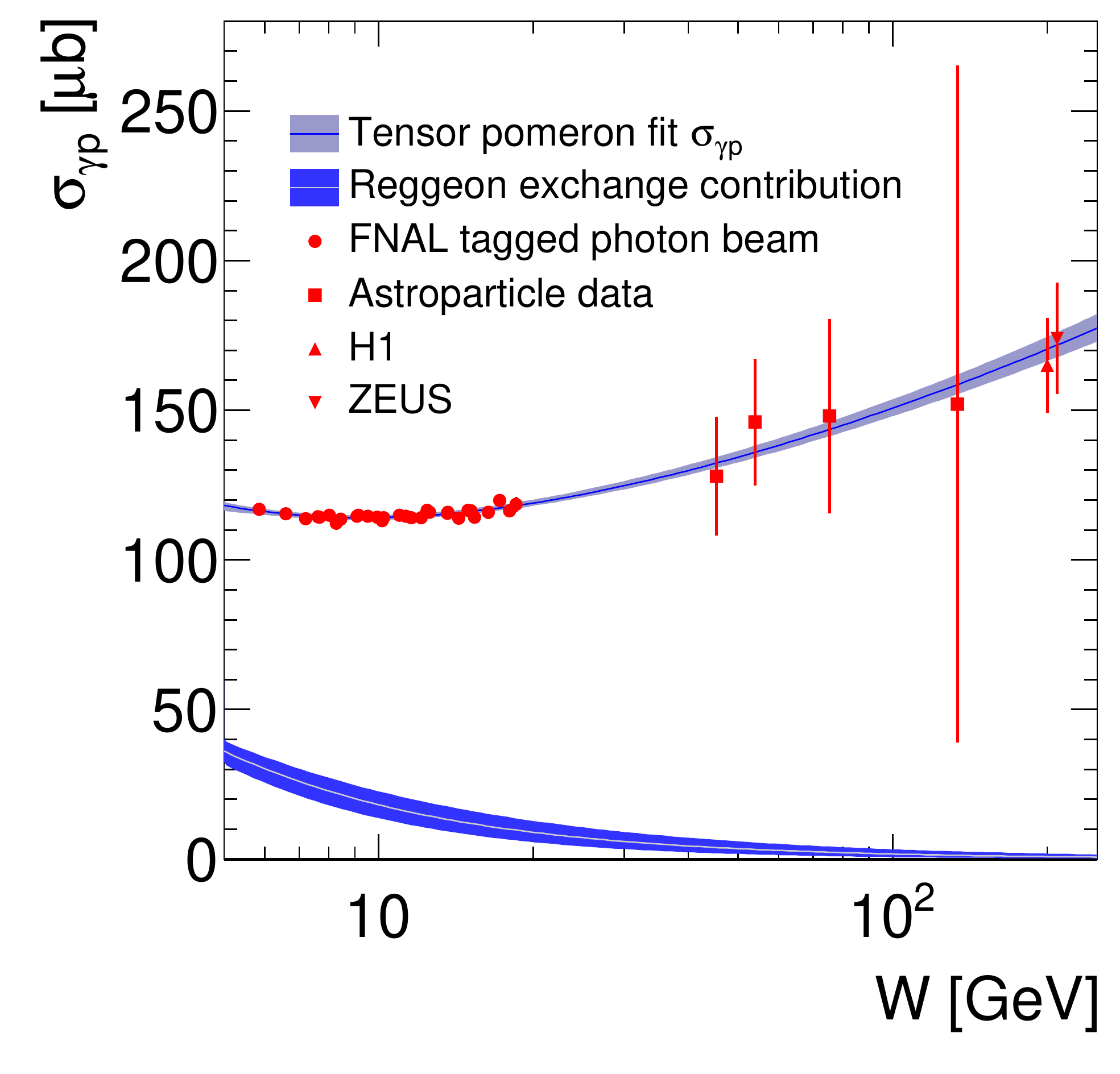}
  \caption{
    Comparison of the global fit to the photoproduction cross sections 
    \cite{Aid:1995bz,Chekanov:2001gw,Vereshkov:2003cp,Caldwell:1978yb}. 
    The reggeon contribution is indicated.
    The experimental uncertainties of the fit are indicated as shaded bands.
   \label{fig4}
   }
\label{fig:comp0}
\end{center}
\end{figure}

Let us now show our fit results starting with photoproduction in fig.\ \ref{fig4}. 
The fit is very satisfactory. The $f_{2R}$ reggeon contribution 
is also indicated. It is found to be important for $W < 30\,\mbox{GeV}$. 

\begin{figure}
\begin{center}
  \includegraphics[width=0.85\textwidth]{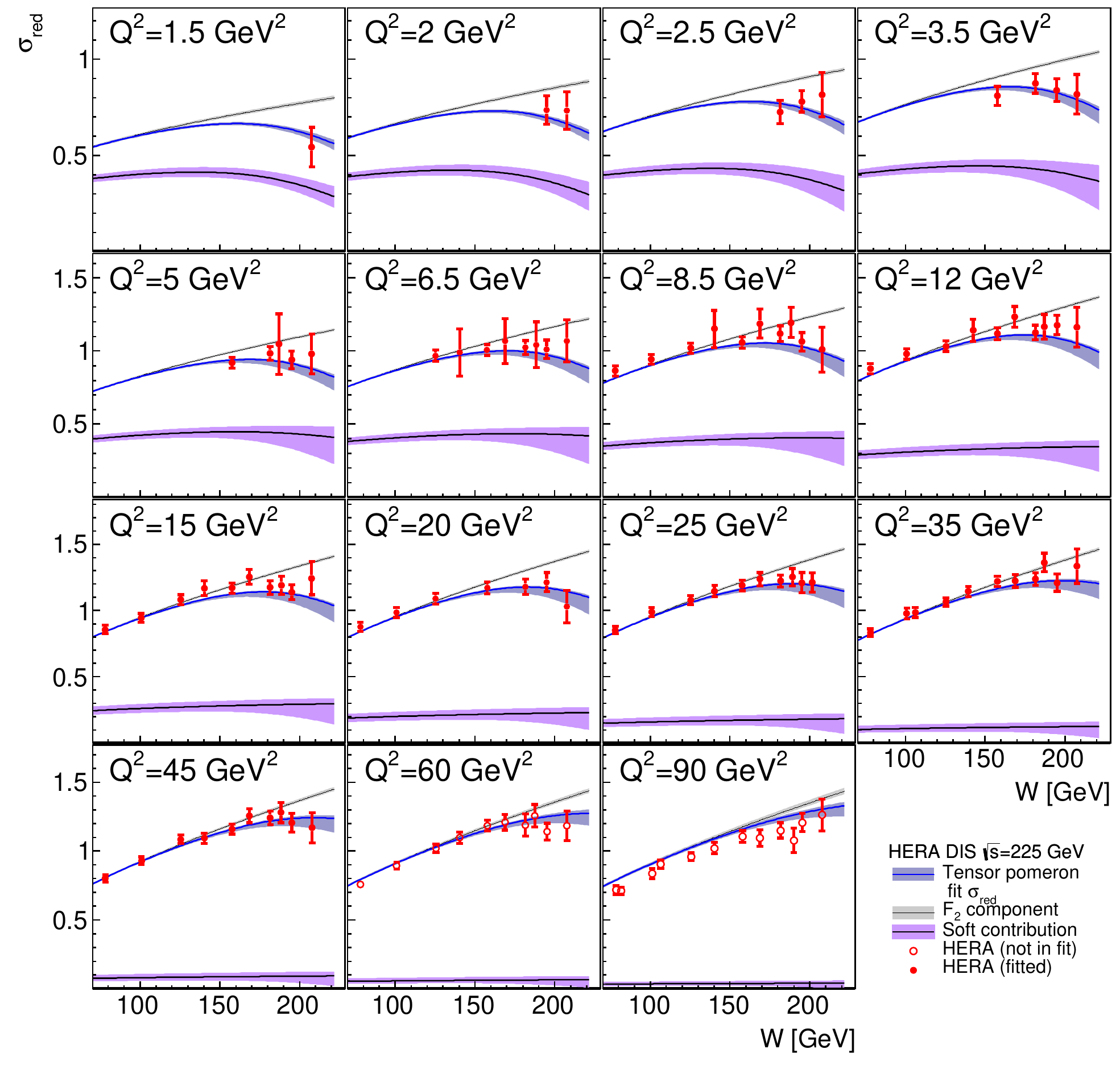}
  \caption{
    Comparison of the fit to DIS cross sections at centre-of-mass
    energy $225\,\text{GeV}$. 
    We also show the soft contribution (soft pomeron plus $f_{2R}$ reggeon) 
    and the contribution of the 
    structure function $F_2$ in the reduced cross section; see \eqref{3.5}. 
    The experimental uncertainties of the fit are indicated as shaded bands.
\label{fig5}  }
\label{fig:comp1}
\end{center}
\end{figure}

\begin{figure}
\begin{center}
  \includegraphics[width=0.85\textwidth]{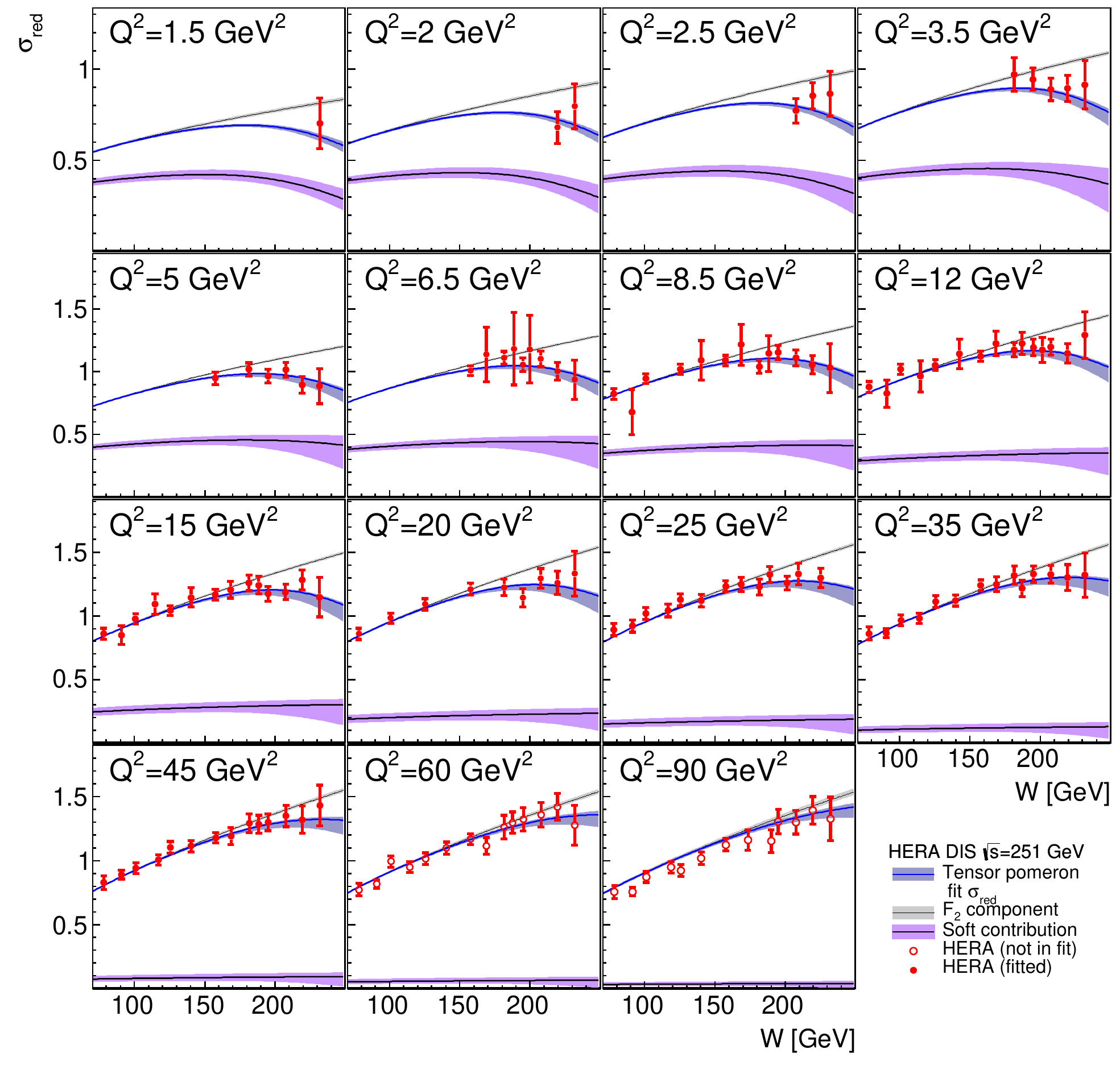}
  \caption{
    Comparison of the fit to DIS cross sections at centre-of-mass
    energy $251\,\text{GeV}$.
    We also show the soft contribution (soft pomeron plus $f_{2R}$ reggeon) 
    and the contribution of the 
    structure function $F_2$ in the reduced cross section; see \eqref{3.5}. 
    The experimental uncertainties of the fit are indicated as shaded bands.
 \label{fig6} }
 \label{fig:comp2}
\end{center}
\end{figure}

\begin{figure}
\begin{center}
  \includegraphics[width=0.85\textwidth]{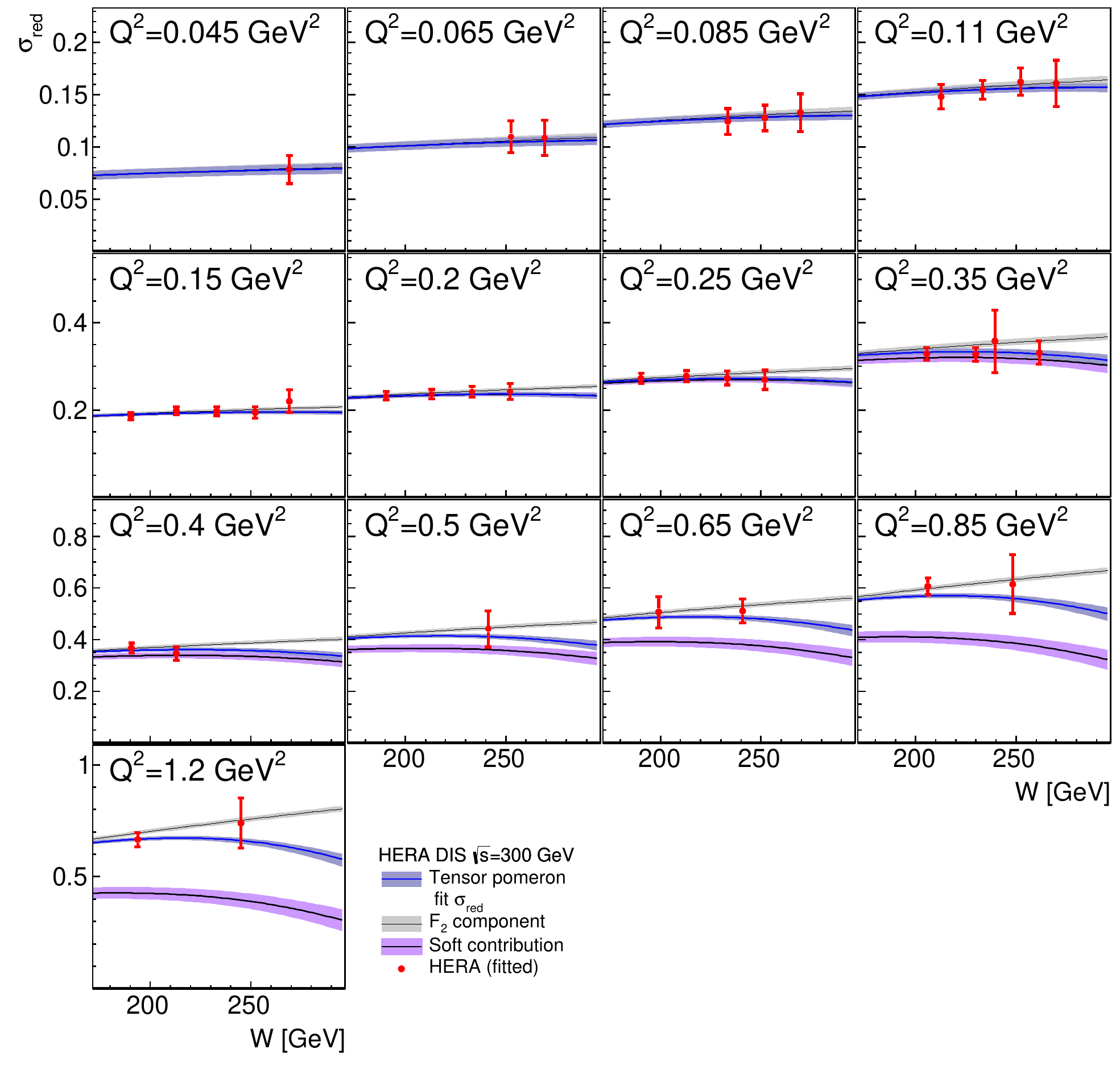}
  \caption{
    Comparison of the fit to DIS cross sections at centre-of-mass
    energy $300\,\text{GeV}$, at low $Q^2<1.5\,\text{GeV}^2$. 
    We also show the soft contribution (soft pomeron plus $f_{2R}$ reggeon) 
    and the contribution of the 
    structure function $F_2$ in the reduced cross section; see \eqref{3.5}. 
    The experimental uncertainties of the fit are indicated as shaded bands.
 \label{fig7} }
 \label{fig:comp3l}
\end{center}
\end{figure}

\begin{figure}
\begin{center}
  \includegraphics[width=0.85\textwidth]{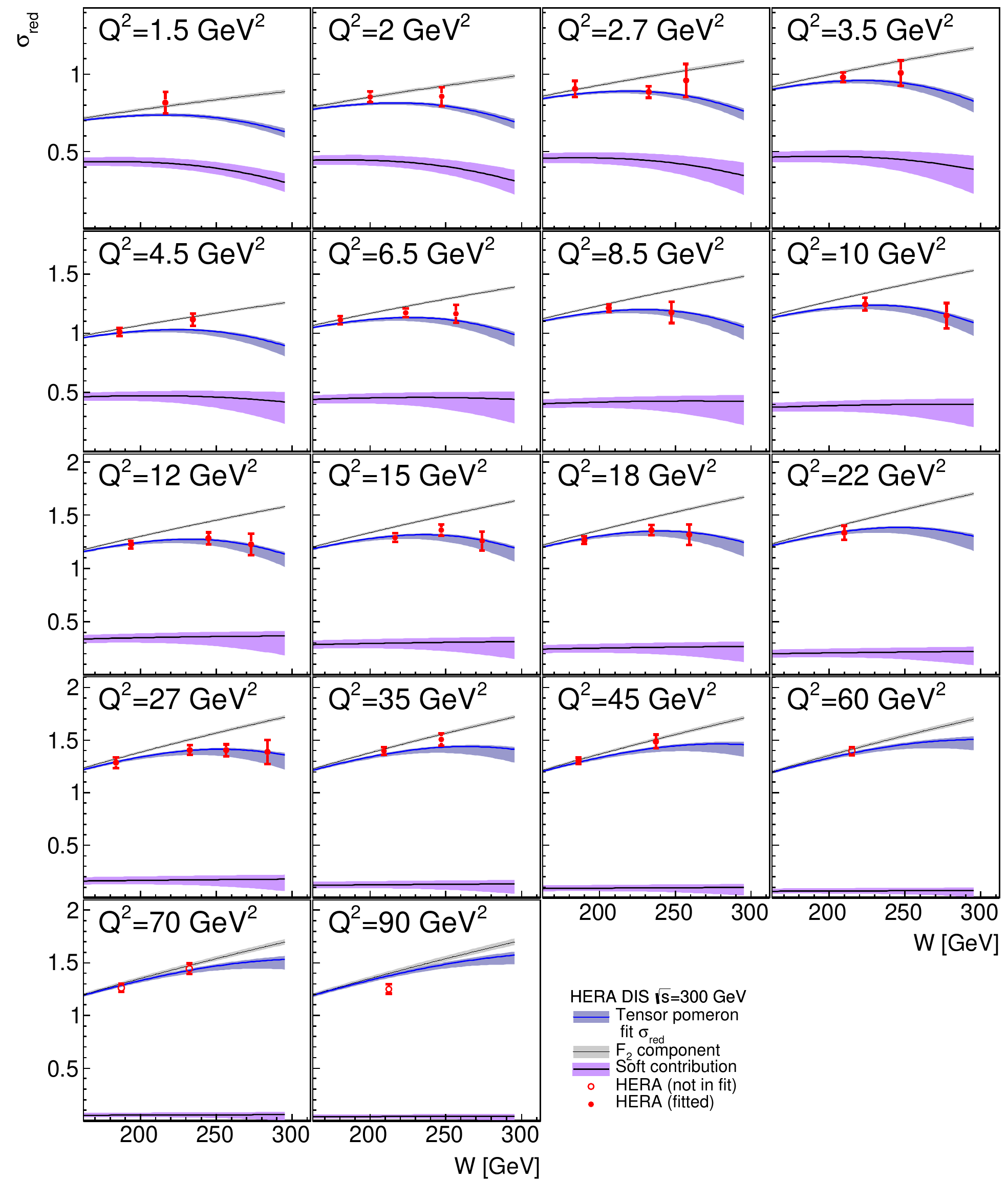}
  \caption{
    Comparison of the fit to DIS cross sections at centre-of-mass
    energy $300\,\text{GeV}$, at high $Q^2\ge1.5\,\text{GeV}^2$.
    We also show the soft contribution (soft pomeron plus $f_{2R}$ reggeon) 
    and the contribution of the 
    structure function $F_2$ in the reduced cross section; see \eqref{3.5}. 
    The experimental uncertainties of the fit are indicated as shaded bands.
\label{fig8}  }
 \label{fig:comp3h}
\end{center}
\end{figure}

\begin{figure}
\begin{center}
  \includegraphics[width=0.85\textwidth]{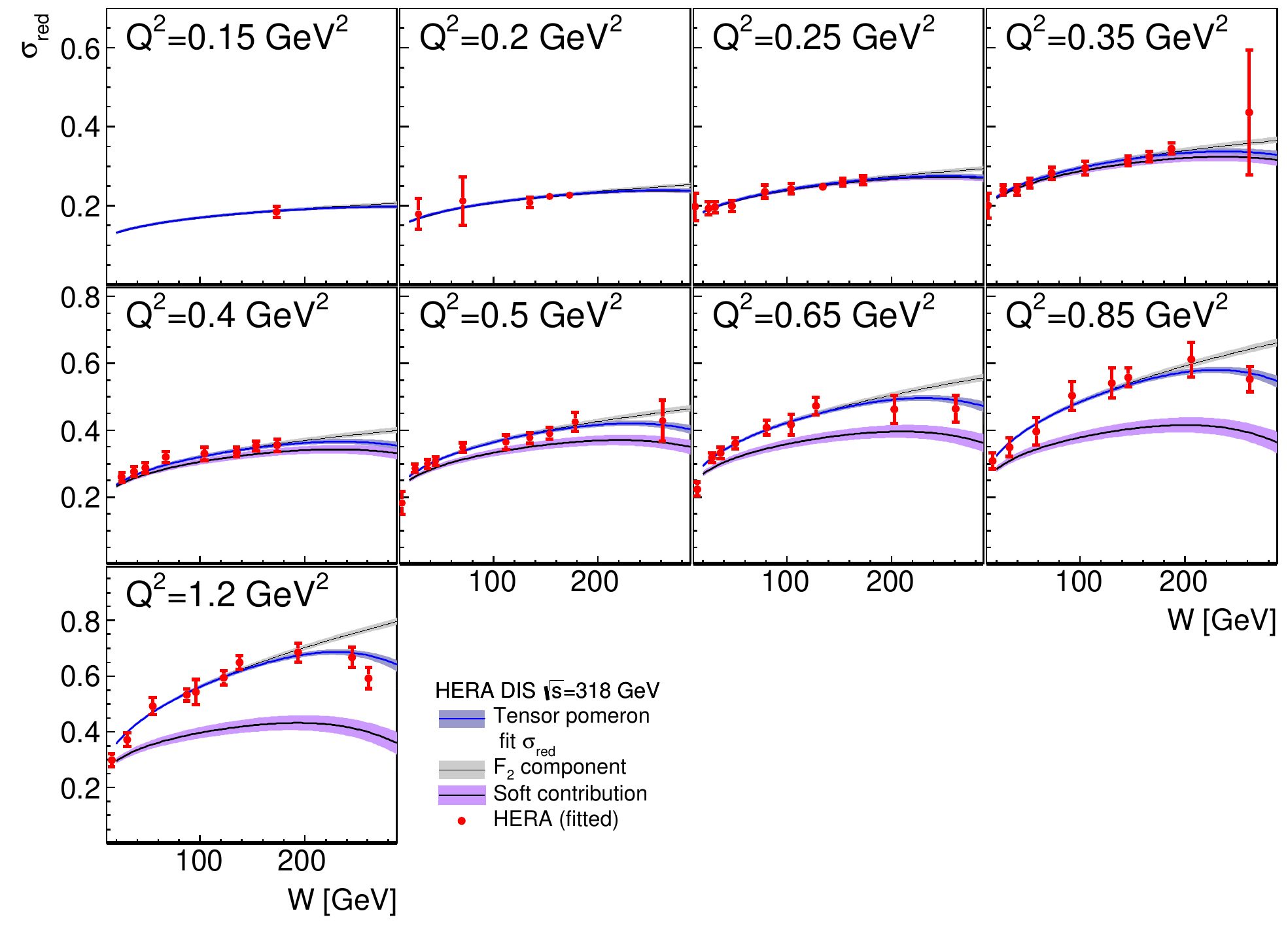}
  \caption{
    Comparison of the fit to DIS cross sections at centre-of-mass
    energy $318\,\text{GeV}$, at low $Q^2<1.5\,\text{GeV}^2$.
    We also show the soft contribution (soft pomeron plus $f_{2R}$ reggeon) 
    and the contribution of the 
    structure function $F_2$ in the reduced cross section; see \eqref{3.5}. 
    The experimental uncertainties of the fit are indicated as shaded bands.
\label{fig9}  }
 \label{fig:comp4l}
\end{center}
\end{figure}

\begin{figure}
\begin{center}
  \includegraphics[width=0.85\textwidth]{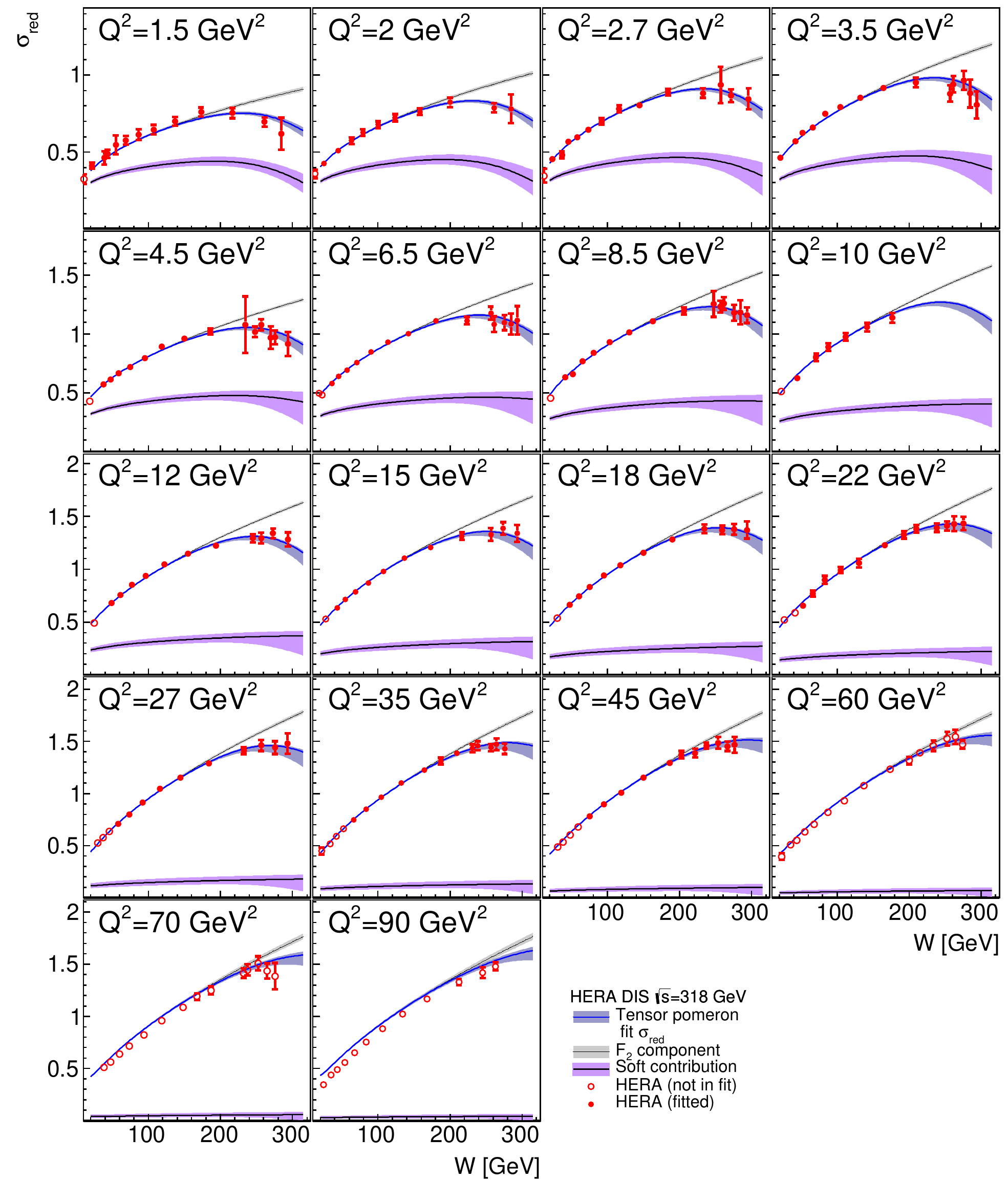}
  \caption{
    Comparison of the fit to DIS cross sections at centre-of-mass
    energy $318\,\text{GeV}$, at high $Q^2\ge1.5\,\text{GeV}^2$.
    We also show the soft contribution (soft pomeron plus $f_{2R}$ reggeon) 
    and the contribution of the 
    structure function $F_2$ in the reduced cross section; see \eqref{3.5}. 
    The experimental uncertainties of the fit are indicated as shaded bands.
\label{fig10} }
 \label{fig:comp4h}
\end{center}
\end{figure}

In figs.\ \ref{fig5} to \ref{fig10} we show our 
fit results for the HERA data. Here we indicate also the soft 
pomeron contribution. The contribution of the $f_{2R}$ 
component for the HERA DIS data which we use ($x<0.01$) is found 
to be very small from the fits and is hardly visible in 
figs.\ \ref{fig5} to \ref{fig10}. 
The quality of our global fit, which has 25 parameters, is assessed 
in table \ref{tab3}, and is overall found to be very satisfactory. 
The experimental uncertainties indicated as 
shaded bands in fig.\ \ref{fig4} and the following figures correspond 
to one standard deviations; see appendix \ref{appfit}. 
\begin{table}[t]\centering
\renewcommand{\arraystretch}{1.2}
  \begin{tabular}{l|r|c}
    dataset & $\chi^2$ & number of points \\
    \hline
    DIS $\sqrt{s}=225\,\text{GeV}$ & $104.98$ & $91$ \\
    DIS $\sqrt{s}=251\,\text{GeV}$ & $113.12$ & $118$ \\
    DIS $\sqrt{s}=300\,\text{GeV}$ & $60.38$ & $71$ \\
    DIS $\sqrt{s}=318\,\text{GeV}$ & $271.82$ & $245$ \\
    \hline
    HERA DIS data, all $\sqrt{s}$ & $553.77$ & $525$ \\ 
    H1 photoproduction & $0.23$ & $1$ \\
    ZEUS photoproduction & $0.03$ & $1$ \\
    cosmic ray data & $0.62$ & $4$ \\
    tagged photon beam & $33.29$ & $30$ \\ 
\hline
all datasets & $587.94$ & $N_{DF}=(561-25)$, probability $6.0\%$ 
\end{tabular}
  \caption{
    Partial $\chi^2$ and number of data points per dataset, goodness
    of fit, number of degrees of freedom and fit probability for our 
    tensor-pomeron fit.
    The partial $\chi^2$ numbers for the individual DIS centre-of-mass
    energies (upper part of the table) do not add up to the number
    quoted for all HERA DIS data. This is expected because 
    correlated uncertainties between the different centre-of-mass
    energies also contribute.
    \label{tab3}
  }
  \label{tab:fitqual}
\end{table}

We now want to discuss in detail the results of our fit. 
We start with the intercepts of the pomerons and of the 
$f_{2R}$ reggeon. From our global fit the soft pomeron ($\mathbbm{P}_1$) 
intercept comes out as 
\be\label{3.13} 
\alpha_1(0) = 1 + \epsilon_1\,, \qquad \epsilon_1= 0.0935 \,({}^{+76}_{-64})\,.
\ee
This is well compatible with the standard value 
$\epsilon \approx 0.08$ to $0.09$ obtained from hadronic reactions; 
see for instance chapters 3 of \cite{Donnachie:2002en} and \cite{Ewerz:2013kda}. 
The value of the $f_{2R}$ intercept is found to be 
\begin{equation}
\label{4.12a}
\alpha_2 (0) = 0.485 \, ({}^{+88}_{-90})
\end{equation}
and is in agreement with the determinations from 
\cite{Donnachie:2002en,Ewerz:2013kda} which quote 
$\alpha_2(0)=0.5475$. 
For the hard pomeron $\mathbbm{P}_0$ we find 
\be\label{4.12b} 
\alpha_0(0) = 1 + \epsilon_0\,, \qquad \epsilon_0= 0.3008 \,({}^{+73}_{-84})\,.
\ee
This is again a very reasonable value. 

Next, let us turn to photoproduction; see fig.\ \ref{fig4}. The photoproduction 
is dominated by soft pomeron exchange in the energy range investigated, 
$6\,\mbox{GeV} < W < 209 \,\mbox{GeV}$. The $f_{2R}$ reggeon 
contribution is important for $W \lesssim 30 \,\mbox{GeV}$ and is needed 
there in order to get a good fit to the data. The hard pomeron $\mathbbm{P}_0$ 
gives only a very small contribution. In fact, 
there is no evidence for a non-zero contribution 
of the hard pomeron to the photoproduction 
cross section in the energy range investigated here. 
At $W= 200\,\mathrm{GeV}$, for instance, the fitted contributions to the 
photoproduction cross section are 
\begin{equation}
\label{4.12c}
\begin{split}
170.4\, {}^{+4.2}_{-4.0} \: \mu{\rm b}& \qquad \mbox{for the soft pomeron } \mathbbm{P}_1\,,
\nonumber\\
0.002\, {}^{+0.086}_{-0.002} \: \mu{\rm b}& \qquad \mbox{for the hard pomeron } \mathbbm{P}_0\,,
\nonumber\\
0.84\, {}^{+0.99}_{-0.58} \: \mu{\rm b}& \qquad \mbox{for the $f_{2R}$ reggeon}. 
\nonumber
\end{split}
\end{equation}
For lower $W$ values the relative contribution of the hard pomeron to 
photoproduction is even smaller due to $\epsilon_0 > \epsilon_1$. 

In figures \ref{fig5}-\ref{fig10} we show the comparison of our global fit 
with the HERA DIS data. Note that in figures \ref{fig5}, \ref{fig6}, \ref{fig8}, 
and \ref{fig10}, we also show the extrapolation of our fit to the region 
$50 \, \mbox{GeV}^2 \le Q^2 \le 90 \, \mbox{GeV}^2$. The HERA data in 
this region are {\em not} included in the fit but still reasonably well 
described by it. In our global fit we have as parameters also 
the pomeron-$\gamma^* \gamma^*$ coupling functions $\hat{a}_j(Q^2)$ and 
$\hat{b}_j(Q^2)$ ($j = 0,1$); see table \ref{tab1} and \eqref{A.10}, 
\eqref{A.11}. The latter are parametrised with the help of cubic splines; 
see appendix \ref{appParam}. In figures \ref{fig11} to \ref{fig14} we show 
the fit results for these functions which are discussed further in 
appendices \ref{appfit} and \ref{appfitres}. Note that above $Q^2 = 50 \,\mbox{GeV}^2$ 
the displayed curves are extrapolations beyond the last spline knot. 
In essence, these functions are extrapolated using simple power laws in $Q^2$; 
see \eqref{D.3}, \eqref{D.5} and \eqref{D.6} in appendix \ref{appParam}. 
\begin{figure}
\begin{center}
  \includegraphics[width=0.75\textwidth]{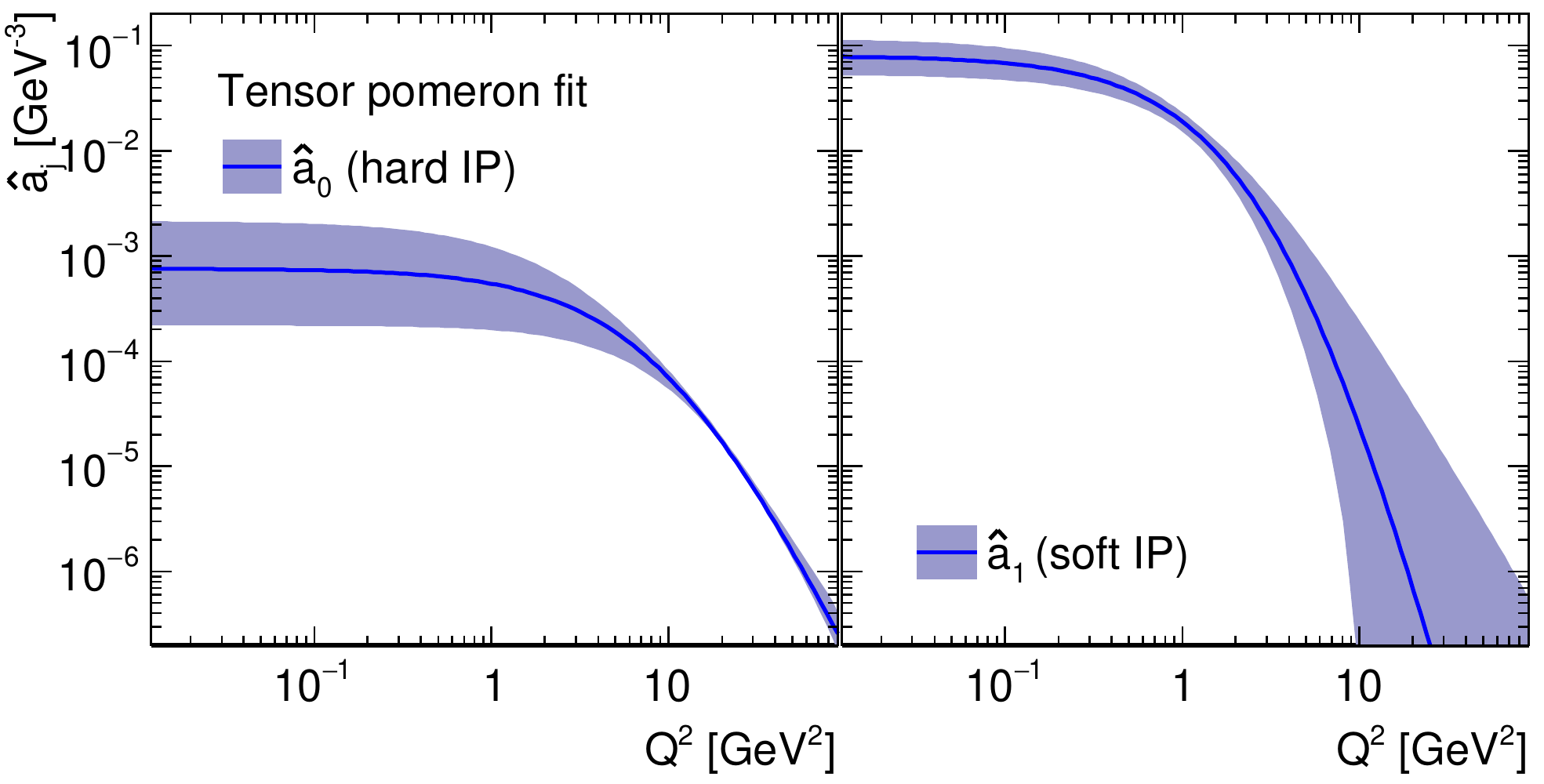}
  \caption{
    The pomeron-$\gamma^*\gamma^*$ coupling functions $\hat{a}_j(Q^2)$ 
    for $j=0$ (hard pomeron) and $j=1$ (soft pomeron); 
    see \eqref{2.13}, \eqref{2.14}, and \eqref{A.11}. 
    The shaded bands indicate the experimental uncertainties.
    \label{fig11}
  }
\end{center}
\end{figure}
\begin{figure}
\begin{center}
  \includegraphics[width=0.75\textwidth]{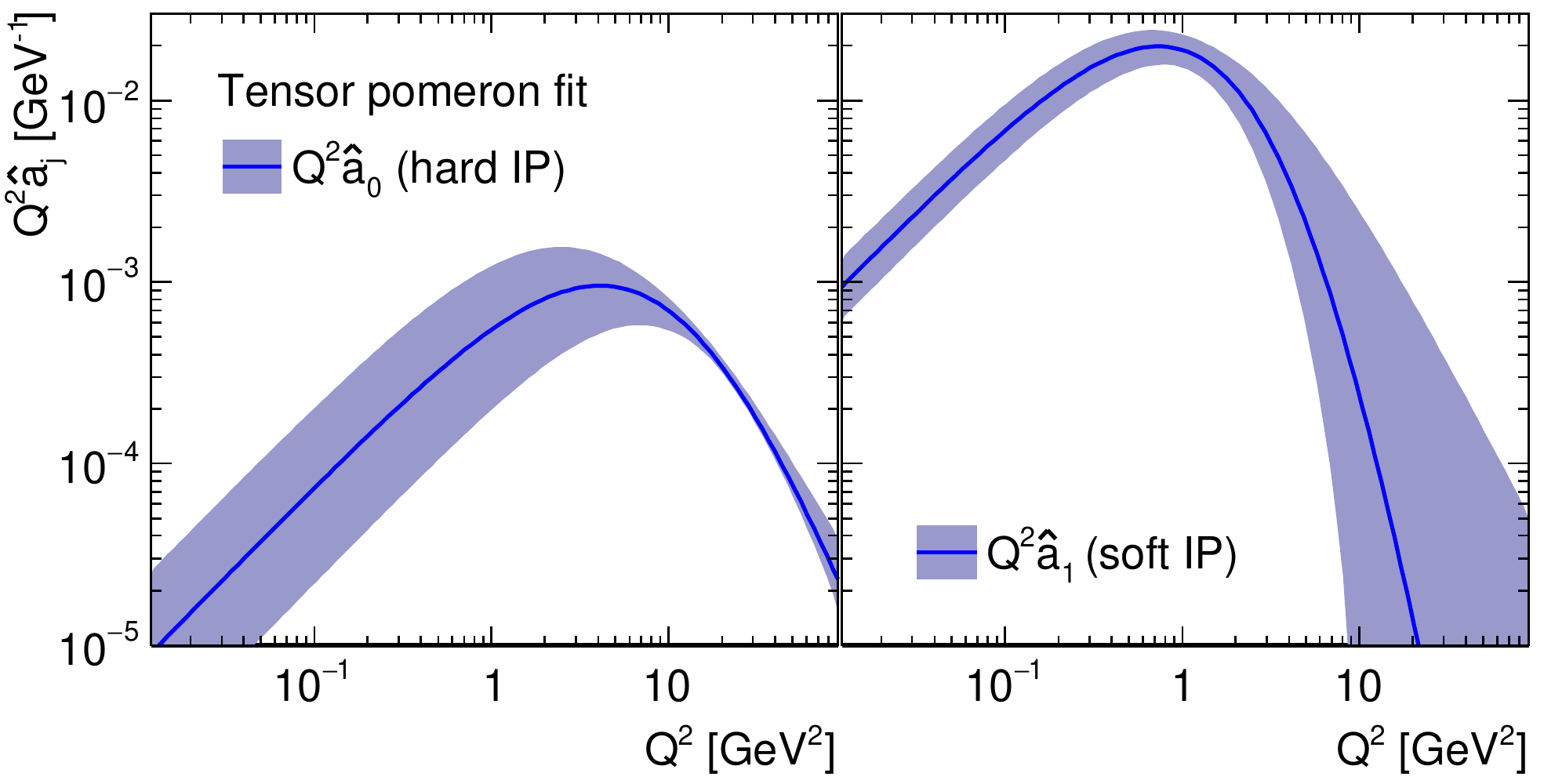}
  \caption{
    The pomeron-$\gamma^*\gamma^*$ coupling functions $Q^2\hat{a}_j(Q^2)$ 
    for $j=0$ (hard pomeron) and $j=1$ (soft pomeron); 
    see \eqref{2.13}, \eqref{2.14}, and \eqref{A.11}. 
    The shaded bands indicate the experimental uncertainties.
    \label{fig12}
  }
\end{center}
\end{figure}
\begin{figure}
\begin{center}
  \includegraphics[width=0.99\textwidth]{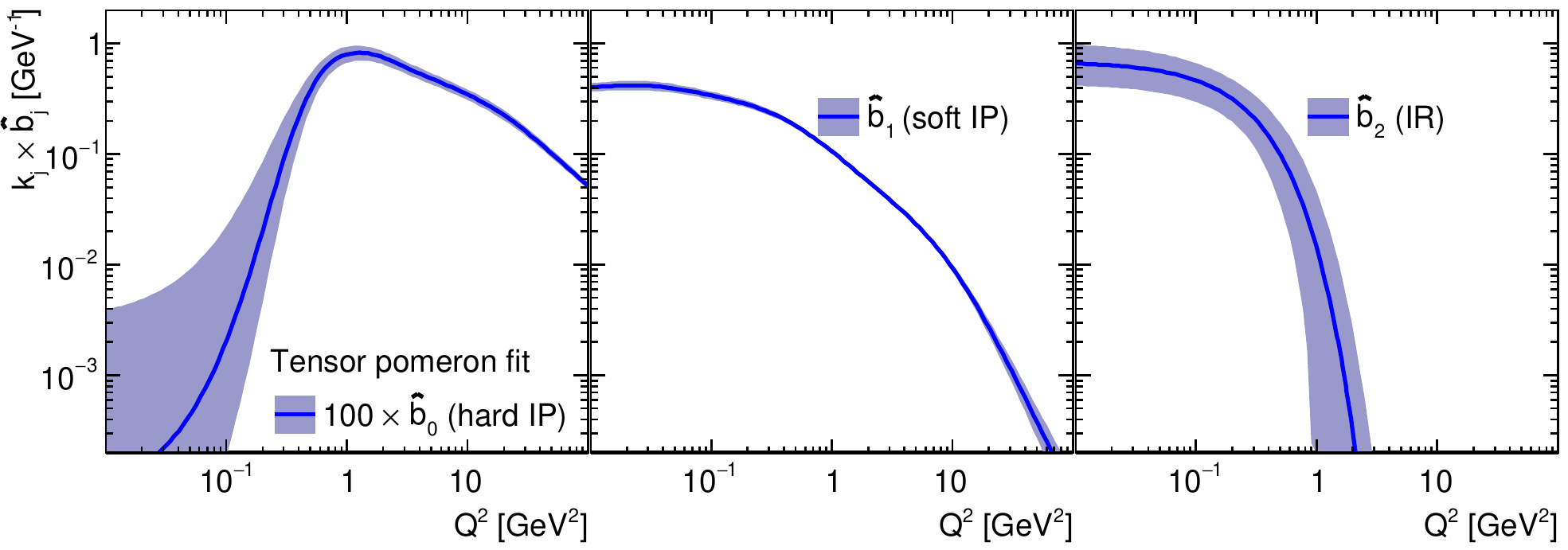}
  \caption{
    The pomeron- and reggeon-$\gamma^*\gamma^*$ coupling functions $\hat{b}_j(Q^2)$ for $j=0$ (hard pomeron), $j=1$
    (soft pomeron), and $j=2$ (reggeon); see \eqref{2.13}, \eqref{2.14}, \eqref{A.11}, and \eqref{A.16}. 
    The shaded bands indicate the
    experimental uncertainties. More precisely, we show the functions $k_j \hat{b}_j(Q^2)$, 
    where $\hat{b}_0(Q^2)$ is scaled up by a factor $k_0=100$ for displaying purposes 
    while the functions $\hat{b}_1(Q^2)$ and $\hat{b}_2(Q^2)$ are not scaled up, that is 
    $k_1 = k_2=1$. 
    \label{fig13}
  }
\end{center}
\end{figure}
\begin{figure}
\begin{center}
  \includegraphics[width=0.99\textwidth]{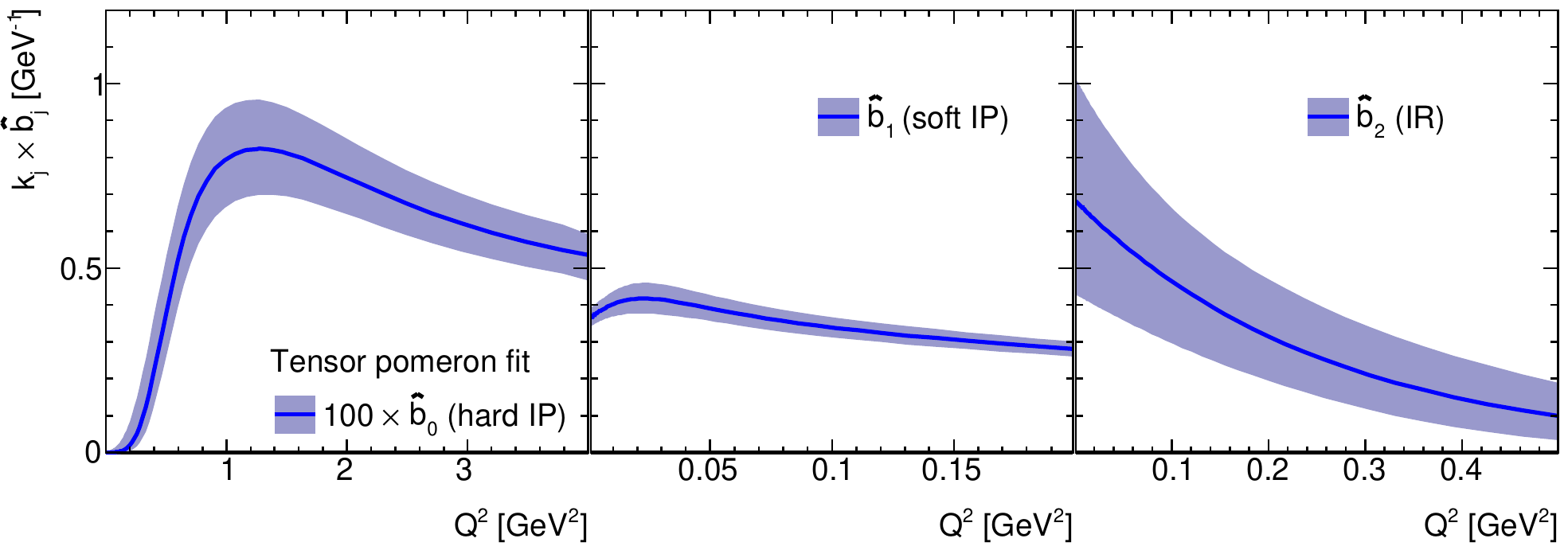}
  \caption{
    The pomeron- and reggeon-$\gamma^*\gamma^*$ coupling functions $\hat{b}_j(Q^2)$ for $j=0$ (hard pomeron), $j=1$
    (soft pomeron), and $j=2$ (reggeon); see \eqref{2.13}, \eqref{2.14}, \eqref{A.11}, and \eqref{A.16}. 
    The shaded bands indicate the
    experimental uncertainties. More precisely, we show the functions $k_j \hat{b}_j(Q^2)$, 
    where $\hat{b}_0(Q^2)$ is scaled up by a factor $k_0=100$ for displaying purposes 
    while the functions $\hat{b}_1(Q^2)$ and $\hat{b}_2(Q^2)$ are not scaled up, that is 
    $k_1 = k_2=1$. 
    \label{fig14}
  }
\end{center}
\end{figure}

Let us now point out some salient features of our global fit to HERA DIS data, 
figures \ref{fig5} to \ref{fig10}. 

We see from figures \ref{fig7} and \ref{fig9} that the soft pomeron 
$\mathbbm{P}_1$ dominates $\sigma_{\rm red}$ for $Q^2 \lesssim 1\,\mbox{GeV}^2$. 
For higher $Q^2$ (figures \ref{fig5}, \ref{fig6}, \ref{fig8}, \ref{fig10}) 
the soft component slowly decreases relative to the hard one. 
For the c.\,m.\ energies $\sqrt{s}$ investigated, the soft and hard 
components are of similar size near $Q^2\approx 5\,\mbox{GeV}^2$.
Dominance of the hard component ($\mathbbm{P}_0$) can only 
be seen for $Q^2 \gtrsim 20 \,\mbox{GeV}^2$. Thus, our fit tells us 
that the soft pomeron ($\mathbbm{P}_1$) contribution is essential 
for an understanding of the HERA data for $Q^2 < 50 \,\mbox{GeV}^2$ 
and $x < 0.01$. 

In figures \ref{fig5} to \ref{fig10} we have also indicated the contribution 
of the structure function $F_2$ alone to $\sigma_{\rm red}$; see \eqref{3.5}. 
At fixed $s$ and $Q^2$, large $W$ corresponds to large $y$; see \eqref{1.2}. 
At large $y$ the negative term $-\tilde{f} \sigma_L$ in $\sigma_{\rm red}$ 
(see \eqref{3.3},\eqref{3.4}) becomes important. The turning away of the data from 
the lines '$F_2$ component' therefore indicates a sizeable contribution 
from the longitudinal cross section $\sigma_L$. Our model gives a good 
description of this feature of the data. 

\begin{figure}
\begin{center}
  \includegraphics[width=0.44\textwidth]{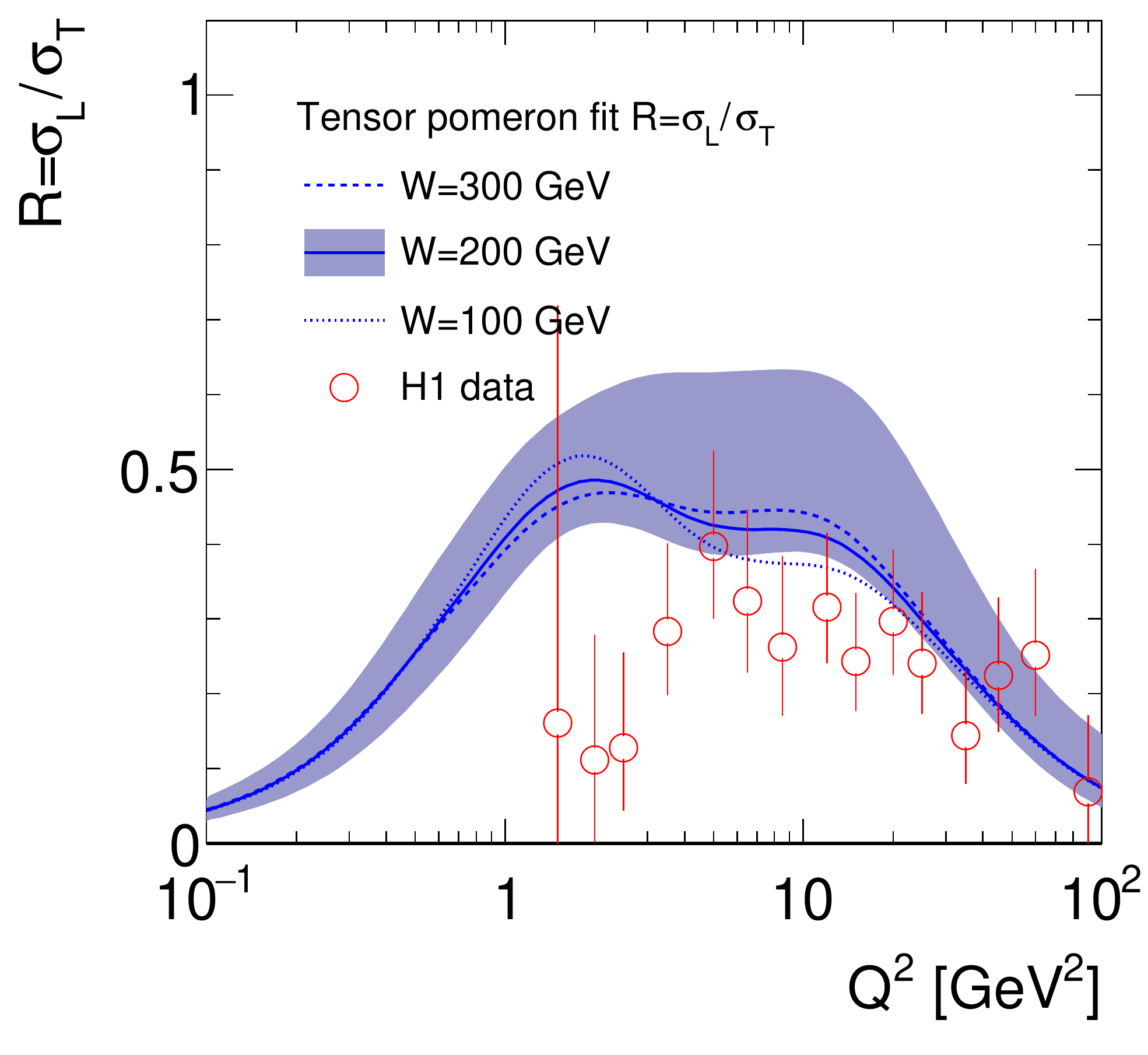}
  \includegraphics[width=0.44\textwidth]{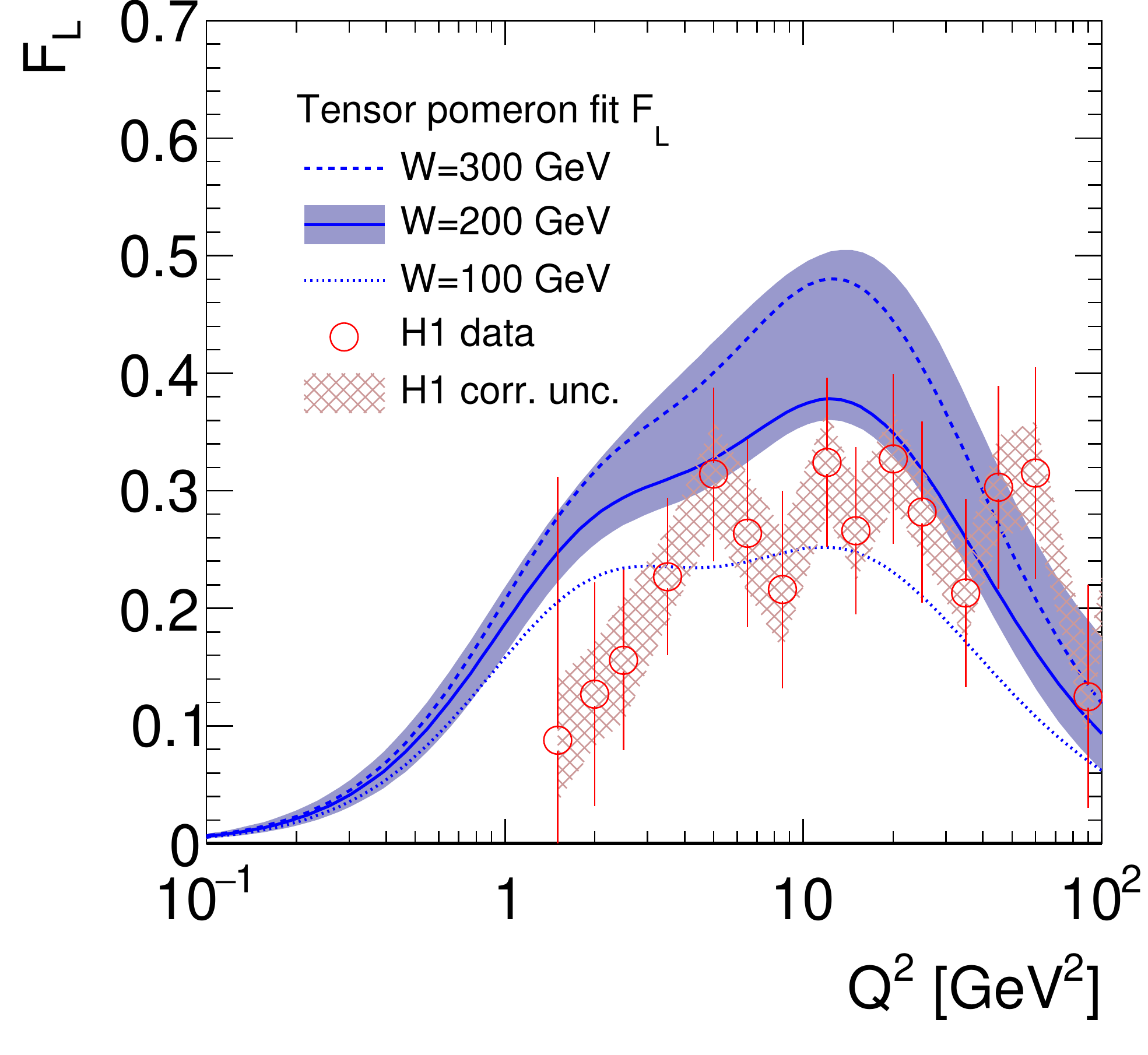}
  \caption{
    The ratio $R=\sigma_L/\sigma_T$ of longitudinal to transverse
    cross sections and the structure function $F_L$ are shown as a
    function of $Q^2$ for three choices of $W$ in
    comparison to data extracted directly from H1 cross section
    measurements at different centre-of-mass energies \cite{Andreev:2013vha}. 
    These data were taken at $W$ near $200\,\text{GeV}$ and are not included in our fit. 
    The error bars correspond to the H1 experimental uncertainties.
    The experimental uncertainties
    of our fit are indicated for $W=200\,\text{GeV}$ as a shaded band.
    For the case of the $F_L$ measurement, the correlated H1 
    uncertainty contribution is shown as a hatched band. 
    \label{fig15}
  \label{fig:FLandR}
  }
\end{center}
\end{figure}

Another way to assess the importance of $\sigma_L$ is to consider the 
ratio 
\be\label{3.15} 
R(W^2,Q^2)= \frac{\sigma_L(W^2,Q^2)}{\sigma_T(W^2,Q^2)} \,.
\ee
Our fit results for $R$ and for $F_L$ \eqref{2.14} are shown in 
fig.\ \ref{fig15}. 
Within the fit ansatz, the ratio $R=\sigma_L/\sigma_T$ of longitudinal to
transverse cross sections depends on $Q^2$ and $W$.
Figure \ref{fig:FLandR} shows the dependence of $R$ and of the structure
function $F_L$ on $Q^2$ at fixed $W$. In
both panels, H1 data \cite{Andreev:2013vha} are shown for comparison with 
our global fit results. 
The H1 data are extracted in a model-independent way directly
from H1 cross sections measured at a fixed $Q^2$ and $x$ but different
centre-of-mass energies. The $W$ corresponding to the H1 data is around 
$200\,\text{GeV}$, the extreme values are $W=232\,\text{GeV}$ at
$Q^2=1.5\,\text{GeV}^2$ and $W=193\,\text{GeV}$ at $Q^2=45\,\text{GeV}^2$.
The same H1 cross section data \cite{Andreev:2013vha} also contribute
strongly to the HERA data combination of DIS cross sections
\cite{Abramowicz:2015mha}, which is used as input to our fit.
Still, the fit predicts $R$ and $F_L$ somewhat above the H1 data.
The H1 $R$ and $F_L$ data however have a sizeable point-to-point
correlated uncertainty, which for $F_L$ is of order $0.045$ as
indicated. Moreover, the determinations of $R$ in the fit or directly
from H1 cross sections probe different aspects of the data.

In the H1 extraction from data, the structure function $F_2$ is a free
parameter for each point in $Q^2$ and $W$, which basically is set by 
the measurements at high centre-of-mass energies
$\sqrt{s}=318\,\text{GeV}$ and $W=200\,\text{GeV}$ (figure
\ref{fig10}). The structure function $F_L$ and the ratio $R$ are
then determined largely by the data points at low $\sqrt{s}=225\,\text{GeV}$
and $W=200\,\text{GeV}$ (figure \ref{fig5}).

In contrast, $F_2$ in our fit is determined largely by data from lower
$W$ and the power exponents $\epsilon_i$. The functions 
$F_L$ and $R$ are then determined from all centre-of-mass energies
together at their respective largest $W$; however, the most precise
data at largest $W$ (figure \ref{fig10}) contribute most.

In appendix \ref{appalt} we present further discussion of the 
ratio $R$ \eqref{3.15}. We show in particular that the rather large 
value of $R$ resulting from the fit is not affected much by making 
different assumptions for the fit parameters. 

\section{Discussion}
\label{sec:discussion}

In this article we developed a two-tensor-pomeron model and used 
it for a fit to data from photoproduction and from HERA deep-inelastic 
lepton-nucleon scattering at low $x$. The c.\,m.\ energy range of these 
data is $6$ to $318\,\mbox{GeV}$, the $Q^2$ range 
$0$ to $50\,\mbox{GeV}^2$. For the theoretical description we also 
included the $f_{2R}$ reggeon exchange which turned out to be 
relevant for energies $\lesssim 30\,\mbox{GeV}$. 
The fit parameters were the intercepts of the two pomerons and of the 
reggeon, and their coupling functions to real and virtual photons. 
The fit turned out to be very satisfactory and allowed us to determine, 
for instance, the intercepts of the hard pomeron ($\mathbbm{P}_0$), 
of the soft pomeron ($\mathbbm{P}_1$) and of the $f_{2R}$ reggeon. 
We obtained very reasonable numbers for these intercepts; see table \ref{tab2}. 
The real photoabsorption cross section $\sigma_{\gamma p}$ is found to 
be dominated by soft pomeron exchange with, at lower energies, a contribution 
from $f_{2R}$ reggeon exchange. Within the errors of our fit a hard pomeron 
contribution is not visible for photoproduction. But as $Q^2$ increases the 
hard pomeron becomes more and more important. Hard and soft pomeron 
give contributions of roughly equal size for $Q^2 \approx 5\,\mbox{GeV}^2$, 
but the soft contribution is still clearly visible for $Q^2 \approx 20\,\mbox{GeV}^2$. 

Our results indicate that in the energy and $Q^2$ range investigated the 
$\gamma^*$-proton absorption cross sections rise with energy as $W^{2\epsilon_1}$ 
for low $Q^2$ and change to $W^{2\epsilon_0}$ for high $Q^2$. 
Here $\epsilon_1 \approx 0.09$ and $\epsilon_0 \approx 0.30$ are 
the intercepts minus one of the pomerons $\mathbbm{P}_1$ and 
$\mathbbm{P}_0$; see table \ref{tab2}. 
It has been realised already a long time ago (see for instance \cite{Gribov:1984tu}) 
that parton densities in hadrons become large in high-energy or low-$x$ scattering. 
This can give rise to parton recombination and saturation, potentially taming the growth 
of cross sections at high energies. At the energies investigated here 
we find no indication that the rise of the $\gamma^*$-proton absorption cross 
sections levels off. The question can be asked if the $\gamma^* p$ 
cross sections could continue to rise indefinitely for higher and higher $W$. 
We note first that there is {\em no} Froissart-like bound for the rise of the 
$\gamma^* p$ cross sections since $\gamma^*$ is not an asymptotic 
hadronic state. Thus, there is no non-linear unitarity relation for the 
$\gamma^* p$ cross sections which would be a prerequisite for the 
derivation of a Froissart-like bound. 
The $\gamma^* p$ cross sections may well stop to rise at higher $W$ due to 
saturation effects, but this will then, in our opinion, not be related to 
the Froissart-Martin-Lukaszuk 
bound \cite{Froissart:1961ux,Martin1966,Lukaszuk:1967zz} 
which applies to hadronic cross sections. We see no rigorous 
theoretical argument against an indefinite rise of the $\gamma^* p$ 
cross sections with $W$. Note that these '$\gamma^* p$ cross sections' 
are in reality current-current correlation functions. The standard 
folklore of quantum field theory (QFT) is that such functions should 
be polynomially bounded which is clearly fulfilled in our case. 
Some time ago, one of us investigated theoretically the low-$x$ 
behaviour of the $\gamma^* p$ cross sections in QCD \cite{Nachtmann:2002yd}. 
There, arguments were given that identify two regimes in low-$x$ DIS, 
one for low $Q^2$ and one for high $Q^2$. It was argued that, in the 
high $Q^2$-region of low-$x$, DIS could be related to a critical 
phenomenon where, for instance, $\epsilon_0$ would be one of the 
critical exponents. In such a picture it would be natural to have a power 
rise with $W$ for the $\gamma^* p$ cross sections $\sigma_T$ and 
$\sigma_L$. But to know the actual behaviour of  $\sigma_T$ and 
$\sigma_L$ for $W$ values higher than available today we will have 
to wait for future experiments. 

We can obtain further support for the view that low-$x$ DIS at high 
enough $Q^2$ can be understood as a critical phenomenon from our 
present results. We see from \eqref{2.18} and \eqref{2.19} and the fit 
results for $\hat{a}_j(Q^2)$ and $\hat{b}_j(Q^2)$ ($j=0,1$) summarised in 
tables \ref{tab4} and \ref{tab:splineparam} that for $Q^2 \gtrsim 20 \,\mbox{GeV}^2$ 
the $\gamma^* p$ cross sections are well represented by simple power laws 
in $Q^2$ and $W^2$: 
\begin{align}
\label{critexp1}
\sigma_T (W^2,Q^2) + \sigma_L (W^2,Q^2) &\propto \hat{b}_0 (Q^2) \, (W^2)^{\epsilon_0}
\nn
\\
&\propto (Q^2)^{-\eta_0} \, (W^2)^{\epsilon_0}\,,
\\
\label{critexp2}
\sigma_L (W^2,Q^2) &\propto Q^2\, \hat{a}_0 (Q^2) \, (W^2)^{\epsilon_0}
\nn
\\
&\propto (Q^2)^{-\delta_0} \, (W^2)^{\epsilon_0}\,.
\end{align}
Here we have from \eqref{D.3}, \eqref{D.5}, \eqref{D.6}, and tables 
\ref{tab4} and \ref{tab:splineparam} 
\be
\label{deltaeta}
\begin{split}
\delta_0 &= 2.51\,({}^{+68}_{-57}) 
\\
\eta_0 &= {} - n_{0,7} = 0.967 (73) \,.
\end{split}
\ee
Such simple power laws \eqref{critexp1} and \eqref{critexp2} were, indeed, 
suggested in \cite{Nachtmann:2002yd}. The quantities $\delta_0$ 
and $\eta_0$ are in this view, together with $\epsilon_0$, critical exponents. 

In our work we have paid particular attention to describing and fitting 
not only the structure function $F_2$, which is proportional to 
 $\sigma_T + \sigma_L$, but the reduced cross section $\sigma_{\rm red}$ 
\eqref{3.3}, \eqref{3.4} which contains all experimentally available 
information on $\sigma_T$ and $\sigma_L$ separately. 
Our fit results for $R=\sigma_L/\sigma_T$ indicate that it is 
rather large, $R \gtrsim 0.4$ for $1\,\mbox{GeV}^2 \lesssim Q^2 \lesssim 10\,\mbox{GeV}^2$ 
even taking the one standard deviation errors into account; see fig.\ 
\ref{fig15}, and also fig.\ \ref{fig:Reps0eps1} in appendix \ref{appalt}. 
We note that such a large value of $R$, taken at face value, presents problems 
for the standard colour-dipole model of low-$x$ DIS. In the framework 
of this model two of us derived a rigorous upper limit of $R \le 0.37248$; 
see \cite{Ewerz:2006vd,Ewerz:2006an} and references therein. 
The derivation of this bound uses only the standard dipole-model relations, 
in particular, the expressions for the photon wave functions at lowest order 
in the strong coupling constant $\alpha_s$ and the non-negativity of the 
dipole-proton cross sections. The then available 
H1 data for $R$ from \cite{Collaboration:2010ry} were compared with this and related bounds 
in \cite{Ewerz:2012az}. A very conservative conclusion from our findings concerning $R$ 
in the present paper is, therefore, as follows. If one wants to be sure to 
be in a kinematic region where the colour-dipole model can be applied 
in the HERA energy range one should limit oneself to $Q^2 \gtrsim 10\,\mbox{GeV}^2$. 
Below $Q^2 \approx 10\,\mbox{GeV}^2$ corrections to the standard 
dipole picture, as listed and discussed e.\,g.\ in \cite{Ewerz:2006vd}, 
may become important. 
There is, however, a strong caveat concerning the $R$ determination from 
our fit to $\sigma_\text{red}$. We use our explicit tensor pomeron model and, 
thus, our $R$ values are not derived in a model-independent way. We cannot 
exclude the possibility that a different model may give somewhat different 
results for $R$ from a fit to $\sigma_\text{red}$. 

The next topic we want to address briefly concerns the twist expansion 
for the structure functions of DIS; see for instance \cite{Yndurain1983}. 
Note that the twist expansion is, in essence, an expansion in inverse 
powers of $Q^2$. Thus, it only makes sense for sufficiently large $Q^2$ 
and, certainly, cannot be extended down to $Q^2=0$. 
It is well known that the leading twist-2 terms correspond to the 
QCD-improved parton picture with parton distributions obeying the 
famous DGLAP evolution equations \cite{Gribov:1972ri,Altarelli:1977zs,Dokshitzer:1977sg}. 
In our framework the question arises how the hard and soft pomeron 
contributions will contribute to leading and higher twists. It is tempting 
to associate, at large enough $Q^2$, the hard pomeron contribution with 
leading twist 2 and the soft pomeron contribution with higher twists. 
Indeed, the latter vanishes relative to the former for large $Q^2$ 
where the ratios of the $\mathbbm{P}_j \gamma^* \gamma^*$ coupling 
functions $\hat{a}_j(Q^2)$ and $\hat{b}_j(Q^2)$ for 
the soft ($j=1$) and hard ($j=0$) pomeron behave as 
\be
\label{ratiosofcouplfcts}
\begin{split}
\frac{\hat{a}_1(Q^2)}{\hat{a}_0(Q^2)} & \propto (Q^2)^{\delta_0-\delta_1} \approx (Q^2)^{-3}\,,
\\
\frac{\hat{b}_1(Q^2)}{\hat{b}_0(Q^2)} & \propto (Q^2)^{n_{1,7} - n_{0,7}} \approx (Q^2)^{-1.2} \,;
\end{split}
\ee
see tables \ref{tab4} and \ref{tab:splineparam}. 
This point of view, as expressed above, is close to what was advocated in \cite{Donnachie:2001zt}. 
Following \cite{Donnachie:2001zt} we would then conclude that higher twist effects -- 
the soft pomeron contribution -- stay important for $x<0.01$ up to 
$Q^2 \approx 20\,\mbox{GeV}^2$. 
Certainly, it will be worthwhile to study in more detail the connection 
of our two-tensor-pomeron model with the description of the HERA 
data using parton distribution functions and with the 
DGLAP and BFKL \cite{Kuraev:1977fs,Balitsky:1978ic} evolution equations. 
But this clearly goes beyond the scope of the present work. 

As we have stated in the introduction it is not our aim here to give a comparison 
of the various theoretical approaches to low-$x$ DIS physics. Let us just briefly 
comment on some recent fits to the HERA low-$x$ data where various 
methods were used. In \cite{Collaboration:2010ry} a so-called $\lambda$-fit 
in which $F_2$ is approximated by a power law in $x$ with a $Q^2$-dependent 
exponent was presented. The ansatz was then extended by adding in this 
exponent a '$\lambda'$ term' proportional to $\ln x$. Furthermore, 
a fit based on DGLAP evolution, as well as dipole model fits were 
presented. In \cite{Abt:2016vjh} a higher-twist ansatz was added to a DGLAP fit. 
Dipole models were used for example in \cite{Luszczak:2016bxd}, and DGLAP 
fits with BFKL-type low-$x$ resummation improvement in \cite{Ball:2017otu} and 
\cite{Abdolmaleki:2018jln}. However, in all these approaches the limit $Q^2 \to 0$, 
that is the photoabsorption cross section, is not included in the considerations. 
Typically, a minimum $Q^2$ of order $3.5 \,\mbox{GeV}^2$ is 
imposed.\footnote{We would like to point out that it is not surprising that dipole 
model fits have difficulties for very low $Q^2$. At low momenta, the use of the 
lowest order photon wave functions becomes questionable. In addition, most 
dipole models (including the ones mentioned above) use Bjorken-$x$ 
as energy variable of the dipole-proton cross section. This means that 
for $Q^2=0$, which implies $x=0$, the dipole-proton cross section is 
constant and, thus, has no energy dependence. Consequently, also the 
total photoabsorption cross section $\sigma_{\gamma p} (W)$ can, in these 
models, have no energy dependence -- in contradiction to experiment; 
see fig.\ \ref{fig4}. Indeed, it has been argued 
in \cite{Ewerz:2004vf,Ewerz:2006vd,Ewerz:2011ph} that in the dipole-proton 
cross section $W$ should be used as the energy variable.} 
In our approach, on the other hand, photoabsorption is treated in the same 
framework as DIS, allowing a detailed investigation of the transition from 
hard to soft scattering. 

\section{Conclusions}
\label{sec:conclusions}

In summary, we have presented a fit, based on a two-tensor-pomeron model, 
to photoproduction and low-$x$ deep-inelastic lepton-nucleon scattering data 
from HERA.  
We have determined the intercepts of the soft and hard pomeron and of 
the $f_{2R}$ reggeon, obtaining very reasonable numbers; see table \ref{tab2}. 

The two-tensor-pomeron model allows us to describe the 
transition from $Q^2=0$ and low $Q^2$, where the real or virtual photon 
acts hadron-like and the soft pomeron dominates, to high $Q^2$, 
the hard scattering regime dominated by the hard pomeron. 
The transition region where both pomerons contribute significantly 
was found to be roughly $0 < Q^2 < 20\,\mbox{GeV}^2$. 
For the photoproduction cross section $\sigma_{\gamma p}(W)$ we 
found no significant contribution from the hard pomeron. Thus, 
$\sigma_{\gamma p}(W)$ is, in the c.\,m.\ energy range 
$6\,\mbox{GeV} < W < 209 \,\mbox{GeV}$, dominated by soft-pomeron 
exchange with a significant $f_{2R}$ contribution for $W < 30\,\mbox{GeV}$. 

In the high-$Q^2$ and low-$x$ regime of DIS we found a good representation 
of the $\gamma^* p$ cross sections $\sigma_T + \sigma_L$ and $\sigma_L$ 
as products of simple powers in $Q^2$ and $W^2$; see \eqref{critexp1}-\eqref{deltaeta}. 
This may suggest that low-$x$ phenomena at high enough $Q^2$ may have 
an interpretation as a critical phenomenon as suggested in \cite{Nachtmann:2002yd}. 

In contrast to our tensor-pomeron model which gives an excellent description 
of the real photoabsorption cross section we found that a vector ansatz 
for the pomeron is ruled out as it gives zero contribution there; see 
section \ref{sec:ComptonAmplitude} and appendix \ref{appVector}. 

We are looking forward to further tests of our two-tensor-pomeron model 
at future lepton-proton scattering experiments in the low-$x$ 
regime, for instance at a future Electron-Ion-Collider \cite{Accardi:2012qut} 
or a Large Hadron Electron Collider LHeC \cite{AbelleiraFernandez:2012cc}. 
In particular, measurements of $\sigma_L$ and $R=\sigma_L/\sigma_T$ 
would be very welcome since these quantities are potentially very promising 
for a discrimination between different models, while at present their 
experimental errors are large.

\section*{Acknowledgments}

The authors thank M.\ Maniatis for providing templates for some of the 
diagrams in this article. O.\,N.\ thanks A.\ Donnachie and P.\,V.\ Landshoff 
for correspondence and M.\ Diehl for discussions. Preliminary results 
of this study were presented by O.\,N.\ at the conference EDS Blois 2017 
in June 2017 in Prague and at the meeting 'QCD -- Old Challenges and 
New Opportunities' at Bad Honnef in September 2017. Thanks go to the 
organisers of these meetings for the friendly and stimulating atmosphere there. 

\appendix

\section{Effective propagators and vertices}
\label{appA}

For the soft pomeron $\mathbbm{P}_1$ we use the effective 
propagator as given in (3.10) and (3.11) of \cite{Ewerz:2013kda}, \\
\vspace*{.2cm}
\hspace*{0.5cm}\includegraphics[width=150pt]{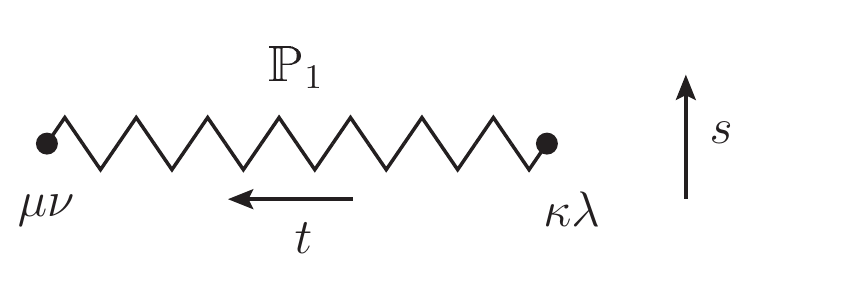}
\be\label{A.1}
i\Delta^{(\mathbbm{P}_1)}_{\mu\nu,\kappa\lambda} (s,t) 
= \frac{1}{4s} \left(g_{\mu\kappa} g_{\nu\lambda} + g_{\mu\lambda} g_{\nu\kappa} 
- \frac{1}{2} g_{\mu\nu} g_{\kappa\lambda} \right) 
\, (-i s \tilde{\alpha}'_1)^{\alpha_1(t)-1} \,.
\ee
The $\mathbbm{P}_1$ trajectory function is taken as linear in $t$, 
\be\label{A.2}
\alpha_1(t) = 1 + \epsilon_1 + \alpha'_1 t \,,
\ee
For the slope parameter $\alpha'_1$ and the parameter $\tilde{\alpha}'_1$ 
multiplying the squared energy $s$ we take the default values from \cite{Ewerz:2013kda}, 
\be\label{A.2a}
\begin{split}
\alpha'_1 &= 0.25 \,\mbox{GeV}^{-2} \,,
\\
\tilde{\alpha}'_1 &= \alpha'_1 \,.
\end{split}
\ee
The intercept parameter $\epsilon_1$ is in our work left free to be fitted. 
From our fits described in section \ref{sec:CompExp} we find (see table \ref{tab2}) 
\be\label{A.3}
\epsilon_1 = 0.0935\, ({}^{+76}_{-64}) \,.
\ee
For the hard-pomeron propagator our ansatz is similar to \eqref{A.1}, \eqref{A.2}, \\
\vspace*{.2cm}
\hspace*{0.5cm}\includegraphics[width=150pt]{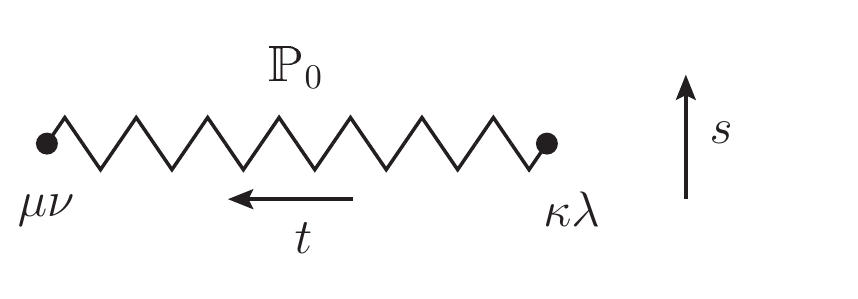}
\be\label{A.4}
i\Delta^{(\mathbbm{P}_0)}_{\mu\nu,\kappa\lambda} (s,t) 
= \frac{1}{4s} \left(g_{\mu\kappa} g_{\nu\lambda} + g_{\mu\lambda} g_{\nu\kappa} 
- \frac{1}{2} g_{\mu\nu} g_{\kappa\lambda} \right) 
\, (-i s \tilde{\alpha}'_0)^{\alpha_0 (t)-1} \,,
\ee
with
\be\label{A.4a}
\alpha_0 (t) = 1 + \epsilon_0 + \alpha'_0 t \,,
\ee
and the parameter $\epsilon_0$ to be determined from experiment.
For $\alpha'_0$ and $\tilde{\alpha}'_0$ we take, for lack of better 
knowledge, the same values as for the soft pomeron, 
\be\label{A.5a} 
\alpha'_0 = \tilde{\alpha}'_0 = 0.25 \,\mbox{GeV}^{-2} \,.
\ee
From the fits in section \ref{sec:CompExp} we get (see table \ref{tab2}) 
\be\label{A.5b}
\epsilon_0 = 0.3008\, ({}^{+73}_{-84}) \,.
\ee

The ansatz for the $\mathbbm{P}_{1} pp$ vertex is given in (3.43) of \cite{Ewerz:2013kda}. 
Making an analogous ansatz for the hard pomeron we get: \\
\vspace*{.2cm}
\hspace*{0.5cm}\includegraphics[height=90pt]{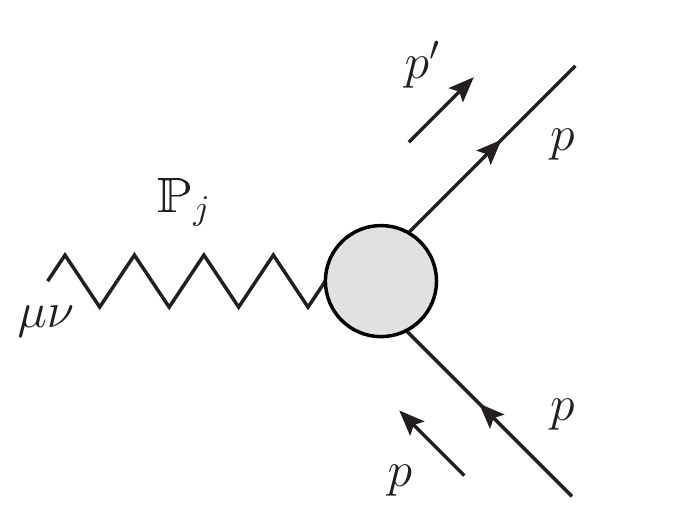} 
\be\label{A.5}
\begin{split}
i \Gamma_{\mu\nu}^{(\mathbbm{P}_j pp)}  (p',p) =& 
-i \,3 \beta_{j pp} F_1^{(j)}[(p'-p)^2]
\\ & \times 
\left\{ \frac{1}{2} \left[ \gamma_\mu (p'+p)_\nu + \gamma_\nu (p'+p)_\mu \right] 
-\frac{1}{4} \, g_{\mu\nu} (\slash{p}' + \slash{p}) \right\} \,, \qquad (j=0,1) \,.
\end{split}
\ee
Here $\beta_{jpp}$ are coupling constants of dimension $\mbox{GeV}^{-1}$ 
and $F_1^{(j)}(t)$ are form factors normalised to 
\be\label{A.6}
F_1^{(j)} (0) =1 \,.
\ee
The standard value for the coupling constant of the soft pomeron to protons is 
\be\label{A.7}
\beta_{1 pp} = 1.87 \, \mbox{GeV}^{-1} \,;
\ee
see (3.44) of \cite{Ewerz:2013kda}. The traditional choice for the form factor $F_1^{(1)} (t)$ 
is the Dirac electromagnetic form factor of the proton even if it is clear that 
this cannot be strictly correct; see the discussion in chapter 3.2 of \cite{Donnachie:2002en}. 
But this is not relevant for our present work where we only need the form factors 
at $t=0$ where they are equal to 1; see \eqref{A.6}. 
For lack of better knowledge we take 
\be\label{A.8an} 
\beta_{0pp} = \beta_{1pp} \,.
\ee
For the processes that we consider in the present paper this gives no restriction for our 
fits since only the products $\beta_{jpp} \hat{a}_j(Q^2)$ and 
$\beta_{jpp} \hat{b}_j(Q^2)$ enter as parameters. 

For our ansatz for the $\mathbbm{P}_j \gamma^* \gamma^*$ vertices we need the 
rank-4 tensor functions defined in (3.18) and (3.19) of \cite{Ewerz:2013kda}, 
\begin{align}\label{A.8a}
\Gamma_{\mu\nu\kappa\lambda}^{(0)} (k_1,k_2) =\,& 
[(k_1\cdot k_2) g_{\mu\nu} - k_{2\mu} k_{1\nu}] 
\left[k_{1\kappa}k_{2\lambda} + k_{2\kappa}k_{1\lambda} - 
\frac{1}{2} (k_1 \cdot k_2) g_{\kappa\lambda}\right] \,,
\\
\label{A.8b}
\Gamma_{\mu\nu\kappa\lambda}^{(2)} (k_1,k_2) = \,
& (k_1\cdot k_2) (g_{\mu\kappa} g_{\nu\lambda} + g_{\mu\lambda} g_{\nu\kappa} )
+ g_{\mu\nu} (k_{1\kappa} k_{2\lambda} + k_{2\kappa} k_{1\lambda} ) 
\nn \\
& - k_{1\nu} k_{2 \lambda} g_{\mu\kappa} - k_{1\nu} k_{2 \kappa} g_{\mu\lambda} 
- k_{2\mu} k_{1 \lambda} g_{\nu\kappa} - k_{2\mu} k_{1 \kappa} g_{\nu\lambda} 
\\
& - [(k_1 \cdot k_2) g_{\mu\nu} - k_{2\mu} k_{1\nu} ] \,g_{\kappa\lambda} \,.
\nn
\end{align}
We have for $i=0,2$
\be\label{A.9a}
\Gamma_{\mu\nu\kappa\lambda}^{(i)} (k_1,k_2) 
= \Gamma_{\mu\nu\lambda\kappa}^{(i)} (k_1,k_2) 
= \Gamma_{\nu\mu\kappa\lambda}^{(i)} (k_2,k_1) 
= \Gamma_{\mu\nu\kappa\lambda}^{(i)} (-k_1,-k_2) \,,
\ee
\be\label{A.9b}
\begin{split}
& k_1^\mu \Gamma_{\mu\nu\kappa\lambda}^{(i)} (k_1,k_2) =0 \,,
\\
& k_2^\nu \Gamma_{\mu\nu\kappa\lambda}^{(i)} (k_1,k_2) =0 \,,
\end{split}
\ee
\be\label{A.9c}
\Gamma_{\mu\nu\kappa\lambda}^{(i)} (k_1,k_2)\, g^{\kappa\lambda} =0 \,. 
\ee
Now we can write down our ansatz for the $\mathbbm{P}_j \gamma^* \gamma^*$ 
vertices in analogy to the $\mathbbm{P} \rho \rho$ vertex in (3.47) of \cite{Ewerz:2013kda}: \\
\vspace*{.2cm}
\hspace*{0.5cm}\includegraphics[height=90pt]{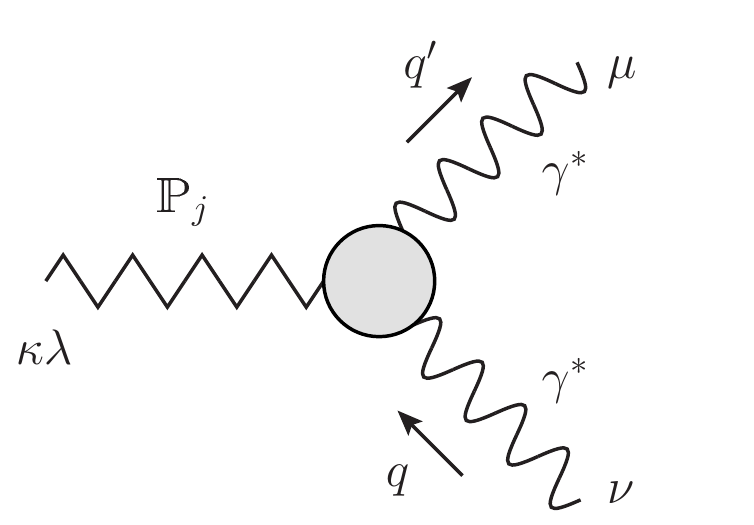} 
\be\label{A.10}
\begin{split}
i \Gamma_{\mu\nu\kappa\lambda}^{(\mathbbm{P}_j \gamma^* \gamma^*)} (q',q) 
=& i
\left[2 a_{j\gamma^*\gamma^*}(q^2, q'^2, t) \Gamma_{\mu\nu\kappa\lambda}^{(0)}(q',-q) 
- b_{j\gamma^*\gamma^*}(q^2, q'^2, t) \Gamma_{\mu\nu\kappa\lambda}^{(2)}(q',-q) \right] \,,
\\
& \qquad\qquad t=(q-q')^2\,, \qquad j=0,1\,.
\end{split}
\ee
Here the coupling parameters $a_{j\gamma^*\gamma^*}$ and 
$b_{j\gamma^*\gamma^*}$ have dimensions $\mbox{GeV}^{-3}$ and 
$\mbox{GeV}^{-1}$, respectively. In our present work only the values of these 
parameters for 
\begin{displaymath}
q^2=q'^2 = - Q^2 \,, \qquad t=0 
\end{displaymath}
enter. Therefore, we set, pulling out also a factor $e^2$, 
\be\label{A.11}
\begin{split}
a_{j\gamma^*\gamma^*} (-Q^2,-Q^2, 0) &= e^2 \hat{a}_j (Q^2) \,,
\\
b_{j\gamma^*\gamma^*} (-Q^2,-Q^2, 0) &= e^2 \hat{b}_j (Q^2) \,,
\\
j=0,1 \,.&
\end{split}
\ee

Our ans\"atze for the effective propagator and the vertices for $f_{2R}$-reggeon 
exchange are as follows. For the $f_{2R}$ propagator we set (see (3.12), (3.13) 
of \cite{Ewerz:2013kda})\\
\vspace*{.2cm}
\hspace*{0.5cm}\includegraphics[width=150pt]{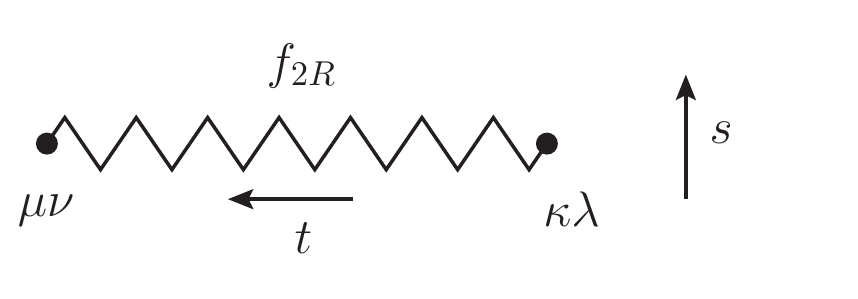} 
\be\label{A.12}
i\Delta^{(f_{2R})}_{\mu\nu,\kappa\lambda} (s,t) 
= \frac{1}{4s} \left(g_{\mu\kappa} g_{\nu\lambda} + g_{\mu\lambda} g_{\nu\kappa} 
- \frac{1}{2} g_{\mu\nu} g_{\kappa\lambda} \right) 
(-i s \tilde{\alpha}'_2)^{\alpha_2 (t)-1} \,,
\ee
\be\label{A.13}
\alpha_2 (t) = \alpha_2 (0)+ \alpha'_2 t \,
\ee
with $\alpha_2(0)$ as fit parameter. For $\alpha'_2$ and $\tilde{\alpha}'_2$ 
we take the default values from (3.13) of \cite{Ewerz:2013kda}:
\be\label{A.13a}
\begin{split}
\alpha'_2 &= 0.9 \,\mbox{GeV}^{-2} \,,
\\
\tilde{\alpha}'_2 &= \alpha'_2 \,. 
\end{split}
\ee
Our fit gives (see table \ref{tab2}) 
\be\label{A.17b} 
\alpha_2(0) = 0.485 \,({}^{+88}_{-90}) 
\ee
which is nicely compatible with the default value from (3.13) of \cite{Ewerz:2013kda}: 
$\alpha_2(0) = 0.5475$. 

The $f_{2R} pp$ vertex is given in (3.49), (3.50) of \cite{Ewerz:2013kda} as\\
\vspace*{.2cm}
\hspace*{0.5cm}\includegraphics[height=90pt]{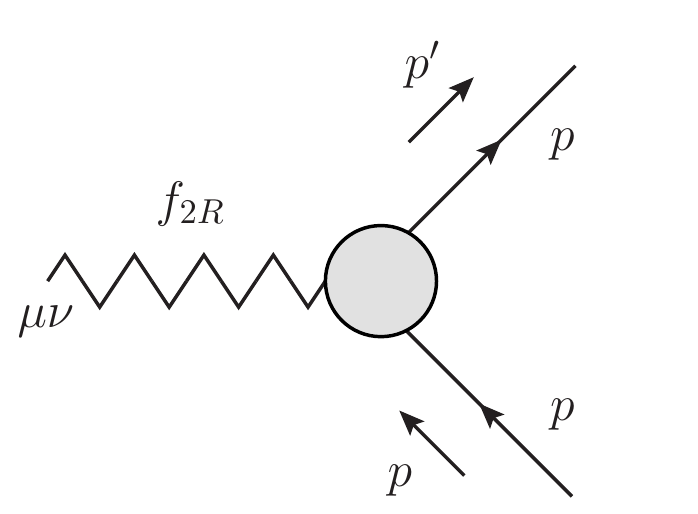} 
\be\label{A.14}
\begin{split}
i \Gamma_{\mu\nu}^{(f_{2R} pp)} (p',p) =& - i g_{f_{2R}pp} \frac{1}{M_0}
F_1[ (p'-p)^2] \\
& \times \bigg\{ \frac{1}{2} [ \gamma_\mu (p'+p)_\nu + \gamma_\nu (p'+p)_\mu ]
- \frac{1}{4} g_{\mu\nu} (\slash{p}' + \slash{p}) \bigg\} \,,
\end{split}
\ee
\be\label{A.15}
g_{f_{2R}pp} = 11.04\,, \qquad M_0 = 1~\text{GeV}\,.
\ee
In our paper we use as coupling parameter 
\be
\label{A.19a}
\beta_{2pp}= \frac{1}{3 M_0} \, g_{f_{2R}pp} = 3.68~\text{GeV}^{-1}\,.
\ee

The ansatz for the $f_{2R}\gamma^{(*)} \gamma^{(*)}$ vertex for real and virtual 
photons will be taken with the same structure as for $f_2\gamma \gamma$ (see (3.39), 
(3.40) of \cite{Ewerz:2013kda}), \\
\hspace*{.7cm}
\begin{tabular}{@{}c@{}}{\includegraphics[height=1.25in]{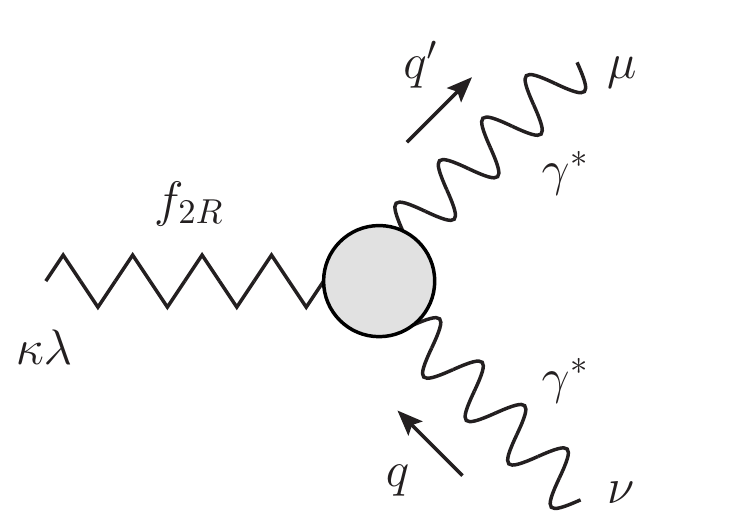}}\end{tabular}
\hspace*{.5cm}
$i \Gamma_{\mu\nu\kappa\lambda}^{(f_{2R} \gamma^* \gamma^*)} (q',q) \,.$

In the present work we need this vertex only for 
\be\label{A.15a}
q'=q\,,\qquad q^2=-Q^2 \le 0\,,
\ee
and our ansatz for this case reads 
\be\label{A.16}
i \Gamma_{\mu\nu\kappa\lambda}^{(f_{2R} \gamma^* \gamma^*)} (q,q) 
= i \,  \left[ 2 e^2 \hat{a}_2 (Q^2) \,\Gamma_{\mu\nu\kappa\lambda}^{(0)}(q,-q) 
- e^2 \hat{b}_2 (Q^2)\,\Gamma_{\mu\nu\kappa\lambda}^{(2)}(q,-q) \right] \,.
\ee

\section{Formulae for a hypothetical vector pomeron}
\label{appVector}

In this appendix we collect the necessary formulae for the 
(hypothetical) vector pomeron couplings to protons and real photons. 
These formulae are used in section \ref{sec:ComptonAmplitude}. 
The $\mathbbm{P}_V pp$ vertex 
and the $\mathbbm{P}_V$ propagator are standard; see e.\,g.\ 
\cite{Donnachie:2002en} and appendix B of \cite{Lebiedowicz:2013ika}. We have \\
\vspace*{.2cm}
\hspace*{0.5cm}\includegraphics[height=90pt]{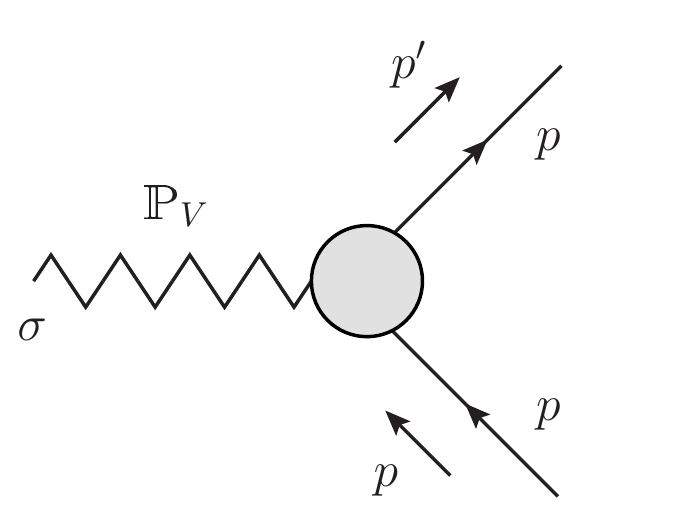} 
\be\label{B.2}
i \Gamma_\sigma^{(\mathbbm{P}_V pp)} (p',p) 
= - i \, 3 \beta_{\mathbbm{P}_V pp} F_1[(p-p')^2] M_0 \gamma_\sigma \,,
\ee
with $\beta_{\mathbbm{P}_V pp} = 1.87 \, \mbox{GeV}^{-1}$, $M_0 = 1\, \mbox{GeV}$, 
and \\
\vspace*{.2cm}
\hspace*{0.5cm}\includegraphics[width=150pt]{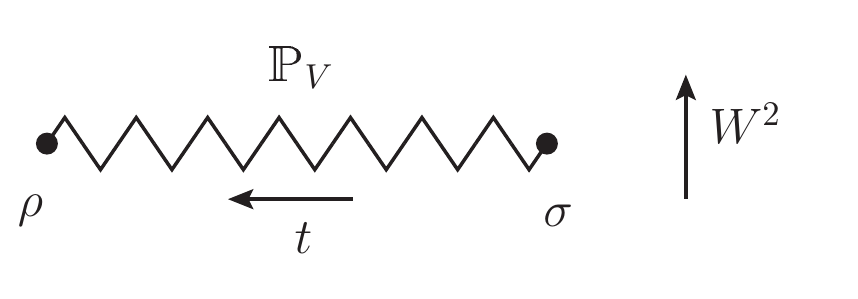}
\be\label{B.3} 
i \Delta_{\rho\sigma}^{(\mathbbm{P}_V)} (W^2,t) = 
\frac{1}{M_0^2} \, g_{\rho \sigma} ( - i W^2 \alpha'_{{\mathbbm{P}_V}})^{\alpha_{\mathbbm{P}_V}(t)-1} \,.
\ee
In \eqref{B.2} $F_1(t)$ is a form factor normalised to $F_1(0)=1$. 
In \eqref{B.3} $\alpha_{\mathbbm{P}_V}(t)$ is the vector pomeron 
trajectory function and $\alpha'_{\mathbbm{P}_V}$ is the slope parameter. 
The numerical values for these quantities play no role in the following and in 
section \ref{sec:ComptonAmplitude}. 
For the $\mathbbm{P}_V\gamma\gamma$ vertex we assume that it 
respects the standard rules of QFT. We have, orienting here 
for simplicity both photons as outgoing, 
\begin{equation}
\label{B.4}
\mbox{
\hspace*{.7cm}
\begin{tabular}{@{}c@{}}{\includegraphics[height=1.35in]{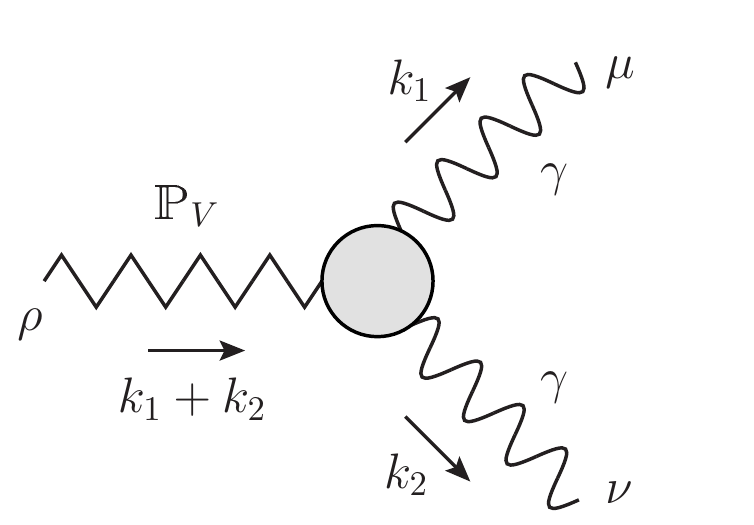}}\end{tabular}
\hspace*{.5cm}
$i \Gamma_{\mu\nu\rho}^{(\mathbbm{P}_V \gamma \gamma)} (k_1,k_2) \,.$
}
\end{equation}
For this vertex function we have the constraints of Bose symmetry for 
the two photons, 
\be\label{B.5}
\Gamma_{\mu\nu\rho}^{(\mathbbm{P}_V \gamma \gamma)} (k_1,k_2) 
= \Gamma_{\nu\mu\rho}^{(\mathbbm{P}_V \gamma \gamma)} (k_2,k_1) \,,
\ee
and of gauge invariance, 
\be\label{B.6}
\begin{split}
k_1^\mu \Gamma_{\mu\nu\rho}^{(\mathbbm{P}_V \gamma \gamma)} (k_1,k_2) &=0\,,
\\
k_2^\nu \Gamma_{\mu\nu\rho}^{(\mathbbm{P}_V \gamma \gamma)} (k_1,k_2) &=0\,.
\end{split}
\ee
The vertex $\mathbbm{P}_V \gamma\gamma$ should also respect 
parity invariance. We have then 14 tensors, constructed from $k_1$, $k_2$ 
and the metric tensor, at our disposal, 
\be\label{B.8}
\begin{split}
&k_{1\mu} k_{1\nu} k_{1\rho}\,, \qquad k_{1\mu} k_{1\nu} k_{2\rho}\,, \qquad 
k_{1\mu} k_{2\nu} k_{1\rho}\,, \qquad k_{1\mu} k_{2\nu} k_{2\rho}\,,
\\
&
k_{2\mu} k_{1\nu} k_{1\rho}\,, \qquad k_{2\mu} k_{1\nu} k_{2\rho}\,, \qquad 
k_{2\mu} k_{2\nu} k_{1\rho}\,, \qquad k_{2\mu} k_{2\nu} k_{2\rho}\,,
\\
& g_{\mu \nu} k_{1 \rho}\,, \qquad g_{\mu \rho} k_{1\nu}\,, \qquad g_{\nu\rho} k_{1\mu} \,,\qquad
g_{\mu \nu} k_{2  \rho}\,, \qquad g_{\mu \rho} k_{2\nu}\,, \qquad g_{\nu\rho} k_{2\mu} \,.
\\
\end{split}
\ee
To construct the most general vertex \eqref{B.4} we have to multiply 
these tensors with invariant functions depending on $k_1^2$, $k_2^2$, 
and $(k_1+k_2)^2$, and take their sum. In the following, however, we 
shall only consider the case $k_1^2=k_2^2$. 
With the requirement \eqref{B.5} we obtain then the following general 
form for $\Gamma^{(\mathbbm{P}_V \gamma \gamma)}$: 
\be\label{B.9}
\begin{split}
\Gamma_{\mu\nu\rho}^{(\mathbbm{P}_V \gamma \gamma)} (k_1,k_2)=& {}\,
A_1 (k_{1\mu} k_{1\nu} k_{1\rho} + k_{2\mu} k_{2\nu} k_{2\rho}) 
+ A_2 (k_{1\mu} k_{1\nu} k_{2\rho} + k_{2\mu} k_{2\nu} k_{1\rho}) 
\\
&+A_3 (k_{1\mu} k_{2\nu} k_{1\rho} + k_{1\mu} k_{2\nu} k_{2\rho}) 
+ A_4 (k_{2\mu} k_{1\nu} k_{1\rho} + k_{2\mu} k_{1\nu} k_{2\rho}) 
\\
&+ A_5 (g_{\mu \nu} k_{1 \rho} + g_{\mu \nu} k_{2  \rho}) 
+ A_6 (g_{\mu \rho} k_{1\nu} + g_{\nu \rho} k_{2\mu}) 
+ A_7 (g_{\nu\rho} k_{1\mu} + g_{\mu \rho} k_{2\nu}) 
\,,
\end{split}
\ee
with coefficient functions 
\be\label{B.10}
A_j = A_j ( k_1^2, (k_1+k_2)^2) \,, \qquad j=1,\dots,7 \,.
\ee
Imposing gauge invariance we find, using \eqref{B.6}, the relations 
\be\label{B.11}
\begin{split}
k_1^2 A_1 + (k_1 \cdot k_2) A_4 + A_5 + A_6 &=0\,,
\\
(k_1 \cdot k_2) A_1 + k_1^2 A_3 &=0\,,
\\
k_1^2 A_2 + (k_1 \cdot k_2) A_4 + A_5 &=0\,,
\\
(k_1 \cdot k_2) A_2 + k_1^2 A_3 + A_7 &=0\,,
\\
(k_1 \cdot k_2)  A_6 + k_1^2 A_7 &=0\,.
\end{split}
\ee

Now we specialise for real photons and assume a general, non-vanishing 
product of their 4-momenta, 
\be\label{B.12}
k_1^2 = k_2^2 = 0 \,, \qquad k_1 \cdot k_2 \neq 0 \,.
\ee
This gives 
\be\label{B.13}
\begin{split}
A_1 &= 0 \,,
\\
A_6 &= 0 \,,
\\
A_5 &= - (k_1 \cdot k_2) A_4 \,,
\\
A_7 &= - (k_1 \cdot k_2) A_2 
\end{split}
\ee
and hence the final form for $\Gamma^{(\mathbbm{P}_V \gamma \gamma)}$: 
\be\label{B.14}
\begin{split}
\Gamma_{\mu\nu\rho}^{(\mathbbm{P}_V \gamma \gamma)} (k_1,k_2) =& {}\,
\hat{A}_2 \left[k_{1\mu} ( k_{1\nu} k_{2 \rho} - (k_1 \cdot k_2) g_{\nu\rho}) 
+ (k_{2\mu} k_{1\rho} - (k_1 \cdot k_2) g_{\mu\rho}) k_{2\nu} \right]
\\
&+ \hat{A}_3 k_{1\mu} k_{2\nu} (k_{1\rho} + k_{2\rho}) 
\\
&+ \hat{A}_4 ( k_{2\mu} k_{1\nu} - (k_1 \cdot k_2) g_{\mu\nu}) (k_{1\rho} + k_{2\rho}) \,,
\end{split}
\ee
where the remaining coefficient functions depend only on $(k_1 + k_2)^2$, 
\be\label{B.154a}
\hat{A}_j = A_j (0,(k_1+k_2)^2) \equiv \hat{A}_j ((k_1+k_2)^2)\,, \qquad j=2,3,4 \,.
\ee
The replacements $k_1 \to q'$ and $k_2 \to -q$ lead to the vertex 
function \eqref{mainB14}. Inserting this in the expression for the Compton 
amplitude corresponding to the diagram in fig.\ \ref{fig16} gives a vanishing 
result; see \eqref{B.15}. 

We note that this type of vertex function \eqref{B.14} would also describe the 
parity conserving decay of a vector particle of spin parity $J^P = 1^-$ 
to two real photons. In accord with the famous 
Landau-Yang theorem \cite{Landau:1948kw,Yang:1950rg}, \eqref{B.14} 
gives zero for the corresponding amplitude. 
Indeed, consider the decay of such a vector particle
\be
\label{B.17}
V(k,\varepsilon) \longrightarrow \gamma(k_1,\varepsilon_1) + \gamma(k_2,\varepsilon_2)\,,
\ee
where 
\be
\begin{gathered}
k_1^2 = k_2^2 =0 \,, \qquad k=k_1 + k_2 \,,
\\
k^2 = m_V^2\,, \qquad k \cdot \varepsilon =0 \,, \qquad k_1 \cdot \varepsilon_1 = k_2 \cdot \varepsilon_2 =0\,.
\end{gathered}
\ee
With \eqref{B.14} we find then 
\be
 \langle \gamma(k_1, \varepsilon_1), \gamma(k_2, \varepsilon_2) | \mathcal{T}| 
V(k,\varepsilon) \rangle = 
\varepsilon^{*\mu}_1 \varepsilon^{*\nu}_2 \,
\Gamma_{\mu\nu\rho}^{(\mathbbm{P}_V \gamma\gamma)} (k_1,k_2) \,\varepsilon^\rho 
=0 \,.
\ee
Note that the Landau-Yang theorem applies to the decay of a massive vector 
particle to two photons. In our present discussion, the vector pomeron exchanged 
in the $t$-channel plays the role of the massive vector particle. 

In conclusion, the same reasoning which leads to the Landau-Yang 
theorem shows that a vector pomeron cannot couple in real 
Compton scattering. But clearly, the behaviour of the total 
$\gamma p$ absorption cross section as measured shows that the 
pomeron does couple in real Compton scattering. 
The tensor pomeron model describes this coupling without 
problems in a satisfactory way; see section \ref{sec:CompExp}, figure \ref{fig4}. 

\section{Parametrisation for coupling functions}
\label{appParam}

\subsection{Reggeon exchange parametrisation}
\label{appParamregge}

For the $f_{2R}$ reggeon, which is expected to contribute only at low $W$ and
low $Q^2$, the following assumptions are made:
\begin{align}
\label{D.1}
  \hat{a}_2(Q^2) & =  0 \,,\\
\label{D.2}
  \hat{b}_2(Q^2) & =  c_2 \exp \left[ -Q^2/d_2\right]\,,
\end{align}
with two fit parameters. The parameter $c_2$
describes the magnitude of the $f_{2R}$ reggeon exchange contribution in
photoproduction. The exponential function containing the parameter 
$d_2>0$ causes the reggeon contribution to vanish 
rapidly with increasing $Q^2$.

\subsection {Pomeron exchange parametrisation}
\label{appParampom}

For the two tensor-pomeron exchanges $\mathbbm{P}_j$, 
$j=0$ and $j=1$, the functions $Q^2\hat{a}_j(Q^2)$ are parametrised as
\be\label{D.3}
Q^2\hat{a}_j(Q^2) =  a_j \frac{Q^2}{m_j^2} \left(\frac{\delta_j+Q^2/m_j^2}{\delta_j+1} \right)^{-1-\delta_j}\,.
\ee
For $\delta_j>0$, this function has a maximum at $Q^2=m_j^2$ with
magnitude $a_j$. For small $Q^2$, the function increases proportionally
to $Q^2$. The parameter $\delta_j>0$ defines the power exponent 
by which the function drops with large $Q^2$.

The functions $\hat{b}_j(Q^2)$ for $j=0$ or $j=1$ are parametrised with
the help of cubic splines $s_j$ with $N=7$ knots each.
Between two knots, 
$z_{j,i}$ and $z_{j,i+1}$, the spline $s_j(z)$ is given by third-order polynomials
\be\label{D.4}
s_{j}(z) = A_{j,i}+ B_{j,i}(z-z_{j,i})+ C_{j,i}(z-z_{j,i})^2+
D_{j,i}(z-z_{j,i})^3\quad\text{for}\quad z_{j,i}\le z\le z_{j,i+1}\,,
\ee
with coefficients $A_{j,i}$, $B_{j,i}$, $C_{j,i}$, $D_{j,i}$
($i=1,\dots, N-1$) and knot positions $z_{j,i}$ ($i=1,\dots, N$).
The function $\hat{b}_j(Q^2)$ is given by
$\exp[s_j(z)]$ using the argument $z=\ln((Q^2+q_{j,0}^2)/M_0^2)$ with $M_0=1\,\mbox{GeV}$.
The offset $q_{j,0}^2$ ensures that $z$ is finite for $Q^2=0$.
The knot positions $z_{j,i}=\log((q^2_{j,i}+q^2_{j,0})/M_0^2)$ are given using fixed
positions in $Q^2$, denoted $q^2_{j,i}$ and ranging from 
$q^2_{j,1}=0$ to $q^2_{j,7}=50\,\text{GeV}^2$. The offset is taken to
be equal to the first nonzero position, $q_{j,0}^2=q_{j,2}^2$.
For the fit, the $2\times 7$ function values
$\hat{b}_{j}(q^2_{j,i})$ are taken as free
parameters.
Given $j$, the $4\times( N-1)$ spline parameters $A_{j,i}$, $B_{j,i}$,
$C_{j,i}$ and $D_{j,i}$ are determined from the fit parameters 
using the usual constraints
on the spline to be continuous up to the second derivatives. The endpoint 
conditions are chosen such that the second derivatives of $s_{j}(z)$
vanish for both $z=z_{j,1}$ and $z=z_{j,7}$.

For predictions at large $Q^2$, the functions $\hat{b}_j(Q^2)$ are
continued for $Q^2>q^2_{j,N}$ using the spline properties at the
endpoint $z_{j,N}$,
\begin{align}
\label{D.5}
\hat{b}_j(Q^2) & = \hat{b}_j(q^2_{j,N})
\left(
  \frac{Q^2+q_{j,0}^2}{q_{j,N}^2+q_{j,0}^2}\right)^{n_{j,N}}\quad\text{for
}Q^2\ge q_{j,N}^2\,,\\
\label{D.6}
\text{where} \quad
n_{j,N} & =  \left.\frac{ds_j}{dz}\right\vert_{z_{j,N}}\,.
\end{align}
Similarly, for cases where $q^2_{j,1}>0$, the function is defined in the region $-q^2_{j,0}<Q^2<q^2_{j,1}$ as
\be\label{D.7}
\hat{b}_j(Q^2)  = \hat{b}_j(q^2_{j,1})
\left(
  \frac{Q^2+q_{j,0}^2}{q_{j,1}^2+q_{j,0}^2}\right)^{B_{j,0}}\quad\text{for
} -q_{j,0}<Q^2< q_{j,1}^2\,.
\ee
A special case is given by $q^2_{j,1}>0$, $q^2_{j,0}=0$ and
$B_{j,0}<0$. In this case $\hat{b}_j(Q^2)\to 0$ for $Q^2\to0$.
In all cases discussed above, the resulting function $\hat{b}_j$ is
defined for all $Q^2>-q_{j,0}$ and is 
continuous up to the second derivative over the full allowed $Q^2$ range.

\section{Fit procedure}
\label{appfit}

A fit with $25$ free parameters is made using the ALPOS package \cite{Alpos}, an
interface to Minuit \cite{James:1975dr}.  The goodness-of-fit function
is defined as
\begin{equation}
\label{d1}
\begin{split}
\chi^2(h)  =&\sum_{i,j}
\left(\log\sigma^{\text{HERA}}_i-\log\sigma_{\text{red}}(Q^2_i,x_i,y_i;h)\right)
\left(V_{\text{HERA}}^{-1}\right)_{ij} \\
& \qquad \times 
\left(\log\sigma^{\text{HERA}}_j-\log\sigma_{\text{red}}(Q^2_j,x_j,y_j;h)\right) \\
 & +\sum_{i,j}
\left(\log\sigma^{\text{PHP}}_i-\log\sigma_{T}(W_i;h)\right)
\left(V_{\text{PHP}}^{-1}\right)_{ij}
\left(\log\sigma^{\text{PHP}}_j-\log\sigma_{T}(W_i;h)\right)\,,
\end{split}
\end{equation}
where $\sigma^{\text{HERA}}_i$ with $i=1,\ldots,525$ are measurements
of reduced cross sections from HERA \cite{Abramowicz:2015mha} and
$Q^2_i$, $x_i$, $y_i$ are the corresponding kinematic variables. 
The prediction $\sigma_{\text{red}}(Q^2_i,x_i,y_i;h)$ depends on the
kinematic variables and on the vector $h$ of the $25$ fit parameters.
The data covariance matrix includes two types of relative uncertainties,
point-to-point uncorrelated, $u_i$, and point-to-point
correlated from a source $k$, $c_{ki}$. The elements of the resulting
covariance matrix are $(V_{\text{HERA}})_{ij} = \delta_{ij}(u_i)^2+\sum_k
c_{ki}c_{kj}$, where $\delta_{ij}$ is the Kronecker symbol.
There are $169$ sources $k$ of correlated uncertainties in 
the HERA data.

A total of $36$ photoproduction data points are included in a similar
manner. The measurements are denoted $\sigma^{\text{PHP}}_i$ with
$i=1,\ldots,36$  and the
corresponding energies
are $W_i$. The predictions are $\sigma_{T}(W_i;h)$.
The covariance matrix $V_{\text{PHP}}^{-1}$ receives
uncorrelated and correlated contributions in analogy to the HERA data discussed above. 
There are two photoproduction
measurements from H1 and ZEUS at high $W$ \cite{Aid:1995bz,Chekanov:2001gw}
and four astroparticle measurements at intermediate $W$
\cite{Vereshkov:2003cp}. These six data points are not correlated to the
other data points. The $30$ low-$W$ data points from Fermilab
\cite{Caldwell:1978yb} have a single correlated contribution in
addition to their uncorrelated uncertainties, a
$0.7\%$ normalisation uncertainty.

The function $\chi^2(h)$ is minimised with respect to $h$ to estimate
the parameters.  For the fit parameters, asymmetric experimental
uncertainties are obtained using the MINOS \cite{James:1975dr}
algorithm. For all other 
quantities shown in this paper, uncertainties are determined as follows.
The HESSE algorithm \cite{James:1975dr} determines the symmetric 
covariance matrix $V$ of the parameter vector $h$ at the minimum
$\hat{h}$ of the log-likelihood function. 
Using an eigenvalue decomposition, the matrix $V$ is written in terms
of dyadic products of orthogonal uncertainty vectors $\delta h_i$, 
$V=\sum_i \delta h_i \,\delta h_i^{T}$. Asymmetric
uncertainties, $+\Delta f_{\text{up}}$ and $-\Delta f_{\text{dn}}$, of a
generic quantity $f(h)$ are then estimated as follows: 
\begin{align}
\label{eqn:errorprop1}
\Delta f_{\text{up}} & =
\sqrt{
  \sum_i{
    \left(
      \max\left[f(\hat{h}+\delta h_i),f(\hat{h}-\delta h_i)\right]
      -f(\hat{h})
    \right)
    ^2}}
\,,
\\
\label{eqn:errorprop2}
\Delta f_{\text{dn}} & = 
\sqrt{
  \sum_i{
    \left(
      \min\left[f(\hat{h}+\delta h_i),f(\hat{h}-\delta h_i)\right]
      -f(\hat{h})
    \right)
    ^2}}
\,.
\end{align}
The uncertainties obtained in this way are termed 'Hessian
uncertainties' or 'one standard deviations' in this paper.

\section{Fit results}
\label{appfitres}

The goodness-of fit found after minimizing and the partial $\chi^2$ numbers 
calculated for individual data sets are summarised in table \ref{tab:fitqual}. 
An acceptable fit probability of $6\%$ is observed. There is no single
dataset which contributes much more than expected to $\chi^2$.
The resulting $25$ parameters at the minimum are summarised in table
\ref{tab4} with their MINOS uncertainties. For technical
reasons, most fit parameters actually are defined as the logarithm of the
corresponding physical quantity.
\begin{table}
  \centering
  {
    \small
  \renewcommand{\arraystretch}{1.5}
\begin{tabular}{|lr|}
\hline
 fit parameter & result \\
\hline
$\epsilon_0$ & $0.3008 (^{+73}_{-84})$\\
$\epsilon_1$ & $0.0935 (^{+76}_{-64})$\\
$\alpha_2(0)$ & $0.485 (^{+88}_{-90})$\\
\hline
$\log(c_2/\text{GeV}^{-1})$ & $-0.38 (^{+36}_{-35})$\\
$\log(d_2/\text{GeV}^{-2})$ & $-1.35 (^{+34}_{-35})$\\
\hline
$\log(a_0/\text{GeV}^{-1})$ & $-6.95 (^{+29}_{-25})$ \\
$\log(m_0^2/\text{GeV}^{2})$ & $1.41 (^{+27}_{-31})$ \\
$\log(\delta_0)$ & $0.92 (^{+24}_{-26})$\\
\hline
$\log(a_1/\text{GeV}^{-1})$ & $-3.92 (^{+18}_{-20})$ \\
$\log(m_1^2/\text{GeV}^{2})$ & $-0.31 (^{+20}_{-19})$ \\
$\log(\delta_1)$ & $1.72 (^{+59}_{-48})$\\
\hline
\end{tabular}
\hspace*{5mm}
\begin{tabular}{|lr|}
\hline
 fit parameter & result \\
\hline
$\log(\hat{b}_0(       0\,\text{GeV}^2)/\text{GeV}^{-1})$ & $-14.2 (^{+30}_{-39})$\\
$\log(\hat{b}_0(     0.3\,\text{GeV}^2)/\text{GeV}^{-1})$ & $-7.02 (^{+69}_{-87})$\\
$\log(\hat{b}_0(       1\,\text{GeV}^2)/\text{GeV}^{-1})$ & $-4.83 (^{+15}_{-16})$\\
$\log(\hat{b}_0(       3\,\text{GeV}^2)/\text{GeV}^{-1})$ & $-5.09 (11)$\\
$\log(\hat{b}_0(      10\,\text{GeV}^2)/\text{GeV}^{-1})$ & $-5.669 (^{+99}_{-101})$\\
$\log(\hat{b}_0(      25\,\text{GeV}^2)/\text{GeV}^{-1})$ & $-6.268 (^{+89}_{-91})$\\
$\log(\hat{b}_0(      50\,\text{GeV}^2)/\text{GeV}^{-1})$ & $-6.899 (^{+78}_{-80})$\\
\hline
$\log(\hat{b}_1(       0\,\text{GeV}^2)/\text{GeV}^{-1})$ & $-1.017 (^{+56}_{-57})$\\
$\log(\hat{b}_1(    0.02\,\text{GeV}^2)/\text{GeV}^{-1})$ & $-0.874 (^{+91}_{-89})$\\
$\log(\hat{b}_1(    0.08\,\text{GeV}^2)/\text{GeV}^{-1})$ & $-1.032 (^{+71}_{-75})$\\
$\log(\hat{b}_1(     0.4\,\text{GeV}^2)/\text{GeV}^{-1})$ & $-1.574 (^{+48}_{-47})$\\
$\log(\hat{b}_1(       2\,\text{GeV}^2)/\text{GeV}^{-1})$ & $-2.871 (^{+34}_{-33})$\\
$\log(\hat{b}_1(      10\,\text{GeV}^2)/\text{GeV}^{-1})$ & $-4.668 (70)$\\
$\log(\hat{b}_1(      50\,\text{GeV}^2)/\text{GeV}^{-1})$ & $-7.87 (29)$\\
\hline
\end{tabular}
  }
  \caption{
    Parameters obtained in the fit to HERA DIS and photoproduction data.
    The uncertainties on the
    least significant digits, determined using the MINOS
    algorithm, are indicated in brackets. Here log is understood as the natural 
    logarithm, that is, to base $e$. 
\label{tab4}
  }
\end{table}
The intercept parameter $\epsilon_1=0.0935 (^{+76}_{-64})$ of the soft pomeron
exchange is compatible with independent extractions, for example with
measurements of the pomeron trajectory from hadronic reactions 
(see \cite{Donnachie:2002en} for a review) and from 
$\rho$ photoproduction data \cite{List:2009pb}.
The spline coefficients characterizing the functions $\hat{b}_j$ are
summarised in table \ref{tab:splineparam}, with their Hessian uncertainties 
(cf.\ \eqref{eqn:errorprop1}, \eqref{eqn:errorprop2}). 
\begin{table}
  \centering
  {
    \small
  \renewcommand{\arraystretch}{1.5}
\begin{tabular}{|l|c|cccc|}
\hline
 $i$ & $q^2_{0,i}$ & $A_{0,i}$ & $B_{0,i}$ & $C_{0,i}$ & $D_{0,i}$\\
\hline
$1$ & $       0\,\text{GeV}^2$ & $-14.2 (39)$ & $12.0 (61)$ & $$0$$ & $-3.5 (22)$\\
$2$ & $     0.3\,\text{GeV}^2$ & $-7.0 (12)$ & $6.9 (31)$ & $-7.4 (46)$ & $2.7 (22)$\\
$3$ & $       1\,\text{GeV}^2$ & $-4.83 (48)$ & $0.37 (46)$ & $-1.09 (66)$ & $0.43 (30)$\\
$4$ & $       3\,\text{GeV}^2$ & $-5.09 (36)$ & $-0.55 (20)$ & $0.11 (23)$ & $-0.060 (90)$\\
$5$ & $      10\,\text{GeV}^2$ & $-5.67 (30)$ & $-0.54 (12)$ & $-0.10 (13)$ & $-0.041 (61)$\\
$6$ & $      25\,\text{GeV}^2$ & $-6.27 (25)$ & $-0.822 (78)$ & $-0.210 (90)$ & $0.102 (44)$\\
\hline
 $7$ & $      50\,\text{GeV}^2$ &\multicolumn{2}{r}{$n_{0,7}=-0.967 (73)$} & & \\
\hline
\end{tabular}\\
\vspace*{5mm}
\begin{tabular}{|l|c|cccc|}
\hline
 $i$ & $q^2_{1,i}$ & $A_{1,i}$ & $B_{1,i}$ & $C_{1,i}$ & $D_{1,i}$\\
\hline
$1$ & $       0\,\text{GeV}^2$ & $-1.02 (18)$ & $0.29 (47)$ & $$0$$ & $-0.17 (27)$\\
$2$ & $    0.02\,\text{GeV}^2$ & $-0.87 (26)$ & $0.04 (12)$ & $-0.35 (57)$ & $0.13 (30)$\\
$3$ & $    0.08\,\text{GeV}^2$ & $-1.03 (24)$ & $-0.28 (22)$ & $0.01 (27)$ & $-0.056 (83)$\\
$4$ & $     0.4\,\text{GeV}^2$ & $-1.57 (16)$ & $-0.59 (11)$ & $-0.23 (12)$ & $0.053 (45)$\\
$5$ & $       2\,\text{GeV}^2$ & $-2.87 (12)$ & $-0.925 (84)$ & $0.02 (11)$ & $-0.090 (58)$\\
$6$ & $      10\,\text{GeV}^2$ & $-4.67 (21)$ & $-1.55 (24)$ & $-0.41 (20)$ & $0.085 (42)$\\
\hline
 $7$ & $      50\,\text{GeV}^2$ &\multicolumn{2}{r}{$n_{1,7}=-2.21 (52)$} & & \\
\hline
\end{tabular}
  }
  \caption{
    Spline parameters characterizing the coupling functions $\hat{b}_j$
    obtained in the fit to HERA and photoproduction data; see \eqref{D.4}. 
    The Hessian uncertainties on the
    two least significant digits are indicated in brackets.
    The quantities $n_{0,7}$ and $n_{1,7}$ determine the large-$Q^2$ 
    behaviour of the coupling functions $\hat{b}_0(Q^2)$ and $\hat{b}_1(Q^2)$, 
    respectively, in the extrapolation region $Q^2 \ge 50 \,\mbox{GeV}^2$; 
    see \eqref{D.5} and \eqref{D.6}. 
  }
  \label{tab:splineparam}
\end{table}
The coupling functions $\hat{a}_j(Q^2)$, $Q^2\hat{a}_j(Q^2)$ and
$\hat{b}_j(Q^2)$ are shown in figures \ref{fig11} to \ref{fig14}. 
The $\hat{a}_j$ are not constrained very well by the
data. The function  $\hat{a}_0$ is poorly known at low
$Q^2\lesssim 2\,\text{GeV}^2$, while $\hat{a}_1$ has
large uncertainty at large $Q^2\gtrsim 5\,\text{GeV}^2$.
The functions $\hat{b}_j$ are much better constrained by data.
The coupling function $\hat{b}_1$ of the soft pomeron is well 
measured over the whole kinematic range investigated here. 
The determination of the coupling function $\hat{b}_0$ of the 
hard pomeron suffers
from increasing experimental uncertainties at very low
$Q^2\lesssim 0.3\,\text{GeV}^2$.
In that kinematic region the DIS cross section is governed by the soft
contribution in the experimentally accessible $W$ range.

It is interesting to observe that the two functions $\hat{b}_j$ each
reach a maximum at some positive $Q^2$ as shown in figure
\ref{fig14}. For $\hat{b}_0$ the maximum is
at $Q^2=1.27\, (^{+29}_{-30})\,\text{GeV}^{2}$ with amplitude 
$\hat{b}_0=0.0082\,(^{+39}_{-36})\,\text{GeV}^{-1}$.
For $\hat{b}_1$ it is at $Q^2=0.0225\,(^{+57}_{-59})\,\text{GeV}^{2}$
with amplitude $\hat{b}_1=0.42 \,(11)\,\text{GeV}^{-1}$.
However, experimental data are sparse in the vicinity of the maximum 
of $\hat{b}_1$, so the experimental evidence for such a maximum is not very strong. 
From the theory point of view such a behaviour of $\hat{b}_0(Q^2)$ 
and $\hat{b}_1(Q^2)$ is easy to understand. $\hat{b}_0(Q^2)$ is 
essentially zero at $Q^2=0$ and must fall with $Q^2$ for large $Q^2$; 
see \eqref{2.13}. Thus it must have a maximum somewhere and it is 
reasonable that this comes out in the $Q^2 \approx 1\,\mbox{GeV}^2$ region. 
For $\hat{b}_1(Q^2)$ we observe that it governs $\sigma_T + \sigma_L$ 
for small $Q^2$; see \eqref{2.11}, \eqref{2.12}. But $\sigma_L$ starts 
proportional to $Q^2$ for $Q^2$ increasing from zero. 
For larger $Q^2$ the soft contribution to $\sigma_T + \sigma_L$ 
will fall with $Q^2$ increasing. Thus, if the initial rise with $Q^2$ 
in $\sigma_L$ is not immediately compensated by a fall in $\sigma_T$ 
we expect a maximum for $\hat{b}_1(Q^2)$. 

The fit results shown in
table \ref{tab4} indicate that the hard pomeron contribution to the
photoproduction cross section, proportional to 
$\hat{b}_0(Q^2=0)$, is compatible with zero, such that
there is no evidence for a non-zero contribution of the hard
component to the photoproduction cross section in the energy range
investigated here. We further observe that the $f_{2R}$ reggeon contributes 
visibly only to the low-$W$ photoproduction data. 

A comparison of the fit results to photoproduction data is shown in 
figure \ref{fig:comp0}. The data are well described by the fit. 

\section{Alternative fits}
\label{appalt}

In this section, alternative fits are studied. In this way we want to check the 
stability of our results under changes of the assumptions entering the fits. 

\subsection{Fit with xFitter}

To cross-check the results obtained with the nominal fit discussed in 
the main text, a fit using the xFitter package \cite{Alekhin:2014irh,xfitterpage} 
is performed. For this purpose, the tensor 
pomeron model, as described in this paper, has been implemented and will 
be included in future releases of the package. 
Similarly to the nominal analysis, the $Q^2$ dependence of the $\hat{b}_j(Q^2)$ 
functions is parametrised using cubic spline functions, however with five 
instead of seven knots compared to the nominal fit. Due to the reduced number 
of spline knots, the total number of free parameters is 21 instead of 25 for the nominal fit.

The fit is performed to the same data sample, with the same kinematic cuts 
as in the nominal analysis. The goodness-of-fits function is taken as in 
\cite{Abramowicz:2015mha}, which differs from the one given in equation 
\eqref{d1} in the treatment of statistical uncertainties, that are considered to follow Poisson 
distribution. Given that for the fitted phase space the statistical uncertainties 
are small compared to the systematic ones, this difference should have a small impact 
on the result. The minimisation is performed using Minuit \cite{James:1975dr} 
while the evaluation of uncertainties uses an improved method introduced 
in \cite{Pumplin:2000vx}.

The fit yields results comparable to the nominal analysis. The quality of 
the fit is good with $\chi^2/N_{DF}= 595/(561-21)$, corresponding 
to a $p$-value of $5\%$. The values of the main parameters are summarised 
in table \ref{tab:xfit}. They are similar to the nominal fit.
\begin{table}
  \centering
  {
    \small
  \renewcommand{\arraystretch}{1.5}
\begin{tabular}{|lr|}
\hline
 fit parameter & result \\
\hline
$\epsilon_0$ & $0.3067 (71)$\\
$\epsilon_1$ & $0.0831 (70)$\\
$\alpha_2(0)$ & $0.394 (78)$\\
\hline
\end{tabular}
\hspace*{5mm}
 \caption{Parameter values obtained in an alternative xFitter fit for the pomeron 
  intercept parameters and the reggeon intercept. \label{tab:xfit}}
}
\end{table}

\subsection{Fit without hard pomeron in photoproduction}

The nominal fit with $25$ parameters indicates that the hard component
$\hat{b}_0(Q^2)$ vanishes for $Q^2\to 0$. A fit is performed where the spline 
knot at $Q^2=0$ is moved to  $q^2_{01}=0.1\,\text{GeV}^2$ and the offset
is set to zero, $q^2_{00}=0$. In the region below the new first knot
$q^2_{01}$, the function $\hat{b}_0(Q^2)$ is extrapolated using
equation \eqref{D.7}. For $Q^2=0$ it is set to zero.
This fit results in a goodness-of-fit $\chi^2=587.90$, very similar to that of the
default $25$-parameter fit presented in table \ref{tab3}. 
There is no significant change to any of the fit parameters.

\begin{boldmath}
\subsection{Studies of the ratio $R$}
\end{boldmath}

\begin{figure}
\begin{center}
   \includegraphics[width=0.7\textwidth]{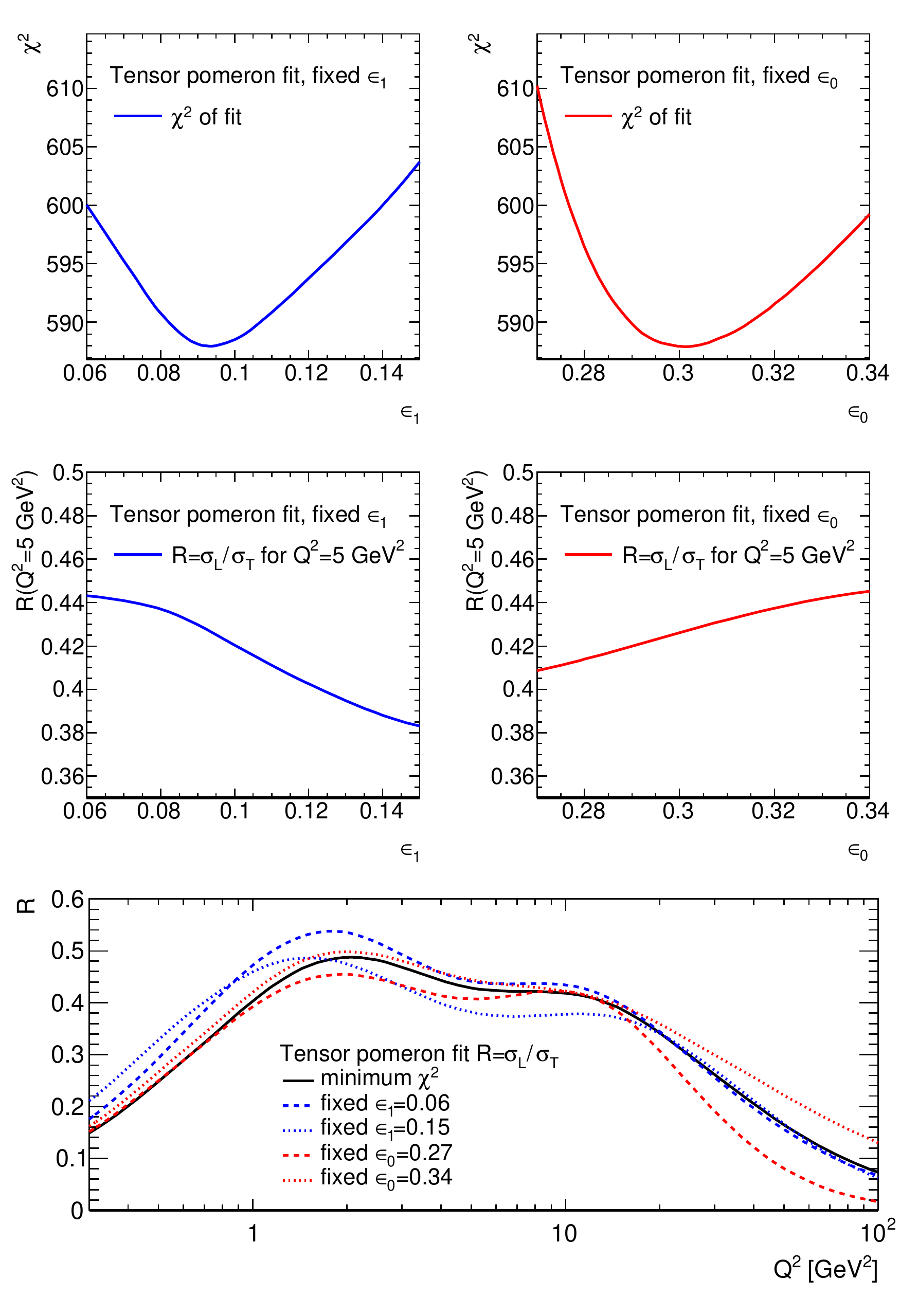}
   \caption{
     The goodness-of-fit $\chi^2$ and the ratio $R=\sigma_L/\sigma_T$
     of longitudinal to transverse 
     cross sections are studied in fits with $24$ free parameters as a
     function of $\epsilon_0$ and $\epsilon_1$.
     The upper panels show $\chi^2$ as a function of $\epsilon_1$
     (left) and $\epsilon_0$ (right).
     The middle panels show $R(Q^2=5\,\text{GeV}^2)$ as a function of
     $\epsilon_1$ (left) and $\epsilon_0$ (right).
     The lower panel shows the $R$ distribution as a function of $Q^2$
     for extreme choices of the $\epsilon_j$. For this study, the
     energy $W$ is set to $200\,\text{GeV}$.
   \label{fig:Reps0eps1}
   }
\end{center}
\end{figure}

The ratio $R$ determined in the $25$-parameter fit is found to be
above $0.4$ in a range of $Q^2$ from about
$1\,\text{GeV}^2$ to about $10\,\text{GeV}^2$. The magnitude of $R$ is 
strongly correlated to the parameters describing the functions $\hat{a}_j(Q^2)$.
However, there are also correlations to other parameters, most notably
to the slopes $\epsilon_j$. Fits with fixed $\epsilon_0$ or
$\epsilon_1$ 
have been performed to study the impact on $\chi^2$ and $R$; see fig.\ 
\ref{fig:Reps0eps1}. 
The scans cover large parameter ranges with a
goodness-of-fit up to and above $\chi^2=600$, corresponding to
parameter variations by more than three standard deviations. The
resulting $R$, however, is not affected by so much.
Thus, with all necessary caution, we think we can say that the 
HERA data, fitted with our two-pomeron model, prefer a relatively 
large value for $R$ in the above $Q^2$ range.


\begin{thebibliography}{99}

%\cite{Adloff:1997mf}
\bibitem{Adloff:1997mf}
  C.~Adloff {\it et al.} [H1 Collaboration],
  {\em A Measurement of the proton structure function $F_2 (x, Q^2)$ at low x and low $Q^2$ at HERA}, 
  Nucl.\ Phys.\ B {\bf 497} (1997) 3
  [hep-ex/9703012].

%\cite{Breitweg:1997hz}
\bibitem{Breitweg:1997hz}
  J.~Breitweg {\it et al.} [ZEUS Collaboration],
  {\em Measurement of the proton structure function $F_2$ and $\sigma_{\rm tot} (\gamma^* p)$ at low $Q^2$ and very low x at HERA}, 
  Phys.\ Lett.\ B {\bf 407} (1997) 432
  [hep-ex/9707025].

%\cite{Devenish:2004pb}
\bibitem{Devenish:2004pb}
  R.~Devenish and A.~Cooper-Sarkar,
  {\em Deep inelastic scattering}, 
  Oxford University Press (2004). 

%\cite{Donnachie:2002en}
\bibitem{Donnachie:2002en}
  A.~Donnachie, H.~G.~Dosch, P.~V.~Landshoff and O.~Nachtmann,
  {\sl Pomeron physics and QCD}, 
  Camb.\ Monogr.\ Part.\ Phys.\ Nucl.\ Phys.\ Cosmol.\  {\bf 19} (2002) 1.
  %%CITATION = CMPCE,19,1;%%

%\cite{ref7}
\bibitem{ref7}
  L.~Caneschi (ed.), 
  {\sl Regge Theory of Low-$p_t$ Hadronic Interactions}, 
  Elsevier Science Publishers B.~V., Amsterdam 1989. 

%\cite{ref8}
\bibitem{ref8}
M.~Haguenauer, B.~Nicolescu, and J.~Tran Thanh Van (eds.), 
Proc.\ XIth International Conference on Elastic and Diffractive Scattering, 
{\sl Towards High Energy Frontiers}, Th\^{e} Gi\^{o}i Publishers, Vietnam, 2006. 

%\cite{Barone:2002cv}
\bibitem{Barone:2002cv}
  V.~Barone and E.~Predazzi, {\em {High-Energy Particle Diffraction},}
  Springer-Verlag, Berlin, Heidelberg, 2002. 

%\cite{Donnachie:1998gm}
\bibitem{Donnachie:1998gm}
  A.~Donnachie and P.~V.~Landshoff,
  {\em Small x: Two pomerons!}, 
  Phys.\ Lett.\ B {\bf 437} (1998) 408
  [hep-ph/9806344].

%\cite{Donnachie:2001he}
\bibitem{Donnachie:2001he}
  A.~Donnachie and P.~V.~Landshoff,
  {\em New data and the hard Pomeron}, 
  Phys.\ Lett.\ B {\bf 518} (2001) 63
  [hep-ph/0105088].

%\cite{Donnachie:2004pi}
\bibitem{Donnachie:2004pi}
  A.~Donnachie and P.~V.~Landshoff,
  {\em Does the hard pomeron obey Regge factorisation?}, 
  Phys.\ Lett.\ B {\bf 595} (2004) 393
  [hep-ph/0402081].

%\cite{Ewerz:2013kda}
\bibitem{Ewerz:2013kda}
  C.~Ewerz, M.~Maniatis and O.~Nachtmann,
  {\em {A Model for Soft High-Energy Scattering: Tensor Pomeron and Vector Odderon},}  
  Annals Phys.\  {\bf 342} (2014) 31 
  [arXiv:1309.3478 [hep-ph]].

%\cite{Dosch:2015jua}
\bibitem{Dosch:2015jua}
  H.~G.~Dosch and E.~Ferreira,
  {\em {Scale Dependent Pomeron Intercept in Electromagnetic Diffractive Processes},} 
  arXiv:1503.06649 [hep-ph].

%\cite{Bolz:2014mya}
\bibitem{Bolz:2014mya}
  A.~Bolz, C.~Ewerz, M.~Maniatis, O.~Nachtmann, M.~Sauter and A.~Sch\"oning,
  {\em {Photoproduction of $\pi^{+} \pi^{-}$ pairs in a model with tensor-pomeron and vector-odderon exchange},} 
  JHEP {\bf 1501} (2015) 151
  [arXiv:1409.8483 [hep-ph]].

%\cite{Lebiedowicz:2013ika}
\bibitem{Lebiedowicz:2013ika}
  P.~Lebiedowicz, O.~Nachtmann and A.~Szczurek,
  {\em {Exclusive central diffractive production of scalar and pseudoscalar mesons; tensorial vs.\ vectorial pomeron},}
  Annals Phys.\  {\bf 344} (2014) 301 
  [arXiv:1309.3913 [hep-ph]].

%\cite{Lebiedowicz:2014bea}
\bibitem{Lebiedowicz:2014bea}
  P.~Lebiedowicz, O.~Nachtmann and A.~Szczurek, 
  {\em {$\rho^0$ and Drell-S{\"o}ding contributions to central exclusive production of $\pi^+
  \pi^-$ pairs in proton-proton collisions at high energies},}
  Phys.\ Rev.\ D {\bf 91} (2015) 074023
  [arXiv:1412.3677 [hep-ph]].

%\cite{Lebiedowicz:2016ioh}
\bibitem{Lebiedowicz:2016ioh}
  P.~Lebiedowicz, O.~Nachtmann and A.~Szczurek, 
  {\em {Central exclusive diffractive production of the $\pi^{+}\pi^{-}$ continuum, scalar and tensor
  resonances in $pp$ and $p \bar{p}$ scattering within the tensor pomeron approach},}
  Phys.\ Rev.\ D {\bf 93} (2016) 054015 
  [arXiv:1601.04537 [hep-ph]].

%\cite{Lebiedowicz:2016zka}
\bibitem{Lebiedowicz:2016zka}
  P.~Lebiedowicz, O.~Nachtmann and A.~Szczurek,
  {\em Exclusive diffractive production of $\pi^{+}\pi^{-}\pi^{+}\pi^{-}$ via 
    the intermediate $\sigma\sigma$ and $\rho\rho$ states in proton-proton 
    collisions within tensor pomeron approach}, 
  Phys.\ Rev.\ D {\bf 94} (2016) 034017
  [arXiv:1606.05126 [hep-ph]].

%\cite{Lebiedowicz:2016ryp}
\bibitem{Lebiedowicz:2016ryp}
  P.~Lebiedowicz, O.~Nachtmann and A.~Szczurek,
  {\em Central production of $\rho^{0}$ in $pp$ collisions with single proton diffractive dissociation at the LHC}, 
  Phys.\ Rev.\ D {\bf 95} (2017) 034036
  [arXiv:1612.06294 [hep-ph]].

%\cite{Klusek-Gawenda:2017lgt}
\bibitem{Klusek-Gawenda:2017lgt}
  M.~Klusek-Gawenda, P.~Lebiedowicz, O.~Nachtmann and A.~Szczurek,
  {\em From the $\gamma \gamma \to p \bar{p}$ reaction to the production of $p \bar{p}$ pairs in ultraperipheral ultrarelativistic heavy-ion collisions at the LHC}, 
  Phys.\ Rev.\ D {\bf 96} (2017) 094029
  [arXiv:1708.09836 [hep-ph]].

%\cite{Lebiedowicz:2018eui}
\bibitem{Lebiedowicz:2018eui}
  P.~Lebiedowicz, O.~Nachtmann and A.~Szczurek,
  {\em Towards a complete study of central exclusive production of $K^{+}K^{-}$ pairs in proton-proton collisions within the tensor Pomeron approach}, 
  Phys.\ Rev.\ D {\bf 98} (2018) 014001
  [arXiv:1804.04706 [hep-ph]].

%\cite{Ewerz:2016onn}
\bibitem{Ewerz:2016onn}
  C.~Ewerz, P.~Lebiedowicz, O.~Nachtmann and A.~Szczurek,
  {\em Helicity in proton-proton elastic scattering and the spin structure of the pomeron}, 
  Phys.\ Lett.\ B {\bf 763} (2016) 382
  [arXiv:1606.08067 [hep-ph]].

%\cite{Adamczyk:2012kn}
\bibitem{Adamczyk:2012kn}
  L.~Adamczyk {\it et al.} [STAR Collaboration],
  {\em Single Spin Asymmetry $A_N$ in Polarized Proton-Proton Elastic Scattering at $\sqrt{s}=$ 200 GeV}, 
  Phys.\ Lett.\ B {\bf 719} (2013) 62
  [arXiv:1206.1928 [nucl-ex]].

%\cite{Nachtmann:1990ta}
\bibitem{Nachtmann:1990ta}
  O.~Nachtmann,
  {\sl Elementary Particle Physics: Concepts and Phenomena}, 
  Springer Verlag, Berlin, 1990. 

%\cite{Hand:1963bb}
\bibitem{Hand:1963bb}
L.~N.~Hand,
{\em Experimental Investigation of Pion Electroproduction}, 
Phys.\ Rev.\  {\bf 129} (1963) 1834.

%\cite{Landau:1948kw}
\bibitem{Landau:1948kw}
  L.~D.~Landau,
  {\em On the angular momentum of a system of two photons}, 
  Dokl.\ Akad.\ Nauk Ser.\ Fiz.\  {\bf 60} (1948) 207.

%\cite{Yang:1950rg}
\bibitem{Yang:1950rg}
  C.~N.~Yang,
  {\em Selection Rules for the Dematerialization of a Particle Into Two Photons}, 
  Phys.\ Rev.\  {\bf 77} (1950) 242.

%\cite{Abramowicz:2015mha}
\bibitem{Abramowicz:2015mha}
  H.~Abramowicz {\it et al.} [H1 and ZEUS Collaborations],
  {\em Combination of measurements of inclusive deep inelastic ${e^{\pm }p}$ scattering cross sections and QCD analysis of HERA data}, 
  Eur.\ Phys.\ J.\ C {\bf 75} (2015) no.12,  580
  [arXiv:1506.06042 [hep-ex]].

%\cite{Aid:1995bz}
\bibitem{Aid:1995bz}
  S.~Aid {\it et al.} [H1 Collaboration],
  {\em Measurement of the total photon-proton cross-section and its decomposition at 200 GeV center-of-mass energy}, 
  Z.\ Phys.\ C {\bf 69} (1995) 27
  [hep-ex/9509001].

%\cite{Chekanov:2001gw}
\bibitem{Chekanov:2001gw}
  S.~Chekanov {\it et al.} [ZEUS Collaboration],
  {\em Measurement of the photon-proton total cross-section at a center-of-mass energy of 209 GeV at HERA}, 
  Nucl.\ Phys.\ B {\bf 627} (2002) 3
  [hep-ex/0202034].

%\cite{Vereshkov:2003cp}
\bibitem{Vereshkov:2003cp}
  G.~M.~Vereshkov, O.~D.~Lalakulich, Y.~F.~Novoseltsev and R.~V.~Novoseltseva,
  {\em Total cross section for photon-nucleon interaction in the energy range $\sqrt{s}$ = 40 GeV - 250 GeV}, 
  Phys.\ Atom.\ Nucl.\  {\bf 66} (2003) 565
   [Yad.\ Fiz.\  {\bf 66} (2003) 591].

%\cite{Caldwell:1978yb}
\bibitem{Caldwell:1978yb}
  D.~O.~Caldwell {\it et al.},
  {\em Measurements of the Photon Total Cross-Section on Protons from 18 GeV to 185 GeV}, 
  Phys.\ Rev.\ Lett.\  {\bf 40} (1978) 1222.

%\cite{Andreev:2013vha}
\bibitem{Andreev:2013vha}
  V.~Andreev {\it et al.} [H1 Collaboration],
  {\em Measurement of inclusive $e p$ cross sections at high $Q^2$ at $\sqrt s =$ 225 and 252 GeV and of the longitudinal proton structure function $F_L$ at HERA}, 
  Eur.\ Phys.\ J.\ C {\bf 74} (2014) 2814
  [arXiv:1312.4821 [hep-ex]].

%\cite{Gribov:1984tu}
\bibitem{Gribov:1984tu}
  L.~V.~Gribov, E.~M.~Levin and M.~G.~Ryskin,
  {\em Semihard Processes in QCD}, 
  Phys.\ Rept.\  {\bf 100} (1983) 1.

%\cite{Froissart:1961ux}
\bibitem{Froissart:1961ux}
  M.~Froissart,
  {\em Asymptotic behavior and subtractions in the Mandelstam representation}, 
  Phys.\ Rev.\  {\bf 123} (1961) 1053.

%\cite{Martin1966}
\bibitem{Martin1966}
  A.~Martin, 
  {\em Extension of the axiomatic analyticity domain of scattering amplitudes by unitarity - I}, 
  Nuovo Cim.\ A {\bf 42} (1966) 930.

%\cite{Lukaszuk:1967zz}
\bibitem{Lukaszuk:1967zz}
  L.~Lukaszuk and A.~Martin,
  {\em Absolute upper bounds for pi pi scattering}, 
  Nuovo Cim.\ A {\bf 52} (1967) 122.

%\cite{Nachtmann:2002yd}
\bibitem{Nachtmann:2002yd}
  O.~Nachtmann,
  {\em Effective field theory approach to structure functions at small $x_{\rm Bj}$}, 
  Eur.\ Phys.\ J.\ C {\bf 26} (2003) 579 
  [hep-ph/0206284].

%\cite{Ewerz:2006vd}
\bibitem{Ewerz:2006vd}
  C.~Ewerz and O.~Nachtmann,
  {\em Towards a nonperturbative foundation of the dipole picture: II. High energy limit}, 
  Annals Phys.\  {\bf 322} (2007) 1670
  [hep-ph/0604087].

%\cite{Ewerz:2006an}
\bibitem{Ewerz:2006an}
  C.~Ewerz and O.~Nachtmann,
  {\em Bounds on Ratios of DIS Structure Functions from the Color Dipole Picture}, 
  Phys.\ Lett.\ B {\bf 648} (2007) 279 
  [hep-ph/0611076].

%\cite{Collaboration:2010ry}
\bibitem{Collaboration:2010ry}
  F.~D.~Aaron {\it et al.} [H1 Collaboration],
  {\em Measurement of the Inclusive $e^{\pm}p$ Scattering Cross Section at High Inelasticity y and of the Structure Function $F_L$}, 
  Eur.\ Phys.\ J.\ C {\bf 71} (2011) 1579
  [arXiv:1012.4355 [hep-ex]].

%\cite{Ewerz:2012az}
\bibitem{Ewerz:2012az}
  C.~Ewerz, A.~von Manteuffel, O.~Nachtmann and A.~Sch\"oning,
  {\em The New $F_L$ Measurement from HERA and the Dipole Model}, 
  Phys.\ Lett.\ B {\bf 720} (2013) 181
  [arXiv:1201.6296 [hep-ph]].

%\cite{Yndurain1983}
\bibitem{Yndurain1983}
  F.~J.~Yndurain, {\em {Quantum Chromodynamics},}
  Springer-Verlag, New York, Heidelberg, 1983. 

%\cite{Gribov:1972ri}
\bibitem{Gribov:1972ri}
  V.~N.~Gribov and L.~N.~Lipatov,
  {\em Deep inelastic $e p$ scattering in perturbation theory}, 
  Sov.\ J.\ Nucl.\ Phys.\  {\bf 15} (1972) 438
   [Yad.\ Fiz.\  {\bf 15} (1972) 781].

%\cite{Altarelli:1977zs}
\bibitem{Altarelli:1977zs}
  G.~Altarelli and G.~Parisi,
  {\em Asymptotic Freedom in Parton Language}, 
  Nucl.\ Phys.\ B {\bf 126} (1977) 298.

%\cite{Dokshitzer:1977sg}
\bibitem{Dokshitzer:1977sg}
  Y.~L.~Dokshitzer,
  {\em Calculation of the Structure Functions for Deep Inelastic Scattering and $e^+ e^-$ Annihilation by Perturbation Theory in Quantum Chromodynamics}, 
  Sov.\ Phys.\ JETP {\bf 46} (1977) 641
   [Zh.\ Eksp.\ Teor.\ Fiz.\  {\bf 73} (1977) 1216].

%\cite{Donnachie:2001zt}
\bibitem{Donnachie:2001zt}
  A.~Donnachie and P.~V.~Landshoff,
  {\em Perturbative QCD and Regge theory: Closing the circle}, 
  Phys.\ Lett.\ B {\bf 533} (2002) 277
  [hep-ph/0111427].

%\cite{Kuraev:1977fs}
\bibitem{Kuraev:1977fs}
  E.~A.~Kuraev, L.~N.~Lipatov and V.~S.~Fadin,
  {\em The Pomeranchuk Singularity in Nonabelian Gauge Theories}, 
  Sov.\ Phys.\ JETP {\bf 45} (1977) 199
   [Zh.\ Eksp.\ Teor.\ Fiz.\  {\bf 72} (1977) 377].

%\cite{Balitsky:1978ic}
\bibitem{Balitsky:1978ic}
  I.~I.~Balitsky and L.~N.~Lipatov,
  {\em The Pomeranchuk Singularity in Quantum Chromodynamics}, 
  Sov.\ J.\ Nucl.\ Phys.\  {\bf 28} (1978) 822
   [Yad.\ Fiz.\  {\bf 28} (1978) 1597].

%\cite{Abt:2016vjh}
\bibitem{Abt:2016vjh}
  I.~Abt, A.~M.~Cooper-Sarkar, B.~Foster, V.~Myronenko, K.~Wichmann and M.~Wing,
  {\em Study of HERA ep data at low Q$^2$ and low $x_{Bj}$ and the need for higher-twist corrections to standard perturbative QCD fits}, 
  Phys.\ Rev.\ D {\bf 94} (2016) 034032
  [arXiv:1604.02299 [hep-ph]].

%\cite{Luszczak:2016bxd}
\bibitem{Luszczak:2016bxd}
  A.~Luszczak and H.~Kowalski,
  {\em Dipole model analysis of highest precision HERA data, including very low $Q^{2}$'s} 
  Phys.\ Rev.\ D {\bf 95} (2017) 014030
  [arXiv:1611.10100 [hep-ph]].

%\cite{Ball:2017otu}
\bibitem{Ball:2017otu}
  R.~D.~Ball, V.~Bertone, M.~Bonvini, S.~Marzani, J.~Rojo and L.~Rottoli,
  {\em Parton distributions with small-x resummation: evidence for BFKL dynamics in HERA data}, 
  Eur.\ Phys.\ J.\ C {\bf 78} (2018) 321 
  [arXiv:1710.05935 [hep-ph]].

%\cite{Abdolmaleki:2018jln}
\bibitem{Abdolmaleki:2018jln}
  H.~Abdolmaleki {\it et al.} [xFitter Developers' Team],
  {\em Impact of low-$x$ resummation on QCD analysis of HERA data}, 
  Eur.\ Phys.\ J.\ C {\bf 78} (2018) 621
  [arXiv:1802.00064 [hep-ph]].

%\cite{Ewerz:2004vf}
\bibitem{Ewerz:2004vf}
  C.~Ewerz and O.~Nachtmann,
  {\em Towards a nonperturbative foundation of the dipole picture: I. Functional methods}, 
  Annals Phys.\  {\bf 322} (2007) 1635
  [hep-ph/0404254].

%\cite{Ewerz:2011ph}
\bibitem{Ewerz:2011ph}
  C.~Ewerz, A.~von Manteuffel and O.~Nachtmann,
  {\em On the Energy Dependence of the Dipole-Proton Cross Section in Deep Inelastic Scattering}, 
  JHEP {\bf 1103} (2011) 062
  [arXiv:1101.0288 [hep-ph]].

%\cite{Accardi:2012qut}
\bibitem{Accardi:2012qut}
  A.~Accardi {\it et al.},
  {\em Electron Ion Collider: The Next QCD Frontier : Understanding the glue that binds us all}, 
  Eur.\ Phys.\ J.\ A {\bf 52} (2016) 268
  [arXiv:1212.1701 [nucl-ex]].

%\cite{AbelleiraFernandez:2012cc}
\bibitem{AbelleiraFernandez:2012cc}
  J.~L.~Abelleira Fernandez {\it et al.} [LHeC Study Group],
  {\em A Large Hadron Electron Collider at CERN: Report on the Physics and Design Concepts for Machine and Detector}, 
  J.\ Phys.\ G {\bf 39} (2012) 075001
  [arXiv:1206.2913 [physics.acc-ph]].

\bibitem{Alpos}
  D.~Britzger {\it et al.}, 
  {\em The ALPOS fit framework}, available at \hfill \linebreak 
  \url{http://www.desy.de/~britzger/alpos/}.

%\cite{James:1975dr}
\bibitem{James:1975dr}
  F.~James and M.~Roos,
  {\em Minuit: A System for Function Minimization and Analysis of the Parameter Errors and Correlations}, 
  Comput.\ Phys.\ Commun.\  {\bf 10} (1975) 343.

%\cite{List:2009pb}
\bibitem{List:2009pb}
  B.~List for the H1 Collaboration, 
  {\em Extraction of the Pomeron Trajectory from a Global Fit to Exclusive $\rho^0$ Meson Photoproduction Data}, 
  in Proceedings of the XVII International Workshop on Deep Inelastic Scattering and Related Subjects DIS 2009, Madrid, Spain, April 26-30, 2009, 
  arXiv:0906.4945 [hep-ex].

%\cite{Alekhin:2014irh}
\bibitem{Alekhin:2014irh}
  S.~Alekhin {\it et al.},
  {\em HERAFitter}, 
  Eur.\ Phys.\ J.\ C {\bf 75} (2015) 304
  [arXiv:1410.4412 [hep-ph]].

\bibitem{xfitterpage}
\url{https://www.xfitter.org/xFitter/}

%\cite{Pumplin:2000vx}
\bibitem{Pumplin:2000vx}
  J.~Pumplin, D.~R.~Stump and W.~K.~Tung,
  {\em Multivariate fitting and the error matrix in global analysis of data}, 
  Phys.\ Rev.\ D {\bf 65} (2001) 014011
  [hep-ph/0008191].

\end{thebibliography}
\end{document}